\documentclass[aps,prb,reprint,superscriptaddress,longbibliography]{revtex4-2}

\usepackage{amsmath}
\usepackage{amssymb}

\usepackage{graphicx}
\usepackage{placeins}
\usepackage{bm}
\usepackage{hyperref}
\usepackage{xcolor}
\usepackage[normalem]{ulem} 
\usepackage{array}
\usepackage{booktabs}

\usepackage{pifont}% http://ctan.org/pkg/pifont
\newcommand{\cmark}{\ding{51}}%
\newcommand{\xmark}{\ding{55}}%

\usepackage{multirow}
\usepackage{comment}
\usepackage{orcidlink}

\renewcommand{\arraystretch}{1.15}

\begin{document}

\title{X-ray magnetic circular dichroism in altermagnets: insights from symmetry}

\author{Hana Schiff \orcidlink{0009-0001-0418-739X}}
\affiliation{Department of Physics and Astronomy, University of California, Irvine, California 92697, USA}

\author{Ishmam Nihal}
\affiliation{Center for Memory and Recording Research, University of California San Diego, La Jolla, CA 92093, USA}
\affiliation{Advanced Light Source, Lawrence Berkeley National Laboratory, Berkeley, California 94720, USA}

\author{John Vinson \orcidlink{0000-0002-7619-7060}} 
\affiliation{Material Measurement Laboratory, National Institute of Standards and Technology, Gaithersburg, Maryland 20899, United States}

\author{Robin Msiska}
\affiliation{Center for Memory and Recording Research, University of California San Diego, La Jolla, CA 92093, USA}
\affiliation{Theoretical Division, Los Alamos National Laboratory, Los Alamos, NM 87545, USA}

\author{Judit Romh\'anyi \orcidlink{0000-0002-4642-7734}}
\affiliation{Department of Physics and Astronomy, University of California, Irvine, California 92697, USA}

\author{David Lederman \orcidlink{0000-0001-8423-5138}}
\affiliation{Department of Physics, University of California, Santa Cruz, California 95064, USA}
\affiliation{Materials Science and Engineering Program, University of California, Santa Cruz, California 95064, USA}

\author{Christoph Klewe}
\affiliation{Advanced Light Source, Lawrence Berkeley National Laboratory, Berkeley, California 94720, USA}

\author{Sujoy Roy}
\affiliation{Advanced Light Source, Lawrence Berkeley National Laboratory, Berkeley, California 94720, USA}

\author{Eric E. Fullerton}
\affiliation{Department of Electrical and Computer Engineering, University of California San Diego, La Jolla, CA 92093, USA}
\affiliation{Center for Memory and Recording Research, University of California San Diego, La Jolla, CA 92093, USA}

\date{\today}

\begin{abstract}
Altermagnets are a unique subset of antiferromagnets possessing ferromagnet-like responses that are odd in the N\'eel vector, making X-ray magnetic circular dichroism (XMCD) a natural probe of ground state altermagnetic order. Because the relativistic ingredient of XMCD at the $L$-edge is supplied predominantly by core-level spin--orbit coupling (SOC), the dichroic signal can reflect the non-relativistic order of the valence electrons. 
Here we develop a complete symmetry framework for XMCD in altermagnets (AMs) preserving the crystalline unit cell. We tabulate the symmetry-allowed optical Hall vector, anisotropic magnetic dipole (AMD), and underlying octupolar components for all magnetic point groups, thereby identifying the experimental geometries that maximize the XMCD signal. We show that only twelve magnetic point groups, none with a rotation axis of order three or higher, are compatible with both altermagnetism and finite XMCD. By extending the Landau theory of collinear AMs, we connect these relativistic constraints to the spin--orbit-free order. We show that the AMD is forced to vanish at zero valence SOC in conventional antiferromagnets, and that every AM sublattice structure allows the Hall vector and AMD at linear or cubic order in the N\'eel vector; whether the resulting XMCD is nonzero depends on the N\'eel vector orientation at finite SOC. The Landau theory thus fixes the N\'eel-vector dependence of the XMCD for all AM candidates, and separates $d$-wave AMs, whose signal is inherited from a non-relativistic AMD, from $g$- and $i$-wave AMs, whose signal requires valence SOC or core--valence multiplet effects. We illustrate these predictions with first-principles Bethe--Salpeter calculations of $L$-edge XMCD for five representative altermagnetic materials and one conventional antiferromagnet.
\end{abstract}

\maketitle

\section{Introduction}\label{sec:intro}
The discovery of antiferromagnetism by Louis N\'eel garnered little attention for practical application at the time, with N\'eel himself noting in his 1970 Nobel lecture that antiferromagnets (AFMs), though of great theoretical interest, ``do not appear to have any practical applications''~\cite{neel1972magnetism}. Decades later, this verdict has been dramatically overturned: AFMs have re-emerged as leading candidates for next-generation spintronic devices. Their zero net magnetization is now recognized as a powerful asset, enabling dense, cross-talk-free memory architectures. The insensitivity of AFMs to external magnetic fields makes AFM devices inherently robust against unwanted external magnetic perturbation. Further, their intrinsic terahertz-frequency spin dynamics promise data processing speeds orders of magnitude faster than their ferromagnetic counterparts. These properties have positioned AFM spintronics as a compelling successor to conventional ferromagnet-based technology~\cite{jungwirth2016antiferromagnetic, dal2024_afmSpintronics,Hou2019_afmSpintronics}. 

The same zero net magnetization that makes AFMs attractive for spintronics applications also makes them difficult to detect and manipulate. Recently, a distinct subset of collinear AFMs termed altermagnets (AMs) was theoretically predicted~\cite{Noda2016, Okugawa2018, Naka2019, Hayami2019, Yuan2020_SpinSplittingAFM, Smejkal2020, smejkal2022emerging, libor2022altermagnet} and subsequently observed experimentally~\cite{fedchenko2024ruo2,krempasky2024altermagnet,reimers2024CrSb,Ding2024_CrSbARPES}. Altermagnets offer the potential to detect and manipulate the N\'eel order while retaining antiferromagnetic advantages \cite{MazinPunchline2022}. The goal of this study is to assist experimental detection of altermagnetism, to advance the search for AM material candidates.

In conventional N\'eel AFMs, opposite-spin sublattices are related by either spatial inversion ($\mathcal{I}$) or a lattice translation. Thus, the compositions of these transformations with time-reversal  ($\mathcal{T}$) are symmetries, and they enforce spin-degenerate bands throughout the Brillouin zone. Collinear AMs break both of these symmetries, with opposite-spin sublattices related by a rotation or mirror operation, enforcing compensation. In the absence of the relativistic spin--orbit coupling (SOC) interaction, these  symmetries generate an even-parity spin splitting throughout the Brillouin zone, driven by exchange interactions. Real materials possess some degree of SOC; but when SOC is weak, the non-relativistic model serves as a good approximation~\cite{libor2022altermagnet, smejkal2022emerging, mcclarty2024landau, schiff2025collinear, song2025altermagnets}. 

Unlike conventional Néel AFMs, the opposite-spin sublattices of AMs are \emph{not} identical, manifest in the rotations and mirrors relating them. This is often a consequence of alternating structural motifs. Such motifs are ubiquitous, appearing through Jahn--Teller distortions, alternating bond lengths, or octahedral tilting. Because the crystal structure sets the magnetic exchange interactions, the AFM order tends to alternate in step with the structural motif. Consequently, many collinear AFMs turn out to abide by altermagnetic symmetries --- even if the resulting spin splitting is not always large~\cite{spaldin2026altermagnet}. Although the term \emph{altermagnet} is recent, this essential structural ingredient, compensated sublattices that are crystallographically inequivalent and not related by inversion or translation, was recognized decades ago \cite{Alben1975}. 

A particularly convenient feature of altermagnets is that they exhibit a range of  responses historically attributed only to ferromagnets, despite their vanishingly small net magnetization, including the anomalous Hall effect, magneto-optical effects such as the Kerr effect, and X-ray magnetic circular dichroism (XMCD)~\cite{Smejkal2020, song2025altermagnets,hariki2025separating,MazinPunchline2022,schiff2025collinear}. These responses all reverse sign when the N\'eel vector is reversed.
Among them, XMCD is a particularly promising probe of the ground state, applicable across all
metallic, semiconducting, and
insulating AM classes~\cite{hariki2024MnTe}. 

The appeal of XMCD rests on a distinctive
feature of the SOC that enters the absorption process. Unlike the anomalous Hall effect and the Kerr effect, whose $\mathcal{T}$-odd response vanishes by symmetry in the absence of valence SOC, the relativistic ingredient required for XMCD is supplied by the strong \textit{core-level} SOC
--- the large $2p_{1/2}$--$2p_{3/2}$ splitting underlying the $L_{2,3}$
edges~\cite{stohr1999xmcd,hariki2025separating,deGroot}. The core and valence levels are naturally separated, and the valence SOC at the $L_{2,3}$ edges is much smaller than that of the core levels. Thus, XMCD predominantly reflects non-relativistic order of the valence electrons~\cite{hariki2024rutile_MnF2_XMCD}. 

We briefly review two routes to predicting an XMCD response. The first route is through the optical Hall vector, $\bm{h}(\omega)$ which is built from the antisymmetric part of the optical conductivity tensor. It represents the finite-frequency, magneto-optical analog of the static anomalous Hall response~\cite{fedchenko2024ruo2}. For circularly polarized light of frequency $\omega$ propagating along the direction $\hat{\bm{k}}$,
\begin{equation}
\mathrm{XMCD}(\omega) \propto 2\,\bm{h}(\omega)\cdot\hat{\bm{k}}.
\end{equation}
Thus, $\mathbf{h}(\omega)$ specifies the light propagation direction which maximizes the XMCD signal~\cite{hariki2025separating,kunes2025xmcd}. The Hall vector is highly sensitive to the sample symmetry. It is a time-odd axial quantity, just like the net moment, and so symmetry will either allow both simultaneously, or neither. In altermagnets, this net moment may be arbitrarily small \cite{Smejkal2020}, while $\bm{h}(\omega)$ remains finite. %It is highly sensitive to the sample symmetry and is always allowed concurrently with a net moment, which may be arbitrarily small, as is the case for some altermagnets 

Through the XMCD sum rules, on the other hand, the integrated dichroic response is governed by the net spin moment, $\langle \bm{S} \rangle$, and the anisotropic magnetic dipole (AMD), $\langle \bm{T}\rangle$. With the quantization direction $z$ taken to be the direction of photon propagation $\hat{\bm{k}},$
\begin{equation}\label{eq:xmcd_ST}
   \frac{1}{\Sigma_{0}} \left(\int_{L_{3}} \mu \,d\omega - 2\int_{L_{2}}\mu\, d\omega \right) \propto \left(\frac{2}{3} \langle S_{z} \rangle + \frac{7}{3} \langle T_{z}\rangle \right),
\end{equation}
where $\mu~\equiv~\mu^{+} - \mu^{-}$, the difference between left ($\mu^{+}$) and right ($\mu^{-})$ polarization absorption, denotes the XMCD spectrum, while $\Sigma_{0}~=~\int_{L_{2,3}}~(\mu^{+} + \mu^{0} + \mu^{-})\,d\omega$, where $\mu^{0}$ denotes the linear polarization absorption, gives the total X-ray absorption \cite{Carra1993_XMCD_ST,Crocombette1996_XMCD_AMD_3dTM,vanDerLaan2014_XMCD_Review,Yamasaki2025_sumRules_multipoleBasis,Wu1994_AMD_Importance,oguchi2004t}. Thus, Eq.~\ref{eq:xmcd_ST} demonstrates that XMCD may be finite even for systems with vanishing net moment, due to the AMD. 

%The quantization axis $z$ in Eq.~\ref{eq:xmcd_ST} is the direction of photon propagation, $\hat{\mathbf{k}}.$ Thus, the integrated XMCD intensity is maximized by aligning the direction of photon propagation along the allowed directions of $\bf{S}$ and $\bm{T}$.

Here, $\bm{S} = \sum_{a} \bm{S}^{(a)}$ and $\bm{T} = \sum_{a} \bm{T}^{(a)}$ are the net spin moment and AMD, obtained as sums over all sites $a$ in the unit cell. At each site, the components $T_{\alpha}^{(a)}$ (for Cartesian components \mbox{$\alpha \in \{x,y,z\}$}) of $\bm{T}^{(a)}$ are expressed in terms of the quadrupole tensor $\bm{Q}^{(a)}$ and the spin operator $\bm{S}^{(a)}$ as 
\begin{equation}\label{eq:T_definition}
T^{(a)}_{\alpha} = \sum_{\beta} Q^{(a)}_{\alpha\beta}S^{(a)}_{\beta} \equiv \sum_{\beta} \mathfrak{T}_{\alpha\beta\beta}^{(a)},
\end{equation}
where $Q_{\alpha\beta} = \delta_{\alpha\beta} - 3\hat{r}_{\alpha}\hat{r}_{\beta}$ is a real-space quadrupole~\cite{Carra1993_XMCD_ST,vanDerLaan2014_XMCD_Review,oguchi2004t,Kurita2024_XMCD_AMD_AFM}, and $\mathfrak{T}^{(a)}_{\alpha\beta\gamma}~\equiv~Q_{\alpha\beta}^{(a)}S_{\gamma}^{(a)}$ may be viewed as an octupole on site $a$. Since $\bm{Q}^{(a)}$ measures the departure of the charge distribution surrounding site $a$ from spherical symmetry, $\langle\bm{T}\rangle$ characterizes net anisotropy of the spin density distribution. %Although AMs may have vanishingly  small $\langle \bm{S} \rangle$ at finite SOC, they may nonetheless sustain a finite $\langle \bm{T}\rangle$ inherited from their non-relativistic symmetries and generate an XMCD response. 
That $\langle \bm{T} \rangle$ can be finite with vanishing $\langle \bm{S} \rangle$ in AMs is intuitive from the inequivalent but symmetry-related environments of their opposite-spin sublattices. This AMD-driven mechanism has been established for collinear AMs under a trigonal crystal field~\cite{sasabe2025t, xie2026fe2o3,ishii2026_Fe2O3_FMDistinctFromAM}, for rutile AMs \cite{hariki2024rutile_MnF2_XMCD,li2026_NiF2_XMCD,Sasabe2023_AMD_XMCD_RuO2}, and has recently been explored in non-collinear AFMs \cite{Kurita2024_XMCD_AMD_AFM}. 

The presence of an XMCD signal has been predicted and used to detect altermagnetism in several materials, including NiF$_2$~\cite{hariki2025separating,li2026_NiF2_XMCD}, RuO$_{2}$ \cite{Sasabe2023_AMD_XMCD_RuO2}, MnF$_{2}$ \cite{hariki2024rutile_MnF2_XMCD}, $\alpha$-MnTe~\cite{hariki2024MnTe, sasabe2025t,Amin2024_MnTe_XMCD_DomainAM}, CrSb~\cite{li2025CrSb}, and $\alpha$-Fe$_2$O$_3$~\cite{xie2026fe2o3,ishii2026_Fe2O3_FMDistinctFromAM}. The conditions on the N\'eel-vector orientation required to observe an XMCD signal in rutile-structure altermagnets and in structures with a trigonal crystal field have also been demonstrated in~\cite{hariki2024rutile_MnF2_XMCD} and ~\cite{sasabe2025t}, respectively. A general description of the dependence of the Hall vector on the N\'eel-vector orientation in AMs has also been derived within the free valence spin (FVS) approximation~\cite{kunes2025xmcd}. %While the weak valence SOC can induce a small ferromagnetic moment, it contributes only marginally to the XMCD spectra of $3d$ elements~\cite{thole1992xmcd,stohr1999xmcd,hariki2024rutile_MnF2_XMCD}. It has been recently demonstrated that this residual ferromagnetic component can be disentangled from the altermagnetic one through its distinct field- and temperature-dependence~\cite{hariki2025separating, ishii2026_Fe2O3_FMDistinctFromAM}.

%While it is established that XMCD can serve as a probe of altermagnetism, no systematic catalogue of the measurement geometries required to probe the order in each magnetic point group (MPG) is currently available. The primary purpose of this study is to provide this catalogue, along with the symmetry allowed forms of $\bm{h}$, $\bm{T}$ and $\mathfrak{T}$, in Sec.~\ref{sec:allowed_Th}. These results are also made available through an interactive web tool, \underline{Ma}gnetic \underline{G}roup \underline{N}avigator for \underline{E}xperimental \underline{X}MCD (\textsc{magnex}) \footnote{\href{https://xmcd-mpg.uc-trace.com/}{https://xmcd-mpg.uc-trace.com/}} to ease future experimental studies of altermagnetism via XMCD by explicitly clarifying the symmetry-imposed requirements for experimental geometries.

While it is established that XMCD can serve as a probe of altermagnetism, no systematic catalogue of the relevant groups, the allowed forms of $\bm{h}$, $\bm{T}$ and $\bm{\mathfrak{T}}$, or the required measurement geometries is available. The primary purpose of this study is to provide this catalogue. In Sec.~\ref{subsec:allowed_terms}, we determine the allowed forms of the AMD and the optical Hall vector in every magnetic point group (MPG). These results are also made available through an interactive web tool, \underline{Ma}gnetic \underline{G}roup \underline{N}avigator for \underline{E}xperimental \underline{X}MCD (\textsc{magnex}) \footnote{\href{https://magnex.uc-trace.com/}{https://magnex.uc-trace.com/}}. Whenever XMCD is allowed by symmetry, a net moment is allowed concurrently, leading to 31 total MPGs relevant to XMCD. We then restrict our attention to altermagnetic orders, and demonstrate in Sec.~\ref{subsec:MPG_AM_XMCD} that only twelve MPGs can describe altermagnets with finite XMCD.

%Further, we identify the symmetry settings relevant to studies of XMCD in altermagnets. In Sec.~\ref{subsec:MPG_AM_XMCD} we demonstrate that only twelve magnetic point groups can describe altermagnets with finite XMCD signal.

Further, this study aims to reconcile the non-relativistic symmetries of altermagnets with their predicted XMCD signals. To this end, Sec.~\ref{sec:LT}  extends the Landau theory techniques developed for collinear altermagnets \cite{schiff2025collinear} to the prediction of XMCD. We find that for every altermagnetic structure allowing a finite XMCD spectrum, the non-relativistic symmetries dictate the dichroic response. The resulting linear and cubic dependences of $\bm{h}$ and $\bm{T}$ on the N\'eel vector additionally predict the angular dependence of the XMCD signal on the N\'eel vector direction. In Sec.~\ref{sec:material_examples}, we exemplify our predictions on six material examples, and verify their validity using first-principles calculation.

Beyond XMCD, Ref.~\cite{Mizumaki2026_XAS_XES_MCD} has proposed using  X-ray emission magnetic circular dichroism (XES-MCD) to directly measure ferroically ordered magnetic octupoles of the kind we derive in Sec.~\ref{sec:allowed_Th} and report for all MPGs in Tab.~\ref{tab:Hall_AMD_Ttensor_full}. In addition to the rank-1 terms probed by XMCD, its sum rule contains the irreducible rank-3 part of $\mathfrak{T}_{\alpha\beta\gamma}$, When the rank-1 (XMCD) contributions vanish, this technique isolates the octupoles.

%%%%%%%%%%%%%%%%%%%%%%%%%%%%%%%%%%%%%%%%%%%%%%%%%%%%  SYMMETRY  %%%%%%%%%%%%%
%%%%%%%%%%%%%%%%%%%%%%%%%%%%%%%%%%%%%%%

\section{Magnetic symmetry constraints on \texorpdfstring{$\bm{T}$}{T} and \texorpdfstring{$\bm{h}$}{h}}\label{sec:allowed_Th}

Here we determine the symmetry-allowed optical Hall vector and ground state AMD in every magnetic point group (Sec.~\ref{subsec:allowed_terms}). From these, we clarify the experimental geometry required to maximize the XMCD signal (Sec.~\ref{subsec:experimental_geometries}). Then, we demonstrate that there are only twelve magnetic point groups simultaneously compatible with altermagnetism and with finite XMCD (Sec.~\ref{subsec:MPG_AM_XMCD}).

%Given that in the experimental setting,  spin-orbit coupling is always present, 
%Here we provide the symmetry-allowed optical Hall vector $\bm{h}(\omega)$ and ground state AMD $\bm{T}$ in every magnetic point group. We first describe the transformation properties of $\bm{T}$ and $\bm{h}$ in Sec.~\ref{subsec:h_T_transformation}, and then derive their allowed contributions in Sec.~\ref{subsec:allowed_terms}. In Sec.~\ref{subsec:experimental_geometries} we outline the experimental geometries maximizing the XMCD signal. In Sec.~\ref{subsec:MPG_AM_XMCD} we demonstrate that there are only twelve magnetic point groups which are compatible both with a finite XMCD response, and with altermagnetic order at zero propagation vector. Nevertheless, we provide a complete catalogue of the XMCD constraints imposed by all magnetic point groups.

\subsection{Transformation properties of \texorpdfstring{$\bm{T}$}{T} and  \texorpdfstring{$\bm{h}$}{h}}\label{subsec:h_T_transformation}

We begin by defining the transformation properties of the net octupole operators $\bm{\mathfrak{T}}$ appearing in Eq.~\ref{eq:T_definition}, $\mathfrak{T}_{\alpha\beta\gamma} = \sum_{a}\mathfrak{T}^{(a)}_{\alpha\beta\gamma}.$ On each sublattice, the octupole is the tensor product between the local quadrupole operator and the spin operator. The quadrupole transforms as a symmetrized rank two tensor, given by the Jahn symbol $[V^{2}]$, while the spin is a rank-one, axial ($e$) and $\mathcal{T}$--odd ($a$) tensor with Jahn symbol $aeV$ \cite{Jahn1949,mtensor}. Thus, $\bm{\mathfrak{T}}^{(a)}$ transforms as the tensor product of these representations, $aeV[V^{2}]$, a rank-three $\mathcal{T}$--odd tensor where the symmetrization applies to the first two indices, $\mathfrak{T}^{(a)}_{\alpha\beta\gamma} = \mathfrak{T}^{(a)}_{\beta\alpha\gamma}$. As a sum of the sublattice $\bm{\mathfrak{T}}^{(a)}$, the full $\bm{\mathfrak{T}}$ shares this transformation law.

Under the action of a magnetic point group symmetry element, whose orthogonal representation matrix is given by $R$, $\mathfrak{T}_{\alpha\beta\gamma}$ transforms as
\begin{equation}\label{eq:Ttensor_transformation}
    \mathfrak{T}_{\alpha\beta\gamma} = \mathcal{P_{T}}
        \,\,\mathrm{det}(R)\sum_{\mu\nu\rho}R_{\alpha\mu}R_{\beta\nu}R_{\gamma\rho} \,\mathfrak{T}_{\mu\nu\rho}, 
\end{equation}
where the determinant appears due to the axial nature of the spin component; if the operation includes time-reversal, then $\mathcal{P_{T}} = -1$, otherwise $\mathcal{P_{T}} = +1$. From Eq.~\ref{eq:T_definition} and Eq.~\ref{eq:Ttensor_transformation} we obtain the transformation law of the AMD components by setting $\gamma = \beta$ and summing over $\beta,$
\begin{align}\label{eq:TVec_TransformationLaw}
  T_{\alpha} \equiv   \sum_{\beta}\mathfrak{T}_{\alpha\beta\beta} &= \mathcal{P_{T}} \,\mathrm{det}(R) \sum_{\beta\mu\nu\rho} R_{\alpha\mu}R_{\beta\nu}R_{\beta\rho}\mathfrak{T_{\mu\nu\rho}} \notag\\
   % &= \mathcal{P_{T}} \mathrm{det}(R) \sum_{\beta\mu\nu\rho} R_{\alpha\mu}(R^{-1})_{\nu\beta}R_{\beta\rho}\mathfrak{T_{\mu\nu\rho}}\notag\\
    &=\mathcal{P_{T}}\,\mathrm{det}(R) \sum_{\mu\nu}R_{\alpha\mu}\mathfrak{T_{\mu\nu\nu}},
\end{align}
where we have used the fact that $R$ is orthogonal, and that $(R^{-1})_{\nu\beta}R_{\beta\rho} = \delta_{\nu\rho}$. Eq.~\ref{eq:TVec_TransformationLaw} demonstrates that the AMD $T_{\alpha}$ transforms as a $\mathcal{T}-$odd axial vector. This is identical to the transformation law for the Hall vector, and so whenever $\bm{h}$ is symmetry-allowed, the AMD is as well. This symmetry also corresponds to the net magnetization, and so only the 31 magnetic point groups allowing for a net magnetic moment will permit a non-zero XMCD response \cite{mtensor}.

Though altermagnets are defined with collinear antiferromagnetic order at zero SOC, in some cases a finite moment can develop once SOC is restored. The appearance of weak ferromagnetism is a consequence of two factors: spin-rotational symmetry breaking of the valence spin interactions, and the orientation of the N\'eel vector. The first condition, supplied by (valence) SOC, breaks degeneracy between the collinear spin axis orientations with respect to the lattice. This SOC effect, alone, is insufficient to generate a net magnetic moment. The N\'eel vector direction determines whether a net moment is allowed by selecting the magnetic point group. The generation of a net moment by SOC from a structure whose compensation is symmetry-enforced at zero SOC has recently been termed spin-orbit magnetism \cite{Liu2026_OrientedSSG,Wang2026_SOM_altermagnets}.

%The rutile altermagnets MnF$_{2}$ and NiF$_{2}$ supply a clear example of this phenomenon. Though the N\'eel vector in MnF$_{2}$ is along the $\bm{c}$ axis while in NiF$_{2}$ it is along the $\bm{a}$ axis, their spin group symmetries are identical, and forbid a net moment. This is because the moment orientation is agnostic to lattice directions at zero SOC. Once SOC is included, the magnetic point group of MnF$_{2}$ is $\mathtt{4'/mmm'}$ while for NiF$_{2}$ it is $\mathtt{m'mm'}$. The latter permits a net moment, while the former does not. So, MnF$_{2}$ remains compensated, while NiF$_{2}$ develops a small in-plane canting angle of $0.9^{\circ}$ \cite{Brown1981_NiF2}. Because the net moment has the same symmetry as $\bm{h}$ and $\bm{T}$, NiF$_{2}$ permits an XMCD signal while MnF$_{2}$ does not; nevertheless, the canting in NiF$_{2}$ is not required to obtain a finite XMCD, as demonstrated in Sec.~\ref{sec:material_examples}. This sensitivity of XMCD to the N\'eel vector orientation in altermagnets has been studied in Ref.~\cite{kunes2025xmcd}. In Sec.~\ref{sec:LT}, we provide a complementary route to predicting XMCD based on Landau theory that does not rely on the FVS approximation used in Ref.~\cite{kunes2025xmcd}.

\subsection{Symmetry-allowed XMCD from \texorpdfstring{$\bm{T}$}{T}}\label{subsec:allowed_terms}

Having established in Sec.~\ref{subsec:h_T_transformation} how the AMD and its octupolar contributions transform, we now find, for every magnetic point group, the symmetry-allowed form of $\bm{h}$ and of $\bm{T}$ in terms of its octupolar contributions $\bm{\mathfrak{T}}$. For the 31 MPGs where XMCD is allowed, these $\bm{h}$ and $\bm{T}$ are summarized in Tab.~\ref{tab:tvec_short}.

For each MPG, we construct the 18-dimensional representation $aeV[V^{2}]$, whose matrices we will denote by $\mathcal{D}_{\bm{\mathfrak{T}}}(g)$, for all group elements $g$. Then, solutions to the system of linear equations $\mathcal{D}_{\bm{\mathfrak{T}}}(g) \, \bm{v} = \bm{v}$ where $\bm{v}$ consists of all $\mathfrak{T}_{\alpha\beta\gamma}$, 
will yield the symmetry-allowed octupoles and any relations between them (e.g. $\mathfrak{T}_{zxx} = \mathfrak{T}_{zyy}$). To obtain the AMD, the $\mathfrak{T}_{\alpha\beta\beta}$ components are extracted and summed over $\beta$. An analogous calculation with the $aeV$ representation is conducted to determine the symmetry-allowed Hall vector $\bm{h}$. 

In Tab.~\ref{tab:tvec_short}, we report the form of $\bm{h}$ and $\bm{T}$ for all magnetic point groups that possess an allowed component. In Tab.~\ref{tab:Hall_AMD_Ttensor_full} we list the allowed $\bm{h}$ and $\bm{T}$ alongside a complete listing of the symmetry-allowed $\mathfrak{T}_{\alpha\beta\gamma}$ components for all magnetic point groups. As expected, 31 of the 122 magnetic point groups allow for a finite XMCD signal, consistent with the known result of magnetic point groups compatible with ferromagnetism \cite{mtensor}. We verify these results using group projector techniques in App.~\ref{app:symm_methods}. These results are also made available through an interactive web tool, \textsc{magnex} \footnote{\href{https://xmcd-mpg.uc-trace.com/}{https://xmcd-mpg.uc-trace.com/}}.

\begin{table}[t!]
\caption{Magnetic point groups with finite Hall vector $\bm{h}$ and AMD $\bm{T}$. The second \& third columns specify the symmetry-adapted forms of $\bm{h}$ and $\bm{T}$, respectively. The experimental geometry maximizing XMCD is $\bm{\hat{k}}\,\|\,\bm{h}.$}\label{tab:tvec_short}
\centering
\footnotesize
\renewcommand{\arraystretch}{1.1}
\begin{tabular}{l l l}
\hline\hline
MPG & Hall $\bm{h}$ & AMD $\bm{T}$ \\
\hline
$\mathtt{1}$, $\mathtt{\bar{1}}$ & $\begin{pmatrix} h_{x} \\ h_{y} \\ h_{z} \end{pmatrix}$ & $\begin{pmatrix} \mathfrak{T}_{xxx} + \mathfrak{T}_{xyy} + \mathfrak{T}_{xzz} \\ \mathfrak{T}_{yxx} + \mathfrak{T}_{yyy} + \mathfrak{T}_{yzz} \\ \mathfrak{T}_{zxx} + \mathfrak{T}_{zyy} + \mathfrak{T}_{zzz} \end{pmatrix}$ \\
$\mathtt{2}$, $\mathtt{m}$, $\mathtt{2/m}$, $\mathtt{m'm2'}$ & $\begin{pmatrix} 0 \\ h_{y} \\ 0 \end{pmatrix}$ & $\begin{pmatrix} 0 \\ \mathfrak{T}_{yxx} + \mathfrak{T}_{yyy} + \mathfrak{T}_{yzz} \\ 0 \end{pmatrix}$ \\
$\mathtt{2'}$, $\mathtt{m'}$, $\mathtt{2'/m'}$ & $\begin{pmatrix} h_{x} \\ 0 \\ h_{z} \end{pmatrix}$ & $\begin{pmatrix} \mathfrak{T}_{xxx} + \mathfrak{T}_{xyy} + \mathfrak{T}_{xzz} \\ 0 \\ \mathfrak{T}_{zxx} + \mathfrak{T}_{zyy} + \mathfrak{T}_{zzz} \end{pmatrix}$ \\
$\mathtt{2'2'2}$, $\mathtt{m'm'2}$, $\mathtt{m'm'm}$ & $\begin{pmatrix} 0 \\ 0 \\ h_{z} \end{pmatrix}$ & $\begin{pmatrix} 0 \\ 0 \\ \mathfrak{T}_{zxx} + \mathfrak{T}_{zyy} + \mathfrak{T}_{zzz} \end{pmatrix}$ \\
\begin{tabular}[c]{l}$\mathtt{4}$, $\mathtt{\bar{4}}$, $\mathtt{4/m}$, $\mathtt{42'2'}$, $\mathtt{4m'm'}$ \\ $\mathtt{\bar{4}2'm'}$, $\mathtt{4/mm'm'}$, $\mathtt{3}$, $\mathtt{\bar{3}}$, $\mathtt{32'}$ \\ $\mathtt{3m'}$, $\mathtt{\bar{3}m'}$, $\mathtt{6}$, $\mathtt{\bar{6}}$, $\mathtt{6/m}$ \\ $\mathtt{62'2'}$, $\mathtt{6m'm'}$, $\mathtt{\bar{6}m'2'}$, $\mathtt{6/mm'm'}$\end{tabular} & $\begin{pmatrix} 0 \\ 0 \\ h_{z} \end{pmatrix}$ & $\begin{pmatrix} 0 \\ 0 \\ 2\mathfrak{T}_{zxx} + \mathfrak{T}_{zzz} \end{pmatrix}$ \\
\hline\hline
\end{tabular}%
%\vspace{-.3cm}
\end{table}

When $h_{\alpha}$ is allowed, the corresponding $T_{\alpha}$ has symmetry-allowed contributions that may arise \emph{either} from (weak) ferromagnetism, or from antiferromagnetic order. This is particularly apparent when the SOC is weak, and we have $\langle \mathfrak{T}^{(a)}_{\alpha\beta\gamma}\rangle \approx \langle Q_{\alpha\beta}^{(a)}\rangle\langle S_{\gamma}^{(a)}\rangle$ \cite{thole1992xmcd,stohr1999xmcd,Carra1993_XMCD_ST,Wu1994_AMD_Importance,Crocombette1996_XMCD_AMD_3dTM,StohrKonig1995_SpinOrbitalAnisotropies_XMCD}. In this case, a finite $\mathfrak{T}_{\alpha\beta\beta}$ for $\beta = \alpha$ may be associated with the ferromagnetic contribution to the AMD, while the terms with $\beta \neq \alpha$ may be interpreted as arising from an AFM order. Thus, it is notable that in \emph{every} instance where $\bm{T}$ is allowed, it contains contributions that may arise from antiferromagnetism.

We remark briefly on the choice of coordinate axes. The form of $\bm{h}$ and $\bm{T}$ obtained through this procedure depends on the choice of coordinate axes for group elements matrices $R$ in Eq.~\ref{eq:Ttensor_transformation} and Eq.~\ref{eq:TVec_TransformationLaw}. Throughout this work, we inherit the axis choices from the \textsc{Bilbao Crystallographic Server}, except in the trigonal and hexagonal groups where the server uses a non-orthogonal basis. For these groups, we apply a basis transformation first, to convert to an orthogonal basis. 

As a simple rule, the group elements denoted with primes in Tab.~\ref{tab:tvec_short} forbid components of $\bm{h}$ and $\bm{T}$ along their axis. In the group $\mathtt{2'}$, for example, the non-trivial group element is taken to be $\mathcal{T} 2_{y}$, forcing $h_{y}$ and $T_{y}$ to vanish. For all MPGs within this row of Tab.~\ref{tab:tvec_short}, the two-fold elements are aligned with the $y$-axis. In a material with one of these MPGs, the two-fold axis might lie along a different direction, and so a basis transformation should be applied to the reported $\bm{h}$ and $\bm{T}$. The web tool \textsc{magnex} performs basis transformations for users \footnote{\href{https://xmcd-mpg.uc-trace.com/}{https://xmcd-mpg.uc-trace.com/}}. 

%The matrices $R$ used to construct the $aeV$ and $aeV[V^{2}]$ representations are sourced from the \textsc{Bilbao Crystallographic Server} \cite{BilbaoIrreps,mtensor}, from which the choice of axes in each group is also inherited. Cartesian axes are used for all groups. 
%Thus, for example, in the group $\mathtt{2'}$, the non-trivial group element is taken to be $\mathcal{T} 2_{y}$. In the hexagonal and trigonal groups, the \textsc{Bilbao} matrices are expressed in a non-orthogonal, oblique hexagonal basis. Thus, a basis transformation is applied to achieve an orthogonal form.
%. In \textsc{magnex}, users can apply basis transformations to these results, for use in different settings.

\subsection{Experimental geometries}\label{subsec:experimental_geometries}

\begin{figure}[t!]
\includegraphics[width=\columnwidth]{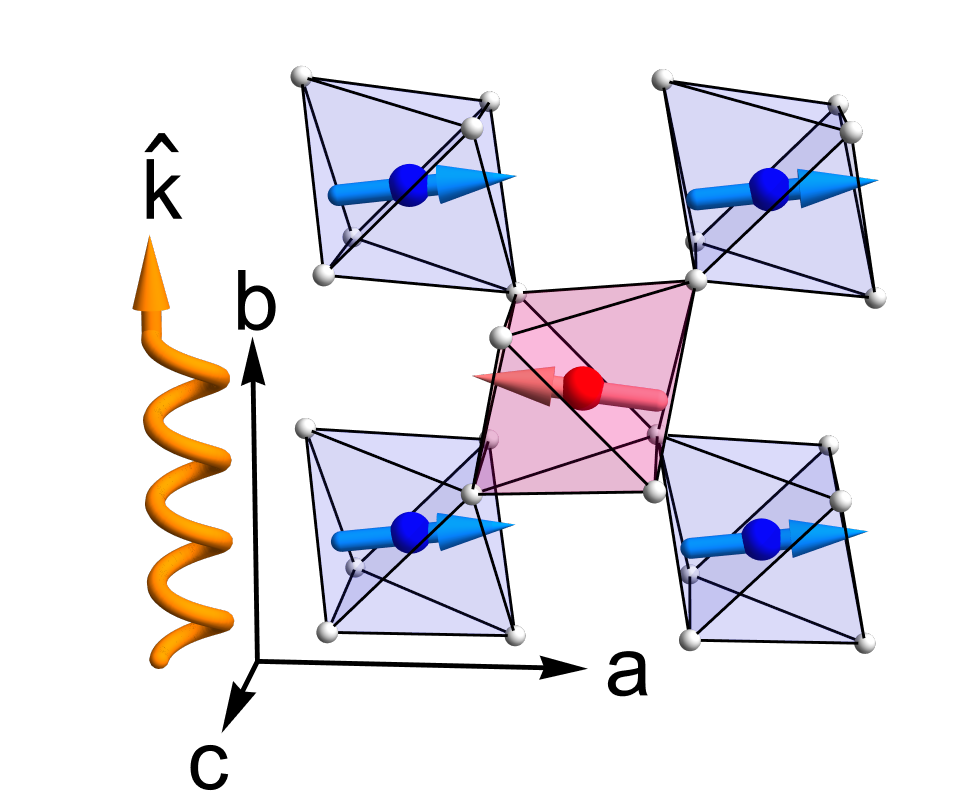}
\caption{The crystallographic and magnetic structure of NiF$_{2}$. Ni atoms are depicted in blue and red, corresponding to the moment directions. F atoms are depicted in white. The MPG is $\mathtt{m'm'm}$, with the unprimed axis along $\bm{b}$, corresponding to the allowed direction of $\bm{h}$ and $\bm{T}$ following Tab.~\ref{tab:tvec_short}. A circularly polarized photon with $\bm{\hat{k}} \,\|\,\bm{h}$ maximizes the XMCD signal. The canting (exaggerated) is along $\bm{b}$, parallel to $\bm{h}$.}  
\label{fig:ExpGeo_NiF2}
\end{figure}

To utilize Tab.~\ref{tab:tvec_short} in experiments, a crystalline and magnetic structure must be assumed or known, and the MPG of the structure identified. If the coordinate axes of the symmetry elements do not coincide with those in Tab.~\ref{tab:tvec_short}, a basis transformation must be performed. Finally, the incident photon direction $\bm{\hat{k}}$ should be aligned with the Hall vector, $\bm{\hat{k}}\,\|\,\bm{h},$ in order to maximize the XMCD signal. For MPGs where $\bm{h}$ is not restricted to one axis, it may point in a different direction than the net moment.

Let us use the rutile altermagnet NiF$_{2}$ as an example. Further examples are provided in Sec.~\ref{sec:material_examples}. %NiF$_{2}$ crystallizes in space group \#136, $\mathtt{P4_{2}/mnm}$ with the magnetic nickel ions sitting on the $2a$ Wyckoff position, which order at a Néel temperature of roughly $T_{\bm{N}} \approx 73.2$~K, and are known to have a weak ferromagnetic canting component \cite{borovikromanov1973_NiF2, Brown1981_NiF2, wei2024_CrystChem_NiF2}.
The crystalline and magnetic structure (with exaggerated canting) is depicted in Fig.~\ref{fig:ExpGeo_NiF2}. We will use a Cartesian frame in which $\bm{\hat{x}}$, $\bm{\hat{y}}$ and $\bm{\hat{z}}$ align with the crystallographic $\bm{a}$, $\bm{b}$ and $\bm{c}$ directions, respectively. The Néel vector (finite below a Néel temperature of $T_{\bm{N}} \approx 73.2$~K) points along $\bm{a}$, and the weak ferromagnetic component lies along $\bm{b}$ \cite{borovikromanov1973_NiF2, Brown1981_NiF2, wei2024_CrystChem_NiF2}. This structure corresponds to the MPG $\mathtt{m'm'm},$ with unprimed two-fold rotation and mirror about $\bm{b}$. This setting differs from that used in Tab.~\ref{tab:tvec_short}, where instead the unprimed elements lie along the crystallographic $\bm{c}-$axis. These settings are related by a $90^{\circ}$ rotation about $\bm{a}$, so the predicted Hall vector direction in Tab.~\ref{tab:tvec_short} must be transformed by this same rotation to obtain $\bm{h} = (0,h_{y},0)^{\mathsf{T}}.$ Thus, in NiF$_{2}$ the Hall vector points along $\bm{b}$, corresponding to the allowed ferromagnetic canting direction, and thus a circularly polarized photon $\hat{\bm{k}}\, \| \,\bm{b}$ will maximize the XMCD signal, as depicted in Fig.~\ref{fig:ExpGeo_NiF2}.

\subsection{Magnetic groups for AMs with non-zero XMCD}\label{subsec:MPG_AM_XMCD}

Here, we show that altermagnets at zero propagation vector ($\bm{q}=\bm{0}$) with finite XMCD always belong to one of twelve magnetic point groups that allow for a net magnetization, and have no axis of order greater than two, corresponding to the first four rows of Tab.~\ref{tab:tvec_short}.

% This may be important, I wasn't sure..
%First, we seek magnetic point groups that support altermagnetism at $\bm{q}=\bm{0}$. In particular, we are interested in the MPGs that arise by starting from a $\bm{q}=\bm{0}$, zero valence SOC AM (a collinear antiferromagnet whose up- and down-sublattices are not related by translation or inversion), and then turning on SOC (which may result in a symmetry-allowed net moment). 

Immediately, we may disqualify any of the 32 MPGs containing pure $\mathcal{T}$ as a symmetry, corresponding to the type-II gray groups, since $\mathcal{T}$--symmetry implies a non-magnetic ground state. There are 90 remaining MPGs: the 32 standard (type I) crystallographic point groups and the 58 black-and-white (type III) groups. As we seek collinear configurations, a valid MPG must preserve an axis up to its orientation; this disqualifies the 11 MPGs based on a cubic point group. For an AM order, we may also exclude the 18 MPGs containing $\mathcal{TI}$. 

Now, we must select those of the remaining 61 MPGs that could simultaneously realize an altermagnetic order and a finite XMCD signal. For this, we separate these 61 MPGs into two classes: 46 MPGs with a principal rotation axis of order three or greater, discussed in Sec.~\ref{subsubsec:high-order-axis}, and the 15 MPGs without any elements of order greater than two, discussed in Sec.~\ref{subsubsec:low-order-axis}. 

\subsubsection{MPGs with principal axis of order three or greater}\label{subsubsec:high-order-axis}
In any (non-cubic) MPG with a principal axis of order three or greater, the preserved collinear orders must have their moments aligned parallel or antiparallel to this axis. Having such a unique axis allows us to define the parity $\lambda(g)=\pm1$, which indicates whether an axial vector along the principal axis, i.e., a magnetic moment, is preserved or reversed by the group element $g$. In addition, we define the time-reversal parity $\tau(g) = \pm1$ of the group elements, indicating whether $g$ is combined with time-reversal ($-1$ for antiunitary $g$) or not ($1$ unitary $g$). We can divide the MPGs with a principal axis of order three or greater into two groups based on their axial and time-reversal parities. We distinguish groups that contain elements for which $\lambda(g)\tau(g) = -1$ and groups without such elements.   %Here, we will demonstrate that sublattice-swapping operations preserve the AM order parameter, while they force $\bm{h}$ to vanish.
\begin{comment}
\begin{figure}[b!]
\includegraphics[width=.9\columnwidth]{Figures/lambda_tau.pdf}
\caption{The effect of $\lambda(g) = \pm1$ and $\tau(g) = \pm1$ on axial vectors (dashed arrows) and $\uparrow/\downarrow$ spins (empty/solid circles) with respect to the axial axis, and on the net time-odd axial vectors, such as $\bm{h}$, $\bm{T}$ and the net moment (solid arrows). Gray panels denote elements with $\lambda(g)\tau(g) = -1$, which must swap $\uparrow/\downarrow$ sublattices.} 
\label{fig:lambda-tau}
\end{figure}
\end{comment}

%The MPG elements $g$ can be sorted according to their time-reversal character, $\tau(g) = \pm1$, and whether they preserve or reverse an axial vector along the principal axis, $\lambda(g) = \pm1$.
%their action on an axial vector along the principal axis. We will call the latter $\lambda(g)$ and the former $\tau(g)$, for each element $g$ of the MPG. When $\lambda(g) = +1$ ($-1$), $g$ preserves (reverses) the orientation of an axial vector along the principal axis. Similarly, $\tau(g) = +1$ ($-1$) denotes elements $g$ which are unitary (antiunitary). 

%These four kinds of elements and their action on an axial axis, the spins, and on a net time-odd axial vector are depicted in Fig.~\ref{fig:lambda-tau}. 

Groups with symmetry elements of $\lambda(g)\tau(g) = -1$ do not preserve $\bm{h}$ or $\bm{T}$: if $\lambda(g)=1$ and $\tau(g) = -1$ the axial vectors are left unflipped, but the group element is combined with time-reversal, assigning a negative sign to $\bm{h}$ and $\bm{T}$. Meanwhile, for $\lambda(g)=-1$ and $\tau(g) = 1$, time-reversal is not involved, but the group element flips the axial vectors, reversing the sign of $\bm{h}$ and $\bm{T}$. %(gray panels in Fig.~\ref{fig:lambda-tau}) forbid net $\bm{h}$ and $\bm{T}$, and map up-spins to down-spins. Thus, a system whose MPG contains such $g$ must simultaneously have its up- and down-spin sublattices swapped for $g$ to be a symmetry. In AMs, such elements are required to exist, because at zero SOC these elements enforce perfectly vanishing net magnetization. 
MPGs containing such $g$ elements can accommodate collinear AMs, preserving this order but enforcing a vanishing net magnetization and $\bm{h},$ thus prohibiting XMCD. 
%Since every element maps the principal axis onto itself, with $\bm{N}$ along this axis each such element survives in the MPG once SOC is restored, combined with $\mathcal{T}$ where needed. So, any AM whose MPG has an axis of order three or greater forbids XMCD.

%Elements for which $\lambda(g)\tau(g) = -1$ require the order to possess opposite-spin sublattices. When $\lambda(g) = -1$ while $\tau(g) = +1$, the axial vector direction is reversed; this can only be remedied if the sublattices are simultaneously reversed, corresponding to an altermagnet. %so long as $g \neq \mathcal{I}$. 

For MPGs with a high-order axis containing only elements with $\lambda(g)\tau(g)=+1$, a net moment and a finite XMCD signal are allowed. However, these groups do not describe an AM order:  $\lambda(g)\tau(g)=+1$ implies that a time-reversal odd axial vector remains invariant under $g$. Thus, there are no $g$ that swap the sublattices, and so at zero SOC compensation is not enforced. Because every element will map the principal axis onto itself, any sublattice-swapping element remains a symmetry once SOC is restored with $\bm{N}$ along this axis, combined with $\mathcal{T}$ where needed. %it possesses no sublattice-swapping elements.  Instead, it 
These MPGs, therefore, accommodate uniform ferromagnetic or ferrimagnetic configurations in which the sublattices are not related by symmetry.

%If $\lambda(g) = +1$ while $\tau(g) = -1$, then $g$ must swap the sublattices to undo the spin-flip on every site. 
%On the other hand, if $\lambda(g)\tau(g) = 1$ for all elements $g$, then we either have a ferromagnet, or we have collinear sublattices that are not related by symmetry, neither of which can arise by turning on SOC starting from a spin-orbit free altermagnet. Since every element maps the principal axis onto itself, once SOC is restored, aligning the moments along this axis ensures each element is preserved up to time-reversal. Thus, we can conclude that the sublattice-exchanging elements $g$ must have $\lambda(g)\tau(g) = -1.$

We conclude that MPGs with principal axes of order three or greater cannot describe compensated AMs with finite XMCD. For each of the 46 axial MPGs lacking $\mathcal{T}$ and $\mathcal{TI},$ we calculate $\lambda(g)\tau(g)$ for all $g$. Any MPG with elements satisfying $\lambda(g)\tau(g) = -1$ can describe an AM, albeit without an XMCD signal. There are 27 such MPGs, summarized in Tab.~\ref{tab:AM_MPG}. The remaining 19 MPGs with high-order axes cannot realize AM order; instead, they show ferro- or ferrimagnetism with naturally arising finite XMCD. These 19 MPGs can be found among the 31 MPGs compatible with XMCD listed in Tab.~\ref{tab:tvec_short}. 

These $\lambda(g)\tau(g)$ values define an irreducible representation (irrep) of the crystal point group. In Sec.~\ref{sec:LT}, this irrep will correspond exactly to that of the spin-orbit free AM order parameter. 

\subsubsection{MPGs with no elements of order greater than two}\label{subsubsec:low-order-axis}

Any possibility of a finite XMCD signal in altermagnets is now committed to MPGs without high-order symmetry axes. In them, there is no unique axis for the collinear moments, and $\lambda(g)$ is ill-defined. Some orientations of the moments may now break sublattice-exchanging elements present at zero SOC. Here, the absence of sublattice-swapping elements does \emph{not} prevent an MPG from describing an AM. The only criterion for AM-compatibility is the exclusion of group elements $\mathcal{T}$ and $\mathcal{TI}$. There are 15 such MPGs in Tab.~\ref{tab:AM_MPG}, of which 12 (denoted in red) are compatible with finite net moment, $\bm{h}$ and $\bm{T}$, and thus a finite XMCD signal. 

\begin{table}[t]
\centering
\caption{The 42 magnetic point groups supporting $\bm q=\bm 0$ collinear
altermagnetism, sorted by crystal system. Groups denoted in \textcolor{red}{red} permit a net moment and Hall vector and AMD, and hence a finite XMCD response. %Type I MPGs are ordinary point groups. Type III MPGs have half of their elements combined with $\mathcal{T}$, while lacking pure $\mathcal{T}$.
}
\label{tab:AM_MPG}
\renewcommand{\arraystretch}{1.2}
\setlength{\tabcolsep}{7pt}
\begin{tabular}{l| c c c |c c c}
\hline\hline
%Triclinic & Monoclinic & Orthorhombic & Tetragonal & Trigonal & Hexagonal \\
%\hline
{}& \multicolumn{3}{c}{axis order max 2} &\multicolumn{3}{c}{axis order greater than 2} \\
\hline
\multirow{4}{*}{\rotatebox[origin=c]{90}{Type I}} &
\textcolor{red}{$\mathtt{1}$}       & \textcolor{red}{$\mathtt{2}$}   & $\mathtt{222}$ & $\mathtt{422}$       & $\mathtt{32}$       & $\mathtt{622}$ \\
& \textcolor{red}{$\mathtt{\bar{1}}$} & \textcolor{red}{$\mathtt{m}$}   & $\mathtt{mm2}$ & $\mathtt{4mm}$       & $\mathtt{3m}$       & $\mathtt{6mm}$ \\
& 
                             & \textcolor{red}{$\mathtt{2/m}$} & $\mathtt{mmm}$ & $\mathtt{\bar{4}2m}$ & $\mathtt{\bar{3}m}$ & $\mathtt{\bar{6}m2}$ \\
& 
                             &                          &                & $\mathtt{4/mmm}$     &                     & $\mathtt{6/mmm}$ \\
\hline
\multirow{8}{*}{\rotatebox[origin=c]{90}{Type III}}& --- & \textcolor{red}{$\mathtt{2^{\prime}}$}            & \textcolor{red}{$\mathtt{2^{\prime}2^{\prime}2}$} & $\mathtt{4^{\prime}}$                  & --- & $\mathtt{6^{\prime}}$ \\
& 
    & \textcolor{red}{$\mathtt{m^{\prime}}$}            & \textcolor{red}{$\mathtt{m^{\prime}m2^{\prime}}$} & $\mathtt{\bar{4}^{\prime}}$            &     & $\mathtt{\bar{6}^{\prime}}$ \\
& 
    & \textcolor{red}{$\mathtt{2^{\prime}/m^{\prime}}$} & \textcolor{red}{$\mathtt{m^{\prime}m^{\prime}2}$} & $\mathtt{4^{\prime}/m}$                &     & $\mathtt{6^{\prime}/m^{\prime}}$ \\
& 
    &                                            & \textcolor{red}{$\mathtt{m^{\prime}m^{\prime}m}$} & $\mathtt{4^{\prime}22^{\prime}}$       &     & $\mathtt{6^{\prime}22^{\prime}}$ \\
& 
    &                                            &                                            & $\mathtt{4^{\prime}m^{\prime}m}$       &     & $\mathtt{6^{\prime}mm^{\prime}}$ \\
& 
    &                                            &                                            & $\mathtt{\bar{4}^{\prime}2^{\prime}m}$ &     & $\mathtt{\bar{6}^{\prime}m^{\prime}2}$ \\
& 
    &                                            &                                            & $\mathtt{\bar{4}^{\prime}2m^{\prime}}$ &     & $\mathtt{\bar{6}^{\prime}m2^{\prime}}$ \\
& 
    &                                            &                                            & $\mathtt{4^{\prime}/mm^{\prime}m}$     &     & $\mathtt{6^{\prime}/m^{\prime}mm^{\prime}}$ \\
%\hline
%2 & 6 & 7 & 12 & 3 & 12 \\
\hline\hline
\end{tabular}
\vspace{-0.5cm}
\end{table}

\begin{comment}
\subsubsection{MPGs for altermagnets with finite XMCD}

Having established the MPGs compatible with altermagnetism, we now ask which additionally allow for a non-zero XMCD signal? As we have seen in Sec.~\ref{subsec:allowed_terms}, 31 MPGs allow XMCD. Of these, only 12 also support altermagnetism. These groups are marked with a dagger in Tab.~\ref{tab:AM_MPG}. None contain a rotation or rotoinversion of order three or greater, because the Hall vector vanishes whenever the N\'eel vector is oriented along the principal rotation axis (of order $\geq 3$), as seen in Sec.~\ref{subsubsec:high-order-axis}. This can be understood intuitively: such groups require both $\bm{N}$ and $\bm{h}$ to lie along the principal axis. However, the AM MPG contains a sublattice-swapping element, i.e. $g$ with $\lambda(g)\tau(g) = -1$, which will force the net Hall vector to vanish. %If $g$ preserves the axis orientation ($\lambda(g) = 1$), then $\tau(g) = -1$ and so $g \bm{h} = -\bm{h}$; if $g$ reverses the axis orientation ($\lambda(g) = -1$), then $\tau(g) = 1$ and $g\bm{h} = -\bm{h}$. 
This result is consistent with the findings of Ref.~\cite{kunes2025xmcd}, and the material-specific settings studied in Refs.~\cite{hariki2024MnTe,sasabe2025t,hariki2024rutile_MnF2_XMCD,xie2026fe2o3,ishii2026_Fe2O3_FMDistinctFromAM,Sasabe2023_AMD_XMCD_RuO2}, providing a complete symmetry framework for efficiently determining whether an AM may be investigated using XMCD.
\end{comment}

%%%%%%%%%%%%%%%%%%%%%%%%%%%%%%%%%%%%%%%
%%%%%%%%%%%%%  LANDAU  %%%%%%%%%%%%%
%%%%%%%%%%%%%%%%%%%%%%%%%%%%%%%%%%%%%%%

\section{Landau Theory}\label{sec:LT}

In this section, we demonstrate that the form of the XMCD in altermagnets at finite SOC is completely determined by symmetries in the spin-orbit free limit. The techniques employed here extend the Landau theory recently developed for collinear altermagnets in Ref.~\cite{schiff2025collinear}, reviewed in Sec.~\ref{subsec:collinear_AM_LT}. In Sec.~\ref{subsec:AMD_zero_SOC}, we derive the condition for linear AMD at zero SOC. Finally, in Sec.~\ref{subsec:couplings} we demonstrate that every AM sublattice configuration possesses an allowed XMCD response at either linear or at cubic order. The dependence of $\bm{T}$ and $\bm{h}$ on the N\'eel vector orientation is encoded in these couplings, and is predicted for every possible altermagnetic sublattice arrangement at $\bm{q}=0$. Our results complement Ref.~\cite{kunes2025xmcd} in determining N\'eel vector dependence of XMCD, reproducing these predictions at finite SOC. However, this Landau theory approach also provides the N\'eel vector dependence of the XMCD even in those cases inaccessible to the FVS approximation.

To utilize the results of this section in an experimental setting, one must presume or know the sublattice pattern of up- and down-spins in the crystal point group. This is described by a one-dimensional, real and inversion-even point group irrep $\Gamma_{\bm{N}}$, identified by the Wyckoff position of the magnetic ions, from Ref.~\cite{schiff2025collinear}. Then, Tab.~\ref{tab:LT_linear_symm},~\ref{tab:LT_linear_anti} and~\ref{tab:LT_cubic} give the dependence $\bm{h}(\bm{N})$, and the free-energy coupling between $\bm{T}$ and $\bm{N}$. The ground state orientation of $\bm{N}$ then specifies the directions of the Hall vector and the AMD. This dependence on $\bm{N}$ predicts an angular dependence of the XMCD signal, exemplified in Sec.~\ref{sec:material_examples}.

\subsection{Landau theory for collinear altermagnets}\label{subsec:collinear_AM_LT}
First, we briefly review the Landau theory for collinear altermagnets that do not enlarge the unit cell ($\bm{q} = \bm{0}$), following Ref.~\cite{mcclarty2024landau,schiff2025collinear}. An AM at zero SOC is a collinear order whose compensation is symmetry-enforced. This restricts the order to a staggered collinear configuration of moments on the lattice, described by a N\'eel vector, $\bm{N}$, which will serve as the primary order parameter. To break Kramers' degeneracy of the opposite-spin bands in momentum space, the sublattices must not be related by translation or inversion. The sublattice structure must be inversion even, and for $\bm{q}=\bm{0}$ altermagnets, this means translations act trivially. These characteristics of $\bm{N}$ can be encoded by a simple symmetry criterion: $\bm{N}$ transforms as a non-trivial, one-dimensional, real and inversion even irreducible representation (irrep), $\Gamma_{\bm{N}}$, of the crystal point group. The Wyckoff positions in each space group that allow for $\Gamma_{\bm{N}}$ are known \cite{schiff2025collinear}.

For altermagnets at zero SOC, the Landau theory is determined by a pseudoprimary multipolar order parameter transforming identically to $\bm{N}.$ These multipoles are of the form $$\bm{O}_{\mu_{1}...\mu_{n}} \equiv \int d^{3}r [r_{\mu_{1}}...r_{\mu_{n}}] \bm{m}(\bm{r}),$$ where $r_{\mu_{i}}$ are Cartesian $x$, $y$ or $z$, $n\geq0$ is an integer, $[...]$ indicates symmetrization, and $\bm{m}(\bm{r})$ is the local magnetization density. When $\bm{O}_{\mu_{1}...\mu_{n}}$ transforms as $\Gamma_{\bm{N}},$ it couples linearly to $\bm{N}$ at zero SOC \cite{mcclarty2024landau,schiff2025collinear}. 

The zero SOC multipole $\bm{O}_{\mu_{1}...\mu_{n}}$ with minimal $n$ holds predictive power for altermagnets at finite SOC. It guarantees the symmetry-allowed components of measurable properties at linear order once SOC is restored. These predictions are tabulated in Ref.~\cite{schiff2025collinear} and can be extended to capture the lowest order XMCD response in real altermagnets based on their spin-orbit free N\'eel vector $\bm{N}.$

\subsection{The AMD at zero SOC}\label{subsec:AMD_zero_SOC}

In the limit of zero valence SOC, the AMD is always forbidden in conventional N\'eel antiferromagnets (referred to here as AFMs), but may be allowed for altermagnets. The magnetic orders in AFMs and AMs are bipartite: the number of sublattices is even; half of the sublattices host up-spins, while the other half host down-spins. A space group element $g$ will map the up-spin sites to the down-spin sites, and vice versa. In AFMs, $g$ can be chosen as inversion or translation, which is forbidden in AMs.

If we denote by $u$ and $d$ an up- and a down-spin sublattice, respectively, then for each $d$ there is a $u$ such that $g u = d$. We can then express the net AMD in terms of the sublattice anisotropic magnetic dipoles,
\begin{align}\label{eq:T_ud_sublattices}
    \langle T_{\alpha} \rangle &= \sum_{a} \langle T_{\alpha}^{(a)} \rangle = \sum_{u} \langle T_{\alpha}^{(u)}\rangle + \sum_{d} \langle T_{\alpha}^{(d)} \rangle \notag\\
    &= \sum_{u}\langle Q_{\alpha\beta}^{(u)}\rangle \langle S_{\beta}^{(u)}\rangle + \sum_{d} \langle Q_{\alpha\beta}^{(d)} \rangle \langle S_{\beta}^{(d)}\rangle.
\end{align}
Because AFMs and AMs are collinear at zero SOC, for every $d$ there is a $u$ such that $\langle S_{\beta}^{(d)}\rangle = -\langle S_{\beta}^{(u)}\rangle.$ Similarly, the quadrupoles satisfy $\langle Q_{\alpha\beta}^{(d)}\rangle~=~R_{\alpha\mu}R_{\beta\nu} \langle Q_{\mu\nu}^{(u)}\rangle,$ where $R$ is the orthogonal part of $g$. We can rewrite Eq.~\ref{eq:T_ud_sublattices} as 
\begin{align}\label{eq:T_u_symmetry}
    \langle T_{\alpha} \rangle &= \sum_{u}  \langle S_{\beta}^{(u)} \rangle \biggl( \langle Q_{\alpha\beta}^{(u)} \rangle - g\langle Q_{\alpha\beta}^{(u)} \rangle \biggr)\notag\\
    &= \sum_{u}  \langle S_{\beta}^{(u)} \rangle \sum_{\mu\nu}\biggl( \delta_{\alpha\mu}\delta_{\beta\nu} - R_{\alpha\mu}R_{\beta\nu} \biggr)\langle Q_{\mu\nu}^{(u)}\rangle.
\end{align}
If all magnetic ions reside on the same crystallographic orbit, then 
\begin{align}\label{eq:T_N_Q}
    \langle T_{\alpha}\rangle &= \frac{N_{\beta}}{2 |u|} \sum_{u\mu\nu}\biggl( \delta_{\alpha\mu}\delta_{\beta\nu} - R_{\alpha\mu}R_{\beta\nu} \biggr)\langle Q_{\mu\nu}^{(u)}\rangle\notag\\
    &=\frac{1}{2} \mathcal{Q}_{\alpha\beta} N_{\beta}.
\end{align}
Here, $N_{\beta}$ is the $\beta$ component of the N\'eel vector, the total staggered magnetization $\bm{N} = \sum_{u} \langle \bm{S}^{(u)}\rangle - \sum_{d}\langle\bm{S}^{(d)}\rangle$, $|u|$ is the number of up-spin sublattices, and we have introduced $\mathcal{Q}_{\alpha\beta} \equiv \frac{1}{|u|}\sum_{u\mu\nu}(\delta_{\alpha\mu}\delta_{\beta\nu} - R_{\alpha\mu}R_{\beta\nu})\langle Q^{(u)}_{\mu\nu}\rangle$. From Eq.~\ref{eq:T_N_Q}, it is clear that the AMD must vanish in AFMs, while it may be non-zero in AMs at zero SOC. When $g$ is a pure translation, $R$ is the identity, and so $R_{\alpha\mu}R_{\beta\nu} = \delta_{\alpha\mu}\delta_{\beta\nu}$. When $g$ is the inversion element, then $R_{\alpha\mu}R_{\beta\nu}~=~(-\mathbb{I})_{\alpha\mu}(-\mathbb{I})_{\beta\nu} = \delta_{\alpha\mu}\delta_{\beta\nu}.$ In both cases, Eq.~\ref{eq:T_N_Q} vanishes. Thus, AMD vanishes at zero valence SOC for AFMs, but may be finite for AMs. The choice of $g$ is not unique; for any other $g'$ it still holds that for every $d$ there is a $u$ such that $g'u = d$, and the remaining derivation proceeds identically. Eq.~\ref{eq:T_u_symmetry} mirrors the relation obtained for $\bm{h}(\omega)$ within the FVS approximation in Ref.~\cite{kunes2025xmcd}, which confirms that the XMCD spectrum at zero SOC vanishes for AFMs.

From Eq.~\ref{eq:T_N_Q}, we also see that at zero SOC, $\Gamma_{\bm{N}}$ must be contained in $[V^{2}]$ for $\bm{T}$ to be finite. Thus, every $d$-wave AM admits XMCD even at zero valence SOC, as discussed in Sec.~\ref{subsec:couplings}.  On the other hand, $g$- and $i$-wave altermagnets have vanishing $\bm{T}$ in the non-relativistic limit. Here the $d$-, $g$- and $i$-wave labels refer to the lowest even-degree polynomial transforming as $\Gamma_{\bm N}$ (degree 2, 4 or 6), so that $d$-wave is equivalent to $\Gamma_{\bm N}\subset[V^2]$; this degree coincides with the multipole rank $n$ only for centrosymmetric point groups, and, e.g., VNb$_3$S$_6$ ($622$, $B_1$) has $n=3$ but is $g$-wave.

\subsection{Three classes of XMCD prediction at finite SOC}\label{subsec:couplings}

At finite SOC, an altermagnet with minimal multipole order $n$ is guaranteed an allowed $\mathcal{T}$-odd tensor at linear order determined by $n$, with specific tensor components determined by $\Gamma_{\bm{N}}$~\cite{schiff2025collinear}. Because $\bm{T}$ and $\bm{h}$ both transform as $aeV$, a finite XMCD response linear in $\bm{N}$ requires this tensor to share irreducible components with $aeV$. Otherwise, since $\bm{T}$ and $\bm{h}$ are odd in $\bm{N}$, the leading response is cubic, $[(aeV)^{3}]$. Every $\bm{q}=\bm{0}$ altermagnetic structure, determined by the crystal point group and $\Gamma_{\bm{N}}$, falls into one of three classes, detailed below. In all three classes, $\bm{h}$ vanishes for $\bm{N}$ along a rotation axis of order three or higher, consistent with Sec.~\ref{subsec:MPG_AM_XMCD}. Examples of each case will be demonstrated on material examples in Sec.~\ref{sec:material_examples}, whose couplings are summarized in Tab.~\ref{tab:material_properties}.
 
\textit{Linear symmetric.} When $\Gamma_{\bm{N}}\subset[V^{2}]$, $\bm{N}$ is guaranteed a linear coupling to $aeV$ through a symmetric tensor, e.g.\ $T_{x}N_{y}+T_{y}N_{x}$~\cite{schiff2025collinear}. This is the coupling of Eq.~\ref{eq:T_N_Q}: the zero-SOC AMD is a linear function of $\bm{N}$ whose coefficient matrix, the quadrupole difference $\mathcal{Q}$, is symmetric and transforms as $\Gamma_{\bm{N}}\subset[V^{2}]$. Thus, this class is the allowed form of the non-relativistic (zero valence SOC) XMCD. It is also the class captured by the FVS approximation~\cite{kunes2025xmcd}. We list all allowed linear couplings for each AM $\Gamma_{\bm{N}}$ in Tab.~\ref{tab:LT_linear_symm}. %\textcolor{red}{Refer to materials sections for examples.} 

%\vspace{0.2cm}
%\onecolumngrid
\begin{table*}[t!] % Standard single-column float placed at the bottom
  \centering
  \makebox[\textwidth][c]{ % Centers a full-page width box across both columns
    \begin{minipage}{\textwidth}
      \centering
      \caption{For each example material and $\bm{N}$ orientation, we list the crystal point group (PG) and the MPG. Then we list the irrep $\Gamma_{\bm{N}}$ defining the spin-orbit free order, the spatial rank $n$ of the minimal multipole, and whether $\Gamma_{\bm{N}}$ is contained in the (anti)symmetric square, which fixes the linear couplings between $\bm{T}$ and $\bm{N}$, and symmetric fourth power of the vector representation, which suffices for a cubic coupling, as in Sec.~\ref{sec:LT}. The Landau column is taken from Tab.~\ref{tab:LT_linear_symm}, Tab.~\ref{tab:LT_linear_anti} and Tab.~\ref{tab:LT_cubic}. Finally, we list non-zero $h_{\alpha}$ predicted in Tab.~\ref{tab:tvec_short}, adapted to the material settings.}
      \label{tab:material_properties}
\renewcommand{\arraystretch}{1.25}
\begin{tabular}{c w{c}{1cm} w{c}{1.5cm} w{c}{1.5cm} w{c}{1.5cm} w{c}{1cm} ccc w{c}{5.5cm} c}
\hline\hline
Material & $\bm{N}$ & PG & MPG & $\Gamma_{\bm{N}}$ & $n$ & $[V^{2}]$ & $\{V^{2}\}$ & $[V^{4}]$ & Landau & $\bm{h}$\\
\hline
MnF$_{2}$ & 001 & $\mathtt{4/mmm}$ & $\mathtt{4'/mm'm}$ & $B_{2g}$ & 2 & \cmark & \xmark & \cmark & \multirow{2}{*}{$N_{x}T_{y} + N_{y}T_{x}$} & $0$\\
NiF$_{2}$ & 100 & $\mathtt{4/mmm}$ & $\mathtt{m'm'm}$ & $B_{2g}$ & 2 & \cmark & \xmark & \cmark & & $h_{y}$\\
\hline
Fe$_{2}$O$_{3}$ & 0001 & $\mathtt{\bar{3}m}$ & $\mathtt{\bar{3}m}$ & $A_{2g}$ & 4 & \xmark & \cmark & \cmark & \multirow{3}{*}{$N_{x}T_{y}-N_{y}T_{x}$} & 0\\
 & $01\bar{1}0$ & $\mathtt{\bar{3}m}$ & $\mathtt{2/m}$ & $A_{2g}$ & 4 & \xmark & \cmark & \cmark & & $h_{x}$\\
 & $2\bar{1}\bar{1}0$ & $\mathtt{\bar{3}m}$ & $\mathtt{2'/m'}$ & $A_{2g}$ & 4 & \xmark &\cmark & \cmark & & $h_{y}$, $h_{z}$\\[0.4em]
Cr$_{2}$O$_{3}$ & 0001 & $\mathtt{\bar{3}m}$ & $\mathtt{\bar{3}\,\!'m'}$ & $A_{2u}$ & 1 & \xmark & \xmark & \xmark & 0 & 0\\
\hline
MnTe & $11\bar{2}0$ & $\mathtt{6/mmm}$ & $\mathtt{mmm}$ & $B_{2g}$ & 4 & \xmark & \xmark & \cmark & \multirow{3}{*}{
\begin{tabular}{c} $T_{z}N_{y}\left(3N_{x}^{2}-N_{y}^{2}\right)$ \\ $N_{z}\left( T_{y}(N_{y}^{2}-N_{x}^{2}) - 2 T_{x}N_{x}N_{y}\right)$ \end{tabular}
} & 0\\
& \multirow{2}{*}{$\bar{1}100$} & \multirow{2}{*}{$\mathtt{6/mmm}$} & \multirow{2}{*}{$\mathtt{m'm'm}$} & \multirow{2}{*}{$B_{2g}$} & \multirow{2}{*}{4} & \multirow{2}{*}{\xmark} & \multirow{2}{*}{\xmark} & \multirow{2}{*}{\cmark} &  & \multirow{2}{*}{$h_{z}$} \\
& & & & & & & & & & \\[0.4em]
VNb$_{3}$S$_{6}$ & $01\bar{1}0$ & $\mathtt{622}$ & $\mathtt{222}$ & $B_{1}$ & 3 &\xmark & \xmark & \cmark & \multirow{3}{*}{\begin{tabular}{c} $T_{z}N_{x}\left(N_{x}^{2}-3N_{y}^{2}\right)$ \\ $ N_{z}\left( T_{x}(N_{y}^{2}-N_{x}^{2})+2 T_{y}N_{x}N_{y}\right)$ \end{tabular}} & 0 \\
& \multirow{2}{*}{$2\bar{1}\bar{1}0$} & \multirow{2}{*}{$\mathtt{622}$} & \multirow{2}{*}{$\mathtt{2'2'2}$} & \multirow{2}{*}{$B_{1}$} & \multirow{2}{*}{3} & \multirow{2}{*}{\xmark} & \multirow{2}{*}{\xmark} & \multirow{2}{*}{\cmark} &  & \multirow{2}{*}{$h_{z}$}\\
& & & & & & & & &  &  \\
\hline\hline
\end{tabular}
    \end{minipage}}
\end{table*}
%\twocolumngrid
 
\textit{Linear antisymmetric.} For the entries of Tab.~\ref{tab:LT_linear_anti}, the guaranteed tensor of Ref.~\cite{schiff2025collinear} ($aV$ for $n=1$, $aeV^{2}$ for $n=3$, and $aeV[V^{2}]$ for $n=4,6$) shares irreducible components with $aeV$, allowing $T_{x}N_{y}-T_{y}N_{x}$ and $\bm{h}\propto\bm{N}\times\hat{\bm{z}}$. Equivalently, $\Gamma_{\bm{N}}\subset\{V^{2}\}$. Because $\mathcal{Q}$ in Eq.~\ref{eq:T_N_Q} is a symmetric matrix, it cannot produce an antisymmetric coupling, and the AMD in this class vanishes at zero valence SOC. The coupling has the Dzyaloshinskii--Moriya form, so $\bm{h}$ is parallel to the weak-ferromagnetic canting. For monoclinic and orthorhombic crystal point groups ($n=1,2$), $\Gamma_{\bm{N}}$ is contained in both $[V^{2}]$ and $\{V^{2}\}$, so the two linear classes coexist (rows marked $^{\ast}$), and both symmetric and antisymmetric linear contributions arise.
 
\textit{Cubic.} Where no $aeV$ component is available, the leading XMCD response is cubic in $\bm{N}$, and these cases are given in Tab.~\ref{tab:LT_cubic}. This requires $\Gamma_{\bm{N}}\otimes aeV$ to share irreps with $ae[V^{3}]$\footnote{Cubic couplings between $\bm{T}$ and $\bm{N}$, or the cubic dependence of $\bm{h}$ on $\bm{N}$ require $aeV$ to be contained in $[(\Gamma_{\bm{N}}\otimes aeV)^{3}] = \Gamma_{\bm{N}}\otimes ae[V^{3}]$.}. The guaranteed tensors of Ref.~\cite{schiff2025collinear} always provide such irreps. In seven cases, indicated by $\tilde{b}_{i}$ coefficients in Tab.~\ref{tab:LT_cubic}, additional terms exist \emph{not} predicted by the guaranteed tensors. Because the FVS Hall vector is strictly linear in $\bm{N}$, it vanishes identically in this class, even when the MPG allows finite $\bm{h}$~\cite{kunes2025xmcd}. Thus, the cases in Tab.~\ref{tab:LT_cubic} require valence SOC for $\bm{T}$, and valence SOC or multipolar core–valence interactions for $\bm{h}$. %Note that Tab.~\ref{tab:LT_cubic} only lists the cubic terms for $\Gamma_{\bm{N}}$ where the lowest-order XMCD is cubic.  

%%%%%%%%%%%%%%%%%%%%%%%%%%%%%%%%%%%%%%%%%%%%%%%%%%%%  MATERIALS  %%%%%%%%%%%%%
%%%%%%%%%%%%%%%%%%%%%%%%%%%%%%%%%%%%%%%

\section{Material Examples}\label{sec:material_examples}

Having laid out the symmetry considerations for XMCD in altermagnets, we illustrate our conclusions with examples. To validate these results, we provide first-principles calculations of the XAS and XMCD spectra, as described in Sec.~\ref{subsec:1st_principles}. We use six materials, two for each of the tetragonal, trigonal and hexagonal systems, in Sec.~\ref{subsec:ex_tetragonal}, Sec.~\ref{subsec:ex_trigonal} and Sec.~\ref{subsec:ex_hexagonal}, respectively. In each case, we demonstrate the symmetry-allowed $\bm{h}$ and $\bm{T}$ as outlined in Sec.~\ref{subsec:allowed_terms}. Then, we identify the spin-orbit free Landau theory, demonstrate the N\'eel vector dependence of the XMCD as in Sec.~\ref{subsec:MPG_AM_XMCD}, and highlight the linear or cubic order at which XMCD appears, following Sec.~\ref{sec:LT}. The symmetry properties of these materials are summarized in Tab.~\ref{tab:material_properties}. Finally, we compare our results to those established in the literature, where relevant. For VNb$_{3}$S$_{6},$ whose XMCD has not yet been studied, we make explicit predictions.  %To validate these results, we provide first-principles calculations of the XAS and XMCD spectra, as described in Sec.~\ref{subsec:1st_principles}.

%To demonstrate the validity of the symmetry-based predictions, we present first principles calculations on six materials, four of which are expected to have finite XMCD in their true ground state configurations. 

\subsection{First-principles computational approach}\label{subsec:1st_principles}

X-ray absorption spectra (XAS) are calculated using the Bethe-Salpeter equation (BSE) approach as implemented in the {\sc ocean} code version 3.3.1 \cite{OCEAN2022}. The BSE treats the equation of motion of a single electron-hole pair excited from the ground state, calculated from density-functional theory using the {\sc Quantum ESPRESSO} code version 7.5 \cite{QE}. We used the Perdew-Becke-Ernzerhof (PBE) exchange-correlation functional \cite{GGA_PhysRevLett.77.3865}, and norm-conserving pseudopotentials from PseudoDojo \cite{PseudoDojo,ONCV_PhysRevB.88.085117}. Structures for each material were taken directly from the Materials Project \cite{Jain2013}. All calculations were done using collinear spin, neglecting valence SOC. Within our approach, the direction of the Néel vector is set as a parameter (see App.~\ref{app:BSeq_approach}). For each material, the isotropic average dipole absorption is presented together with the calculated XMCD for specific photon directions. 
Complete details of the {\sc ocean} calculations including input files and spectra are available \cite{OCEAN-repo}.

An advantage of the BSE approach is the lack of free parameters beyond the lattice vectors and atomic positions. 
In addition, the code does not make use of symmetries; the selection rules for electron-hole and electron-photon interactions are enforced through numerical integration, providing an independent check on the predictions based on symmetry and magnetic groups. However, the BSE has limitations at the 3{\it d} transition metal L edges we investigate here \cite{PhysRevB.86.195135, deGroot}. The so-called multiplets, which can be thought of as secondary {\it d-d} or charge-transfer excitations, are beyond the lowest-order perturbations treated within our BSE approach. These excitations lead to additional structure at both L$_3$ and L$_2$ that is absent in our calculations. 
%\clearpage

\begin{figure*}[t!]
\caption{The XAS and magnetic structures for (a) MnF$_2$ with $\bm{N}\,\|\,[001]$; (b) fictitious MnF$_2$ with re-oriented N\'eel vector $\bm{N}\,\|\,[100]$; (c) NiF$_2$ for $\bm{N}\,\|\,[100]$ (neglecting canting). Figures (d), (e) and (f) depict the corresponding XMCD spectra for photon propagation $\bm{\hat{k}}$ along three cartesian directions, coinciding with $\bm{a}$, $\bm{b}$ and $\bm{c}$.} 
\label{fig:Rutiles}
\includegraphics[width=\textwidth]{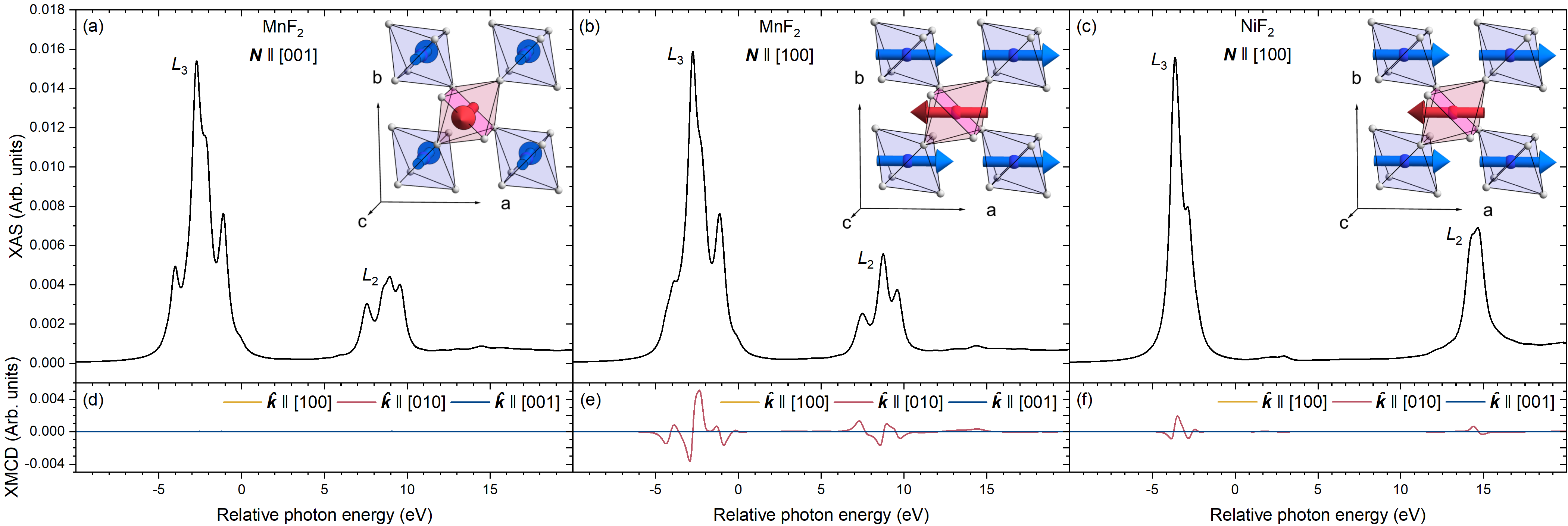}
%\vspace{-4cm}
\end{figure*}

%\vspace{.5cm}
\subsection{Tetragonal symmetry rutile: NiF$_{2}$ and MnF$_{2}$}\label{subsec:ex_tetragonal}

\begin{comment}
\begin{figure}[h!]
\includegraphics[width=\columnwidth]{figs/inset_figs/MnF2_L001_inset.png}
\includegraphics[width=\columnwidth]{figs/inset_figs/MnF2_L100_inset.png}
\includegraphics[width=\columnwidth]{figs/inset_figs/NiF2_L100_inset.png}
\caption{Top) MnF$_2$ for $\bm{N}$=[001] with respect to the three cartesian directions. Middle) Fictitious MnF$_2$ for $\bm{N}$=[100] with respect to the three cartesian directions. Bottom) NiF$_2$ for $\bm{N}$=[100] } 
\label{fig:Rutiles}
\end{figure}
\end{comment}

We begin our discussion with the rutile altermagnets, whose crystal structure has point group symmetry $\mathtt{4/mmm}.$ We will use NiF$_{2}$ (\href{https://next-gen.materialsproject.org/materials/mp-559798}{mp-559798} \cite{Jain2013}, \href{https://cryst.ehu.es/magndata/index.php?this_label=0.36}{\textsc{magndata}-0.36} \cite{Gallego2016_MAGNDATA}) and MnF$_{2}$ (\href{https://next-gen.materialsproject.org/materials/mp-560902}{mp-560902} \cite{Jain2013}, \href{https://cryst.ehu.es/magndata/index.php?this_label=0.15}{\textsc{magndata}-0.15} \cite{Gallego2016_MAGNDATA}) as canonical examples, differing principally by their N\'eel vector orientations, and also in their net moments. The latter has $\bm{N}\,\|\,\bm{c}$ and remains compensated, while in the former $\bm{N}\,\|\,\bm{a}$ and it develops a net moment. First, we introduce the crystallographic and magnetic symmetries of these two systems.
%\FloatBarrier

For both, the crystal space group is $\mathtt{P4_{2}/mnm}$ (\#136), with the magnetic ions occupying the $2a$ Wyckoff position. The magnetic structures are shown in the insets of Fig.~\ref{fig:Rutiles}. The four-fold rotations and roto-inversion along $\bm{c}$, the two-fold rotations aligned with $\bm{a}$ and $\bm{b}$, and the mirrors perpendicular to $\bm{a}$ and $\bm{b}$ are paired with fractional translations along the unit cell body diagonal. MnF$_{2}$ has a N\'eel temperature of roughly $67~\mathrm{K}$, whereas in NiF$_{2}$ it is $73~\mathrm{K}$ \cite{deHaas1940_MnF2NiF2_MagProperties,Stout1942_MnF2,Stout1949_MnF2_Mag,Griffel1950_MnF2Structure,Stout1954_MagAnisoNiF2,Alikhanov1960_NiF2,Brown1981_NiF2,Baur1958_RutileDiflourides,Baur1971_RutileStructures,Yamani2010_MnF2_NeutronTeaching}. The MPG of NiF$_{2}$ is $\mathtt{m'm'm}$ \cite{Moriya1960_ThrMag_NiF2,Chen2025_WithMaterialSupp}, while for MnF$_{2}$ it is $\mathtt{4'/mm'm}$ \cite{dzialoshinskii1958_piezo_MnF2_Fe2O3,Chen2025_WithMaterialSupp}.

We can immediately see from Eq.~\ref{eq:TVec_TransformationLaw} that in MnF$_{2}$, the net AMD must vanish. The two-fold rotation about $\bm{c}$ forces $T_{x}$ and $T_{y}$ to vanish, and the four-fold rotation composed with $\mathcal{T}$ causes the remaining $T_{z}$ to vanish. Because $\bm{T}$ and $\bm{h}$ transform identically, the same argument applies to the Hall vector. The MPG $\mathtt{4'/mm'm}$ is able to describe an altermagnetic order, and so it appears in Tab.~\ref{tab:AM_MPG} \footnote{In Tab.~\ref{tab:AM_MPG}, the standard setting $\mathtt{4'/mm'm}$ is reported. The MPG of MnF$_{2}$ has the $\bm{a}$ and $\bm{b}$ axes rotated by $45^{\circ}$ with respect to the standard setting.}. However, it forbids a net time-odd axial vector such as $\bm{h}$, $\bm{T}$ or the net magnetization. As such, it does not appear in Tab.~\ref{tab:tvec_short}, and it does not produce an XMCD spectrum along any direction, as confirmed by first-principles calculation in Fig.~\ref{fig:Rutiles}(d).

On the other hand, in NiF$_{2}$, the antiunitary (primed) two-fold rotations along $\bm{a}$ and $\bm{c}$ force $T_{x}$ and $T_{z}$ to vanish, but the component $T_{y}$ along $\bm{b}$ survives. Consequently, $\mathtt{m'm'm}$ appears in both Tab.~\ref{tab:tvec_short} and Tab.~\ref{tab:AM_MPG}. Note that, as described in Sec.~\ref{subsec:experimental_geometries}, in Tab.~\ref{tab:tvec_short} the standard setting is used, where the $\bm{c}$ axis two-fold rotation is unitary (un-primed). To transform into the setting of NiF$_{2}$, one can rotate the standard setting by $90^{\circ}$ about the $\bm{a}$ axis. From this result, the XMCD experiment realized with the photon propagation direction $\bm{\hat{k}}\,\|\,\bm{b}$ should yield a non-zero spectrum, while for other directions the spectrum should vanish. This is demonstrated in the spectra of Fig.~\ref{fig:Rutiles}(f). Notably, when the N\'eel vector in MnF$_{2}$ is artificially set along the $\bm{a}$ axis, its symmetry also corresponds to $\mathtt{m'm'm}$ and an XMCD signal is allowed. This is confirmed by first-principles calculation in Fig.~\ref{fig:Rutiles}(e). We also highlight that $\mathtt{m'm'm}$ allows a net magnetic moment to develop, which is observed in NiF$_{2}$; indeed, the antiferromagnetic order is canted by roughly $0.9^{\circ}$ \cite{Alikhanov1960_NiF2,Stout1954_MagAnisoNiF2,Moriya1960_ThrMag_NiF2,borovikromanov1973_NiF2,Brown1981_NiF2}. This phenomenon has recently been called spin-orbit magnetism, in which the non-relativistic symmetries forbid a net moment, while relativistic symmetries permit it \cite{Liu2026_OrientedSSG,Wang2026_SOM_altermagnets}. Because this net moment is not required for finite XMCD, it has been neglected in the first principles calculation in Fig.~\ref{fig:Rutiles}(f). 

We now show that these predictions from the relativistic MPGs arise naturally in the Landau theory language. At zero (valence) SOC, real-space transformations from the crystal point group do not transform any degree of freedom in the spin space. At most, they can swap or preserve the up- and down-spin sublattices. For both NiF$_{2}$ and MnF$_{2}$, the sublattice-swapping pattern of the crystal point group elements is given by the $\Gamma_{\bm{N}} = B_{2g}$ irrep of $\mathtt{4/mmm}$. The four-fold elements, and the twofold rotations (mirrors) along (perpendicular to) $\bm{a}$ and $\bm{b}$ swap sublattices, while the diagonal two-fold rotations and mirrors preserve them; at zero SOC, the sublattice-swapping elements force the net moment to vanish.

To understand the limit of vanishing valence spin-orbit coupling, we check whether $\Gamma_{\bm{N}} = B_{2g}$ is contained in $[V^{2}]$ of the crystallographic point group $\mathtt{4/mmm}$. The six-dimensional symmetric vector square $[V^{2}]$ decomposes as $2A_{1g} \oplus B_{1g}\oplus B_{2g} \oplus E_{g}$, which contains $B_{2g}$, corresponding to non-zero $\mathcal{Q}_{xy}$ in the non-relativistic limit (see Eq.~\ref{eq:T_N_Q}). Indeed, precisely these components appear in the first-principles calculation Eq.~\ref{eq:Tij_MnF2}, and from Eq.~\ref{eq:T_N_Q}, give rise to non-vanishing $\bm{T}$ whenever $\bm{N}$ lies in the $\bm{ab}$-plane.

When SOC is restored, $B_{2g} \subset [V^{2}]$ means that $\bm{h}$ and $\bm{T}$ appear as linear order, symmetric effects. We see from Tab.~\ref{tab:LT_linear_symm} that  $\bm{h}\propto(N_{y},N_{x},0)^{\mathsf{T}}$ and $T_{x}N_{y} +T_{y}N_{x}$ describes the linear coupling between $\bm{T}$ and $\bm{N}$. Both the AMD and the Hall vector vanish for $\bm{N}\,\|\,\bm{c}$ as in MnF$_{2}$, while $\bm{N}\,\|\,\bm{a}$ as in NiF$_{2}$ reproduces the non-zero $h_{y}$ predicted in Sec.~\ref{sec:allowed_Th}. In combination with the symmetric form of $\bm{h}$, this approach reproduces the N\'eel vector dependence of $\bm{h}$ reported in Ref.~\cite{kunes2025xmcd}. These symmetry properties for MnF$_{2}$ and NiF$_{2}$ are summarized in Tab.~\ref{tab:material_properties}. 

%\textcolor{red}{Still need to compare with other predicted spectra and XMCD measurements!!!}

\begin{figure*}[t!]
\caption{The XAS and magnetic structures for (a) Fe$_2$O$_{3}$ with $\bm{N}\,\|\,[2\bar{1}\bar{1}0]$; (b) Fe$_2$O$_{3}$ for $\bm{N}\,\|\,[01\bar{1}0]$; and (c) Fe$_2$O$_{3}$ for $\bm{N}\,\|\,[0001]$. Figures (d), (e) and (f) depict the corresponding XMCD spectra for photon propagation $\bm{\hat{k}}$ along three cartesian directions. The spectra for $\bm{\hat{k}}\,\|\,[01\bar{1}0]$ in (d) and $\bm{\hat{k}}\,\|\,[2\bar{1}\bar{1}0]$ in (e) are multiplied $(\times5)$ for visibility.
%Top) Fe$_2$O$_3$ for $\bm{N}$=[001] with respect to the three cartesian directions (all three are zero to the numerical noise of the calculation). Middle) Fe$_2$O$_3$ for $\bm{N}$=[100] (the XMCD signal of $\bm{\hat{k}}$=[010] is non-zero, but much smaller than [001]). Bottom) Fe$_2$O$_3$ for $\bm{N}$=[010] shows a small but non-zero signal for $\bm{\hat{k}}$=[100]. 
}
\includegraphics[width=\textwidth]{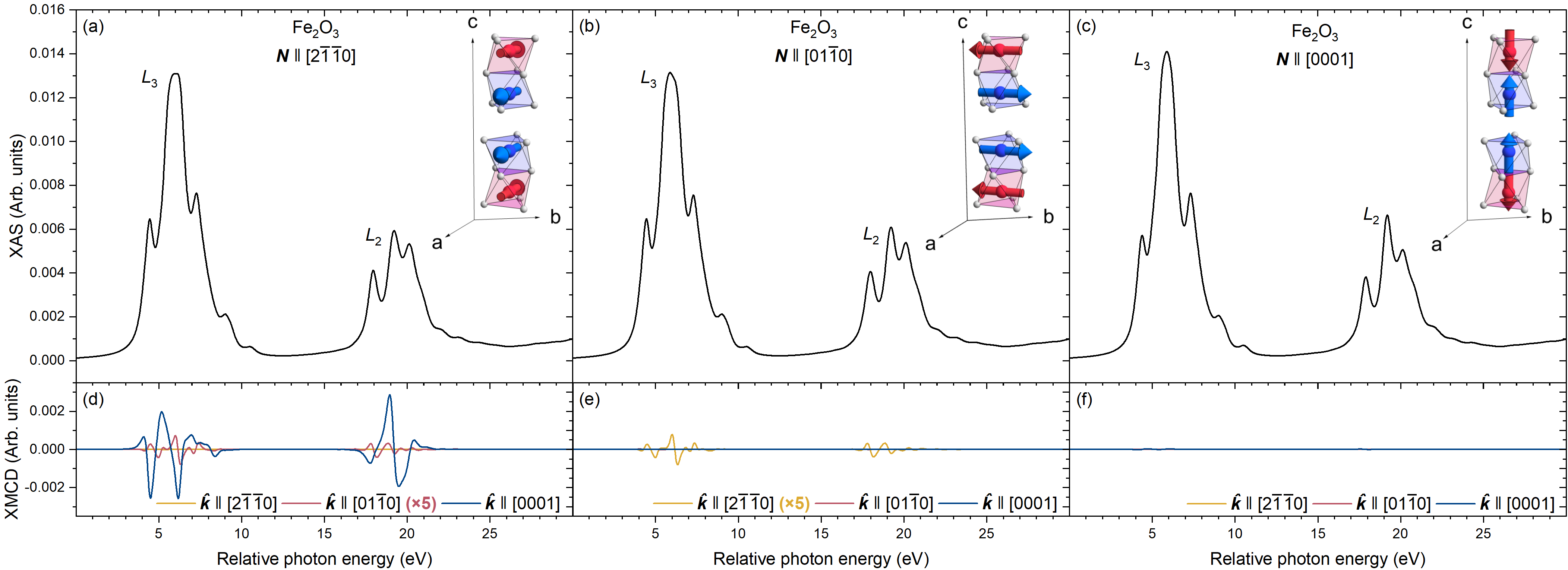}
\label{fig:Fe2O3}
\end{figure*}

The predictions outlined here align with existing literature. In Ref.~\cite{hariki2024rutile_MnF2_XMCD}, XMCD spectra are numerically simulated in MnF$_{2}$, finding $\bm{h}$ vanishes for $\bm{N}\,\|\,\bm{c}$, also consistent with Ref.~\cite{Mizumaki2026_XAS_XES_MCD}, but is finite for in-plane $\bm{N}$. They also establish symmetry relationship between $\bm{h}$ and $\bm{N}$ that is generalized in Ref.~\cite{kunes2025xmcd} and reproduced by our linear order terms in the Landau theory. Further, Ref.~\cite{hariki2025separating} theoretically predicted and experimentally measured (in Ref.~\cite{li2026_NiF2_XMCD}) the possibility of separating altermagnetic and ferromagnetic contributions to the XMCD in NiF$_{2}$ on the basis of their differing contributions to $\bm{h}$, the component from $\bm{N}$ being consistent with our results here. We briefly note also that Ref.~\cite{yang2026_strain_MnF2} finds the same symmetry constraint on anomalous Hall and anomalous magneto-optic transport properties: they must vanish when $\bm{N}\,\|\,\bm{c}$, and the symmetry analysis for RuO$_{2}$, explored in \cite{Sasabe2023_AMD_XMCD_RuO2}, is identical to the two rutiles studied here.

%\clearpage

\subsection{Trigonal symmetry corundum: Fe$_{2}$O$_{3}$ and Cr$_{2}$O$_{3}$}\label{subsec:ex_trigonal}

Next, we explore two examples with trigonal symmetry, hematite (Fe$_{2}$O$_{3}$, $T_{\bm{N}}\approx 950~\mathrm{K}$ \cite{Shull1951_Fe2O3,hill2008_neutron_Fe2O3}, \href{https://next-gen.materialsproject.org/materials/mp-19770}{\mbox{mp-19770}} \cite{Jain2013}, \href{https://cryst.ehu.es/magndata/index.php?this_label=0.65}{\textsc{magndata}-0.65} \cite{Gallego2016_MAGNDATA}) and chromia (Cr$_{2}$O$_{3}$, $T_{\bm{N}}\approx307~\mathrm{K}$ \cite{brockhouse1953_Cr2O3,odell1970_Cr2O3_electrodynamics,street2014_increasing_Cr2O3_TN}, \href{https://next-gen.materialsproject.org/materials/mp-19399}{mp-19399} \cite{Jain2013}, \href{https://cryst.ehu.es/magndata/index.php?this_label=0.59}{\textsc{magndata}-0.59} \cite{Gallego2016_MAGNDATA}), both of which have space group symmetry $\mathtt{R\bar{3}c}$ (\#167) with their magnetic ions occupying the $12c$ Wyckoff position. Here, the threefold axes are parallel to $\bm{c}$, the twofold rotation axes are parallel to $\bm{a}$ and its $120^{\circ}$ rotated counterparts, and the mirrors are perpendicular to these twofold axes. The twofold rotations and mirrors are paired with half-translations along $\bm{c}$ \cite{dzialoshinskii1957_WeakFM_hematite,dzialoshinskii1958_piezo_MnF2_Fe2O3}.

\begin{figure}[b!]
\includegraphics[width=\columnwidth]{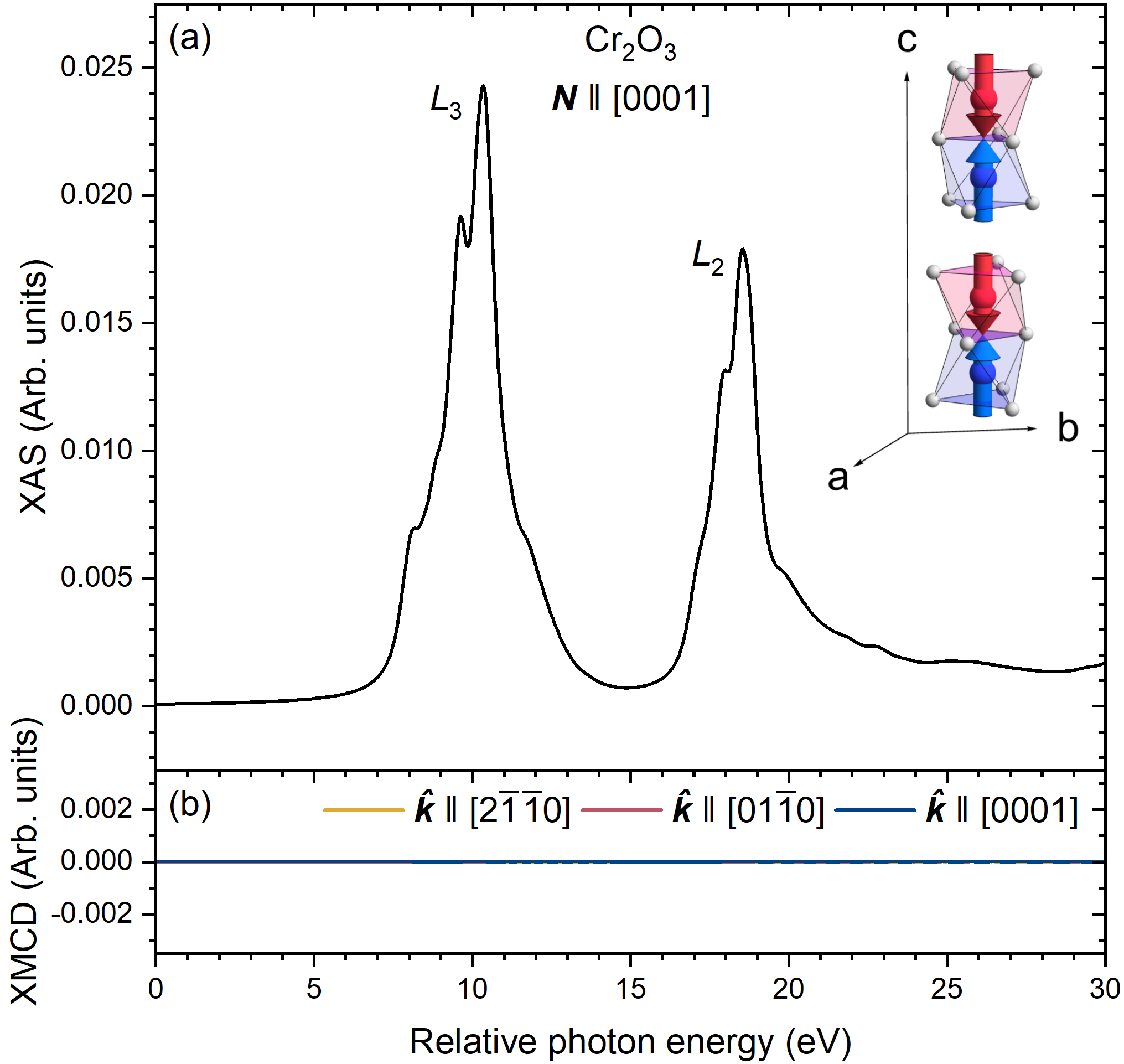}
\caption{For Cr$_2$O$_3$ with $\bm{N}\,\|\,[0001]$, (a) the XAS spectrum and (b) the vanishing XMCD for photon propagation $\bm{\hat{k}}$ along three cartesian directions.} 
\label{fig:Cr2O3}
\end{figure}

Both of these materials have crystal point group $\mathtt{\bar{3}m}$, but they differ in their sublattice structure, and in some cases in their N\'eel vector orientations. Chromia has $\bm{N}\,\|\,\bm{c}$, and along the $\bm{c}$ direction, the moments within one unit cell take on a $+-+-$ pattern (with $\pm$ referring to up- and down-spin directions). This magnetic structure, shown in the inset of Fig.~\ref{fig:Cr2O3}, leads to $\mathtt{\bar{3}\,\!'m'}$ as the MPG for chromia \cite{brockhouse1953_Cr2O3,dzialoshinskii1957_WeakFM_hematite}. Hematite, on the other hand, is known to have two dominant phases separated by the Morin transition ($T_{M} \approx 260~\mathrm{K}$). Above $T_{M}$, the N\'eel vector lies in the basal plane and admits canting that leads to a weak ferromagnetic moment. In this case, depending on whether the N\'eel vector lies along the $[01\bar{1}0]$ or the $[2\bar{1}\bar{1}0]$ directions, the MPG is $\mathtt{2/m}$ or $\mathtt{2'/m'},$ respectively. Below $T_{M},$ $\bm{N}\,\|\,\bm{c}$ and remains compensated, with MPG $\mathtt{\bar{3}m}$ \cite{dzialoshinskii1958_piezo_MnF2_Fe2O3,Lovesey2022_NeutronsMPG_Fe2O3,Chen2025_WithMaterialSupp,luo2025_Fe2O3_MOKE,ishii2026_Fe2O3_FMDistinctFromAM}. Both above and below $T_{M},$ the moments take on a $+--+$ sublattice pattern \cite{Shull1951_Fe2O3,hill2008_neutron_Fe2O3,dzialoshinskii1958_piezo_MnF2_Fe2O3,Lovesey2022_NeutronsMPG_Fe2O3,Verbeek2024_Fe2O3_ferroMultipole}, and the corresponding magnetic structures are depicted in the insets of Fig.~\ref{fig:Fe2O3}.

In the following, we will refer to the $x$-components as those along $\bm{a}$ (along the $[2\bar{1}\bar{1}0]$ direction), the $z$-components as those along $\bm{c}$ (the $[0001]$ direction), and the $y$-components as those perpendicular to both $\bm{a}$ and $\bm{c}$ in a right-handed Cartesian coordinate system (along $[01\bar{1}0]$) \footnote{This frame differs from that of Tab.~\ref{tab:LT_linear_anti} and App.~\ref{app:LT_couplings}, in which the twofold axes of $\bar{3}m$ are taken along $\hat{y}$; the two are related by a $90^\circ$ rotation about $\bm{c}$. The linear coupling $N_xT_y - N_y T_x$ is invariant under this rotation, while the cubic invariant $N_y(3N_x^2 - N_y^2)T_z$ of the $\hat{y}$-twofold frame becomes $N_x(N_x^2 - 3N_y^2)T_z$ up to an overall sign.}. When $\bm{N}\,\|\,\bm{c}$ in both Fe$_{2}$O$_{3}$ and Cr$_{2}$O$_{3}$, the XMCD must vanish, from Eq.~\ref{eq:TVec_TransformationLaw}. In both cases, the threefold rotations about $\bm{c}$ forbid the in-plane components of $\bm{h}$ and $\bm{T}$, while the twofold rotations perpendicular to $\bm{c}$ forbid the component along $\bm{c}.$ While $\mathtt{\bar{3}m}$ describing altermagnetic Fe$_{2}$O$_{3}$ appears in Tab.~\ref{tab:AM_MPG}, it does not appear in Tab.~\ref{tab:tvec_short} as it cannot support XMCD. On the other hand, $\mathtt{\bar{3}\,\!'m'}$ describing Cr$_{2}$O$_{3}$ cannot describe an altermagnet, nor can it support XMCD, and so it does not appear in either table. These conclusions are consistent with the spectra in Fig.~\ref{fig:Fe2O3}(f) and Fig.~\ref{fig:Cr2O3}(b).

When $\bm{N} \,\|\, \bm{a}$ in Fe$_{2}$O$_{3}$, the primed twofold rotation of $\mathtt{2'/m'}$ is parallel to $\bm{N}$ and it forbids the $x$-component of both the Hall vector and the AMD. The remaining $y$- and $z$-components are allowed. In Tab.~\ref{tab:tvec_short}, the group axes, inherited from the \textsc{Bilbao Crystallographic Server}, take the $y$-axis to be primed. A rotation by $90^{\circ}$ about $\bm{c}$ takes the \textsc{Bilbao} setting into that of hematite, recovering our solution, and matching the prediction of Fig.~\ref{fig:Fe2O3}(d).

When $\bm{N}\perp\bm{a}$, along the $[01\bar{1}0]$ direction, the twofold axis of $\mathtt{2/m}$ is perpendicular to $\bm{N}$, and it forbids the $y-$ and $z-$components of $\bm{h}$ and $\bm{T}$, while the $x$-components survive. This corresponds to the prediction in Fig.~\ref{fig:Fe2O3}(e), and the results of Tab.~\ref{tab:tvec_short} once the \textsc{Bilbao} setting, whose twofold rotation is along the $y$-axis, is rotated by $90^{\circ}$ about $\bm{c}$.

With the direct MPG results established, let us again demonstrate that the Landau theory reproduces these results. Here, unlike our two rutile examples, hematite and chromia differ significantly, due to their different spin sublattice patterns. At zero (valence) SOC, hematite's N\'eel vector transforms according to the $A_{2g}$ irrep of $\mathtt{\bar{3}m},$ while in chromia it transforms as $A_{2u}.$ 

Notably, in chromia, $\bm{N}$ is odd under spatial inversion and thus it is not altermagnetic. So, it cannot couple to inversion-even quantities, like net axial vectors. This is the reason for the zero in chromia's Landau column in Tab.~\ref{tab:material_properties}. It is the only such example we shall consider, but serves as an important example: non-altermagnetic collinear AFMs cannot possess an XMCD signal.

We turn our attention now to the Landau theory for hematite, with $\Gamma_{\bm{N}} = A_{2g}$. The symmetric vector square of $\mathtt{\bar{3}m}$ decomposes as $[V^{2}] = 2A_{1g}\oplus2E_{g},$ which does not contain $A_{2g}.$ As a result, hematite has no XMCD in the limit of zero (valence) SOC, consistent with the predictions of the FVS approximation \cite{kunes2025xmcd}. Next, we check the antisymmetric vector square's decomposition, $\{V^{2}\} = A_{2g} \oplus E_{g}$; it contains $A_{2g}$, and gives rise to a linear, antisymmetric coupling $N_{x}T_{y}-N_{y}T_{x}$, and $\bm{h}\propto (N_{y}, - N_{x},0)^{\mathsf{T}}$ from Tab.~\ref{tab:LT_linear_anti}, and summarized in Tab.~\ref{tab:material_properties}. We recover the N\'eel vector dependence of the XMCD response: $\bm{N} \,\|\,\bm{c}$ yields vanishing XMCD, while in-plane $\bm{N}$ allows it. This N\'eel vector dependence is \emph{not} captured by the FVS approximation, and can be attributed to the effects of SOC or multipolar core-valence exchange. It also reproduces features of Fig.~\ref{fig:Fe2O3}(d)--(f). For $\bm{N}\,\|\,\bm{a}$, there is a finite, albeit small response when the photon propagation direction $\bm{\hat{k}}$ is parallel to $[01\bar{1}0]$. When $\bm{N}\perp\bm{a}$ in the basal plane, there is a small signal for $\bm{\hat{k}} \,\|\,\bm{a}$.

It remains then to understand the XMCD arising from $\bm{\hat{k}}\,\|\,\bm{c}$. It does not arise from the linear order, but rather from the cubic order, and thus is not captured in the FVS approximation. There are four cubic couplings, only one of which leads to finite $h_{z}$ and $T_{z}$:  $h_{z} \propto N_{x}(N_{x}^{2}-3N_{y}^{2}) $, and the free energy invariant $T_{z}N_{x}(N_{x}^{2}-3N_{y}^{2})$. It is not included in Tab.~\ref{tab:LT_cubic}, as this table only reports scenarios where the linear contributions vanish. We see that $h_{z}$ vanishes whenever $\bm{N}\perp\bm{a}$. The in-plane N\'eel vector can be parameterized by $\bm{N} = N(\cos\phi,\sin\phi,0)^{\mathsf{T}}$ with $\phi$ measured from the $\bm{a}$ axis. Then, we have $\bm{h}\propto \cos(3\phi)\bm{\hat{z}}$, which vanishes for $\phi = (2m+1)\frac{\pi}{6}$ ($\bm{N}$ along an axis perpendicular to $\bm{a}$ and its symmetry-related directions), and it is maximal whenever $\phi = m\frac{\pi}{3}$ (along $\bm{a}$ and its symmetry-related directions). 

%\textcolor{red}{Still need to compare with other predicted spectra and XMCD measurements!!!}

The results discussed here agree with the existing literature on Fe$_{2}$O$_{3}$; we are not
aware of bulk XMCD measurements on Cr$_{2}$O$_{3}$ in zero field. Density-Functional Theory (DFT) + multiplet ligand field theory calculations on Fe$_{2}$O$_{3}$ in
Ref.~\cite{xie2026fe2o3} find, in our frame, finite $h_{y}$ and $h_{z}$ for $\bm{N}\,\|\,\bm{a}$
(with $h_{y}$ giving only a very weak XMCD spectrum), a very small $h_{x}$ for
$\bm{N}\perp\bm{a}$ in the basal plane, and a vanishing Hall vector for $\bm{N}\,\|\,\bm{c}$.
Experimentally, the appearance of XMCD above $T_{M}$ was first observed in Ref.~\cite{Kuiper1993_Fe2O3_XMCD}, where it is recognized that for $\bm{N}\,\|\,\bm{c}$, the dichroism vanishes. Ref.~\cite{yamamoto2026_Fe2O3_XMCD} observes a finite XMCD for
$\bm{\hat{k}}\,\|\,\bm{c}$ above $T_{M}$ that vanishes below $T_{M}$, where $\bm{N}\,\|\,\bm{c}$.
Ref.~\cite{ishii2026_Fe2O3_FMDistinctFromAM}, measuring at room temperature with
$\bm{\hat{k}}\,\|\,\bm{c}$, finds a finite XMCD in the zero-field $\mathtt{2'/m'}$ state that is
suppressed when an in-plane field rotates $\bm{N}$ toward the $\mathtt{2/m}$ configuration, consistent with the node of $h_{z}\propto\cos(3\phi)$. Finally, Ref.~\cite{GalindezRuales2025_Fe2O3}
reports an anomalous Hall conductivity $\sigma_{xy}$, the dc counterpart of $h_{z}$, in
$1\%$ Ti-doped films that is absent in the easy-axis phase and appears only once a field-induced
spin flop brings $\bm{N}$ into the basal plane, with a sign that alternates between Hall bars
rotated by $60^{\circ}$; their XMCD-Photoemission Electron Microscopy (XMCD-PEEM) measurements likewise show no significant XMCD in the low-temperature collinear phase.

%\begin{itemize}
    %\item Ref.~\cite{xie2026fe2o3}: DFT$+$DMFT calculations. $\bm{N}\,\|\,[010],$ finite XMCD for $\bm{\hat{k}} \,\|\,[001] \,\&\, [100]$ (with $[100]$ being weak). With $\bm{N}\,\|\,\bm{c}$ $\bm{h}$ vanishes, with $\bm{N}\,\|\,[100]$ it exists along $h_{y}$, though \emph{very} small. Our frame is rotated with respect to theirs by $90^{\circ}$ about $\bm{c}$.
    %\item Ref.~\cite{ishii2026_Fe2O3_FMDistinctFromAM}: Measured \& calculated. Agree with the MPGs and vanishing vs. non-vanishing XMCD along $\bm{\hat{k}}\,\|\,\bm{c}.$ Fairly good qualitative agreement with their spectra at $0T$ (Fig.~2(a)). Our basis is rotated $30^{\circ}$ from theirs.
    %\item Ref.~\cite{GalindezRuales2025_Fe2O3}: agree with the MPGs. Thin film measurements $\sim 330K$, measuring $h_{z}$. Not super connected with our predictions, and I feel like their XMCD spectra are not very clean?
    %\item Ref.~\cite{yamamoto2026_Fe2O3_XMCD}: qualitatively good agreement with their calculation and experiment (Fig.~2(c,e)) which have $\bm{\hat{k}}\,\|\,\bm{c}$ and $T_{M}<T<T_{\bm{N}}$. Our frames agree.
%\end{itemize}

\FloatBarrier

\subsection{Hexagonal symmetry: MnTe and VNb$_{3}$S$_{6}$}\label{subsec:ex_hexagonal}

\begin{figure}%[h!]
\includegraphics[width=\columnwidth]{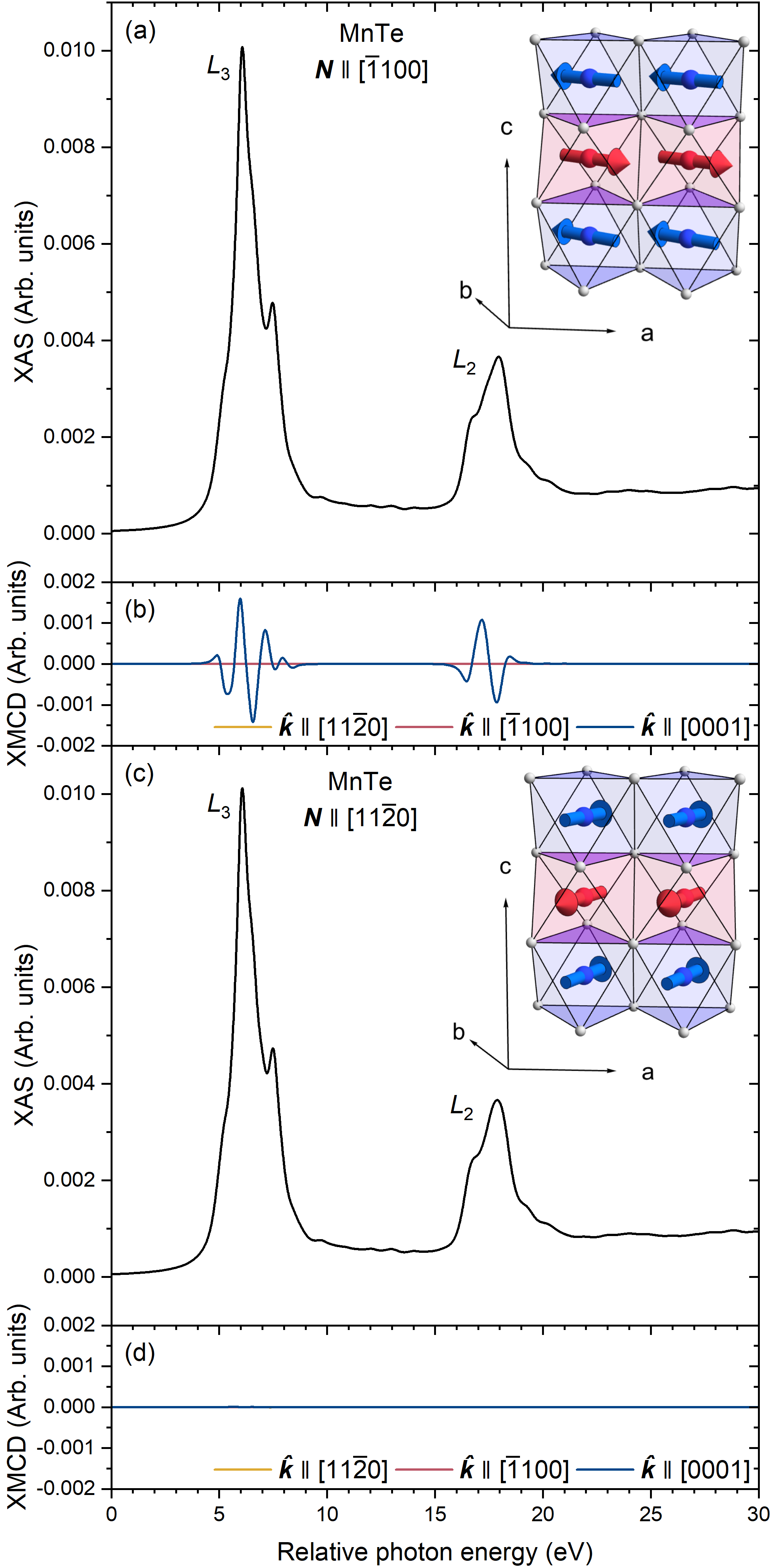}
\caption{MnTe XAS spectra for (a) $\bm{N}\,\|\,[\bar{1}100]$ and (c) $\bm{N}\,\|\,[11\bar{2}0]$. The corresponding XMCD spectra along three cartesian directions are shown in (b) and (d), respectively.}  
\label{fig:MnTe}
\end{figure}

Finally, we consider two example materials with hexagonal crystal symmetry, MnTe ($T_{\bm{N}}\approx 307~\mathrm{K}$ \cite{Kelley1939_MnTe_TN,Squire1939_MnTe_TN,Serres1947_MnTe_TN,Uchida1956_MnTe_TN,Komatsubara1963_MnTe_TN_MagOrder}, \href{https://next-gen.materialsproject.org/materials/mp-404}{\mbox{mp-404}} \cite{Jain2013}, \href{https://cryst.ehu.es/magndata/index.php?this_label=0.800}{\textsc{magndata}-0.800} \cite{Gallego2016_MAGNDATA}) and VNb$_{3}$S$_{6}$ ($T_{\bm{N}}\approx 50~\mathrm{K}$ \cite{Lu2020_VNb3S6_CrystStruct,Wang2023_VNb3S6_TN_AHE,Ray2025_VNb3S6_neutrons,Lygouras202_VNb3S6_MagStruct,Hall2021_VNb3S6_MagStruct}, \href{https://next-gen.materialsproject.org/materials/mp-15958}{mp-15958} \cite{Jain2013}, \href{https://cryst.ehu.es/magndata/index.php?this_label=0.712}{\textsc{magndata}-0.712} \cite{Gallego2016_MAGNDATA}). The space group for MnTe is $\mathtt{P6_{3}/mmc}$ (\#194) \cite{Kunitomi1964_MnTe_neutron} with the Mn ions residing on the $2a$ Wyckoff position. The space group contains sixfold screw axes along the $\bm{c}$ direction, inversion, two-fold axes along $\bm{a}$ and its $120^{\circ}$ rotated counterparts, as well as twofold rotations perpendicular to these that are paired with a half-translation along $\bm{c}$. The N\'eel vector lies within the basal plane, with an antiferromagnetic pattern along $\bm{c}$ and a ferromagnetic pattern within the plane \cite{Komatsubara1963_MnTe_TN_MagOrder,Greenwald1953_MnTe_Structure,Hirone1957_MnTe_NiAs_MagStructures,Kunitomi1964_MnTe_neutron,Kriegner2017_MnTe_MagStruct,Betancourt2023_MnTe_AHE_MagStruct}. Depending on whether the in-plane N\'eel vector lies along $[11\bar{2}0]$ or $[\bar{1}100]$, the magnetic point group is either $\mathtt{mmm}$ or $\mathtt{m'm'm},$ respectively \cite{Betancourt2023_MnTe_AHE_MagStruct,Chen2025_WithMaterialSupp,Liu2026_MnTe_MPGs_StrainAHE}. The latter is the experimentally actualized ground state. These structures are depicted in the insets of Fig.~\ref{fig:MnTe}.

\begin{figure}%[h!]
\includegraphics[width=\columnwidth]{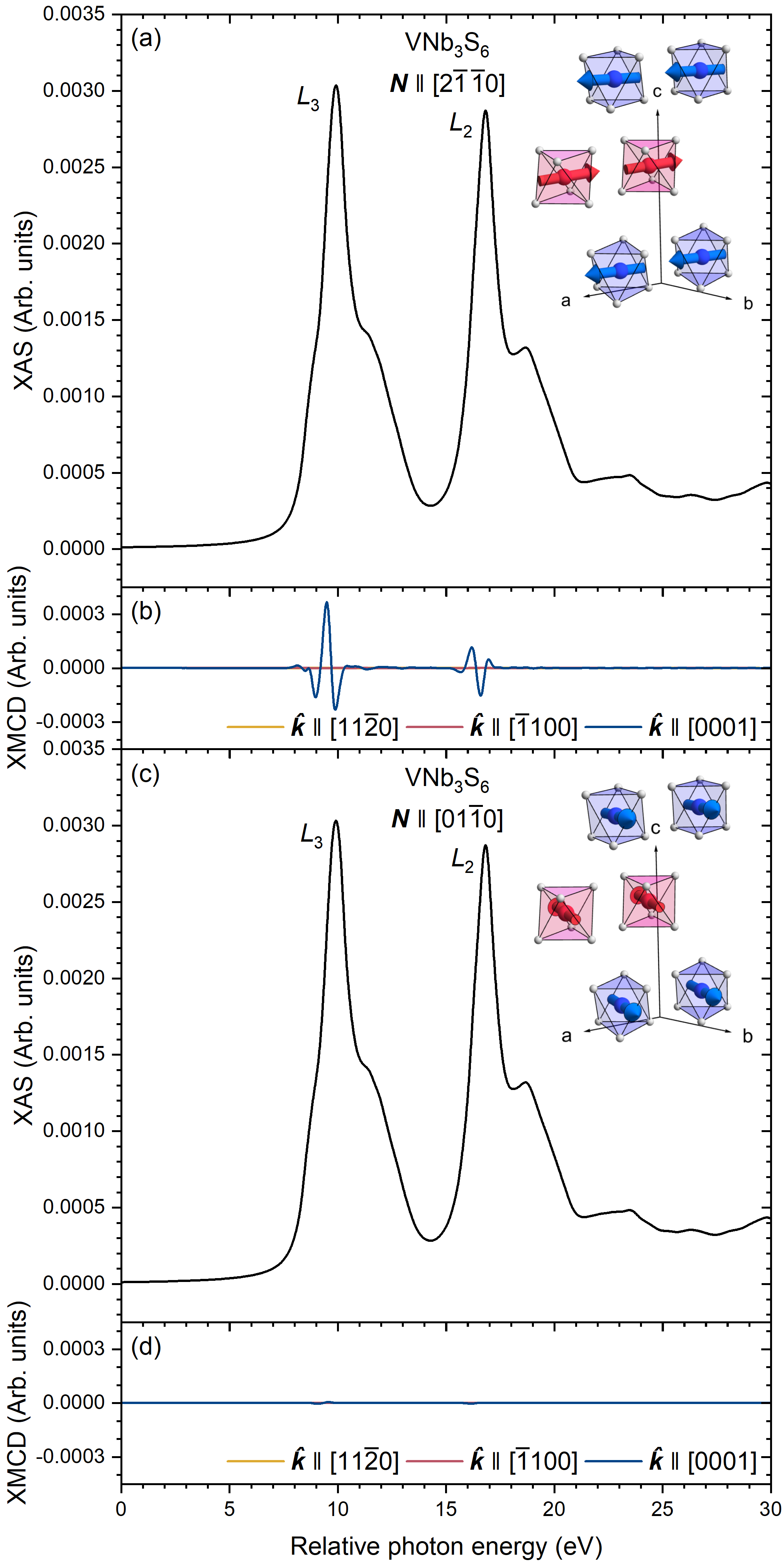}
\caption{VNb$_3$S$_6$ XAS spectra for (a) $\bm{N} \,\|\,[2\bar{1}\bar{1}0]$ and (c) $\bm{N}\,\|\,[01\bar{1}0]$. The corresponding XMCD spectra along three cartesian directions are shown in (b) and (d), respectively.}  
\label{fig:VNb3S6}
\end{figure}

For VNb$_{3}$S$_{6}$, the space group is $\mathtt{P6_{3}22}$ (\#182), with the magnetic V ions occupying the $2c$ Wyckoff position \cite{Lu2020_VNb3S6_CrystStruct,Ray2025_VNb3S6_neutrons,Hall2021_VNb3S6_MagStruct,Wang2023_VNb3S6_TN_AHE,Lygouras202_VNb3S6_MagStruct}. The space group is non-centrosymmetric, and has sixfold screw axes along $\bm{c}$ and twofold rotations along $\bm{a}$ and its $120^{\circ}$ rotated counterparts, as well as two-fold axes perpendicular to these paired with half-translations along $\bm{c}$. The N\'eel vector lies within basal plane, with moments alternating along $\bm{c}$ but aligned within the planes \cite{Lu2020_VNb3S6_CrystStruct,ghosh2025_VNb3S6_Raman_MagStruct,Ray2025_VNb3S6_neutrons,Lygouras202_VNb3S6_MagStruct,Hall2021_VNb3S6_MagStruct}. If the N\'eel vector lies along $[01\bar{1}0]$ the MPG is $\mathtt{222}$, while if it lies along $[2\bar{1}\bar{1}0]$, its MPG is $\mathtt{2'2'2}$ \cite{Lygouras202_VNb3S6_MagStruct,Ray2025_VNb3S6_neutrons,Hall2021_VNb3S6_MagStruct}. These magnetic structures are shown in the insets of Fig.~\ref{fig:VNb3S6}. Because these examples have different parent space groups, we will analyze them separately. Symmetry data for these materials are summarized in Tab.~\ref{tab:material_properties}. 

Let us begin with MnTe. With $\bm{N}\,\|\,[11\bar{2}0],$ the magnetic point group is given by $\mathtt{mmm}$. All elements are unitary, and it contains twofold axes along $[11\bar{2}0]$, $[\bar{1}100]$ and $\bm{c}$, mirrors perpendicular to these axes, and inversion. From Eq.~\ref{eq:TVec_TransformationLaw}, we can see that $\bm{T}$ and $\bm{h}$ must vanish: any two of the three twofold rotations are sufficient to forbid all three components of the AMD and the Hall vector. This is confirmed by the XMCD spectra in Fig.~\ref{fig:MnTe}. When $\bm{N}\,\|\,[\bar{1}100]$, the magnetic symmetry is given by $\mathtt{m'm'm}$, where the $\bm{c}$-axis twofold rotation, inversion and mirror perpendicular to $\bm{c}$ are unitary, whereas the $[\bar{1}100]$ and $[11\bar{2}0]$ twofold rotations and the mirrors perpendicular to these axes are paired with $\mathcal{T}$. Referring to the $x$-, $y$- and $z$-components as those along $[11\bar{2}0]$, $[\bar{1}100]$ and $\bm{c}$ respectively,  rotations and mirrors (unitary or not) forbid the $x$- and $y$-components of $\bm{T}$ and $\bm{h}$, while $T_{z}$ and $h_{z}$ survive, matching the XMCD spectrum in Fig.~\ref{fig:MnTe}, where only the scenario with photon propagation direction $\bm{\hat{k}}\,\|\,\bm{c}$ yields a non-zero signal. These results are also consistent with Tab.~\ref{tab:tvec_short} and Tab.~\ref{tab:AM_MPG}. 

Now we consider the MPG constraints imposed on $\bm{T}$ and $\bm{h}$ in VNb$_{3}$S$_{6}$. When $\bm{N}\,\|\,[01\bar{1}0]$, the MPG $\mathtt{222}$ is unitary, and consists of three twofold axes along $\bm{c}$, $[01\bar{1}0]$ and $[2\bar{1}\bar{1}0]$. From Eq.~\ref{eq:TVec_TransformationLaw}, any two of these elements are sufficient to forbid all three components the Hall vector and the AMD, which is consistent with the vanishing XMCD observed in Fig.~\ref{fig:VNb3S6}. When $\bm{N}\,\|\,[2\bar{1}\bar{1}0]$, the magnetic symmetry is given by $\mathtt{2'2'2}$, which contains a unitary twofold rotation along $\bm{c}$, and two antiunitary twofold rotations along $[01\bar{1}0]$ and $[2\bar{1}\bar{1}0]$. In this case, the components of $\bm{T}$ and $\bm{h}$ along $\bm{c}$ (which we call here the $z$-components) are allowed by symmetry, consistent with the observation of finite XMCD in Fig.~\ref{fig:VNb3S6}. These results coincide with the predictions of Tab.~\ref{tab:tvec_short} and Tab.~\ref{tab:AM_MPG}.

Having laid out the direct constraints from the MPGs, we will once again show that the Landau theory reproduces these results, following the procedure outlined in Sec.~\ref{sec:LT}. Beginning with MnTe, $\Gamma_{\bm{N}} = B_{2g}$ describes the N\'eel vector's transformations under the crystal point group $\mathtt{6/mmm}$ at zero SOC, regardless of the orientation of $\bm{N}$. The symmetrized and antisymmetrized vector square representations decompose as 
\begin{align*}
    [V^{2}] &= 2A_{1g} \oplus E_{1g}\oplus E_{2g}\\
    \{V^{2}\} &= A_{2g} \oplus E_{1g},
\end{align*}
and we see that \emph{neither} contains $B_{2g}.$ Thus, XMCD has no contribution at vanishing (valence) SOC and core--valence multipole exchange. Once these are restored it cannot appear at the linear order. Instead, because $$[V^{4}] = 3A_{1g} \oplus B_{1g}\oplus B_{2g} \oplus 2E_{1g} \oplus 3E_{2g},$$ which contains $B_{2g}$, both the Hall vector and the AMD appear at the cubic order. From Tab.~\ref{tab:LT_cubic}, we see that the coupling between $\bm{N}$ and $\bm{T}$ is given by two linearly independent invariants in the free energy,
\begin{align*}
    N_{z}\left(2T_{x}N_{x}N_{y} + T_{y}(N_{x}^{2}-N_{y}^{2}) \right)\quad \mathrm{and}\quad N_{y}T_{z}(3N_{x}^{2}-N_{y}^{2}),
\end{align*}
while the Hall vector has two symmetry-distinct cubic contributions,
\begin{align*}
    \bm{h}\propto\begin{pmatrix}
        2N_{x}N_{y}N_{z}\\
        N_{z}(N_{x}^{2}-N_{y}^{2})\\
        0
    \end{pmatrix} \quad \mathrm{and}\quad 
    \bm{h}\propto \begin{pmatrix}
        0\\0\\
        N_{y}(3N_{x}^{2}-N_{y}^{2})
    \end{pmatrix}.
\end{align*}
The first contribution to $\bm{T}$ and $\bm{h}$ requires $\bm{N}$ to have both an out-of-plane component and an in-plane component, i.e. $N_{z}\neq 0$ and $(N_{x},N_{y})\neq(0,0)$. The second contribution appears whenever $\bm{N}$ has a component in the basal plane, but vanishes if $N_{y}^{2} = 3N_{x}^{2}$ or if $N_{y} = 0$. We can parameterize this in-plane N\'eel vector as $\bm{N} = N(\cos\phi,\sin\phi,0)^{\mathsf{T}}$, such that $\phi = 0$ corresponds to the $[11\bar{2}0]$ direction \footnote{Because this frame is related to that of Tab.~\ref{tab:LT_cubic} by a sixfold rotation, under which the couplings change at most by an overall sign, the form of $\bm{h}$ is unchanged.}. Then, we find $\bm{h} \propto \sin(3\phi)\bm{\hat{z}},$ which vanishes whenever $\phi = m\frac{\pi}{3}$ for integer $m$, namely for $\bm{N}\,\|\,[11\bar{2}0]$ and all of its symmetry-related directions. This is consistent with the direct MPG analysis, and with the spectrum depicted in Fig.~\ref{fig:MnTe}. On the other hand, $\bm{h}$ is maximal when $\phi = (2m+1)\frac{\pi}{6},$ that is, for $\bm{N}\,\|\,[\bar{1}100]$ and its symmetry-related directions. This again matches the results of the direct MPG calculation, and the spectrum found in Fig.~\ref{fig:MnTe}. Note that both cubic contributions vanish whenever $\bm{N}\,\|\,\bm{c}$.

Finally, we examine the Landau theory for VNb$_{3}$S$_{6}$. Here, $\Gamma_{\bm{N}} = B_{1}$ describes the transformation of the N\'eel vector under the crystal point group $\mathtt{622}$ at zero SOC. Similar to the case of MnTe, we can see that $B_{1}$ is not contained in either $[V^{2}]$ or $\{V^{2}\}$, as
\begin{align*}
    [V^{2}]&= 2A_{1} \oplus E_{1} \oplus E_{2}\\
    \{V^{2}\} &= A_{2} \oplus E_{1},
\end{align*}
and as a result the XMCD for VNb$_{3}$S$_{6}$ vanishes in the limit of zero (valence) SOC, as in the FVS approximation. There are no linear contributions to the XMCD, but because $$[V^{4}] = 3A_{1} \oplus B_{1} \oplus B_{2} \oplus 2E_{1} \oplus 3E_{2},$$ which contains $B_{1}$, the Hall vector and AMD both appear at cubic order. Two cubic invariants between $\bm{N}$ and $\bm{T}$ may appear in the free energy,
$$N_{z}\left(T_{x}(N_{x}^{2}-N_{y}^{2})-2T_{y}N_{x}N_{y}\right)\quad \mathrm{and}\quad N_{x}T_{z}\left(N_{x}^{2}-3N_{y}^{2}\right),$$
and the Hall vector has two independent cubic expansions in the $\bm{N}$ components,
\begin{align*}
    \bm{h}\propto\begin{pmatrix}
        N_{z}(N_{x}^{2}-N_{y}^{2})\\-2N_{x}N_{y}N_{z}\\0
    \end{pmatrix} \quad \mathrm{and} \quad 
    \bm{h}\propto\begin{pmatrix}
        0\\0\\N_{x}(N_{x}^{2}-3N_{y}^{2})
    \end{pmatrix}.
\end{align*}
Again similar to the case of MnTe, the first contribution to both the Hall vector and the AMD requires $N_{z}\neq 0$ together with a non-zero in-plane component of $\bm{N}$. The second contribution only requires an in-plane N\'eel vector, which we may parameterize as $\bm{N} = N(\cos\phi,\sin\phi,0)^{\mathsf{T}}$ with $\phi$ measured from the $[2\bar{1}\bar{1}0]$ axis. This yields $\bm{h}\propto \cos(3\phi)\bm{\hat{z}},$ which vanishes for $\phi = (2m+1)\frac{\pi}{6}$, i.e. whenever $\bm{N}\,\|\,[01\bar{1}0]$ or any of the axes related to it by symmetry, and it is maximal whenever $\phi = m\frac{\pi}{3},$ with $\bm{N}\,\|\,[2\bar{1}\bar{1}0]$ and its related directions. Both terms vanish whenever $\bm{N}\,\|\,\bm{c}$. Again, these results are consistent with the calculated spectra of Fig.~\ref{fig:VNb3S6}, and with the direct MPG predictions. These cubic contributions to $\bm{h}$ and $\bm{T}$ are summarized for both MnTe and VNb$_{3}$S$_{6}$ in Tab.~\ref{tab:material_properties}.

We find these predictions to be in agreement with the literature on XMCD in MnTe. DFT+Dynamical Mean Field Theory Anderson impurity model calculations in Ref.~\cite{hariki2024MnTe} find a finite XMCD for $\bm{\hat{k}}\,\|\,\bm{c}$ when
$\bm{N}\,\|\,[1\bar{1}00]$, and a vanishing XMCD for $\bm{N}\,\|\,[11\bar{2}0]$. For $\bm{\hat{k}}\,\|\,\bm{N}$
the single-site contributions are large but cancel between the two sublattices, consistent with
$h_{x}=h_{y}=0$; the vanishing of both in-plane spectra in Fig.~\ref{fig:MnTe} implies a vanishing XMCD for every in-plane $\bm{\hat{k}}$, including
$\bm{\hat{k}}\,\|\,\bm{N}$. They demonstrate that the effect of valence SOC is marginal, while core--valence multipole interactions are significant in the $\bm{\hat{k}}\,\|\,\bm{c}$ and $\bm{N}\,\|\,[1\bar{1}00]$ geometry. This is consistent with the cubic contribution to $\bm{h}$, which is absent in the FVS approximation. The same work measures a finite XMCD with $\bm{\hat{k}}\,\|\,\bm{c}$ in epitaxial
films, whose sign reverses with the direction of the cooling field. Using XMCD-PEEM at normal
incidence, Ref.~\cite{Amin2024_MnTe_XMCD_DomainAM} images this signal and describes its variation as $\cos(3\phi')$, with
$\phi'$ measured from $[1\bar{1}00]$; since $\phi'=\phi+90^{\circ}$, this is identical to our
$h_{z}\propto\sin(3\phi)$. Transmission XMCD imaging with $\bm{\hat{k}}\,\|\,\bm{c}$ in Ref.~\cite{Yamamoto2025_MnTe_XMCD} also shows a threefold contrast dependence on the orientation of $\bm{N}$, with three sign reversals
across $180^{\circ}$ domain walls and six around winding textures, consistent with the $\sin(3\phi)$ prediction for an
in-plane rotation of $\bm{N}$. On bulk single crystals, Ref.~\cite{Takegami2025_MnTe_XMCD} finds an XMCD amplitude that agrees with the single-domain calculation of Ref.~\cite{hariki2024MnTe}. They note this XMCD cannot distinguish the three $120^{\circ}$-related domains, again consistent with the $\sin(3\phi)$ dependence; the circular dichroism in RIXS used there to separate them does not require $\mathcal{T}$ breaking and lies outside the present analysis. Finally, within the one-electron model of Ref.~\cite{sasabe2025t}, a finite $\langle T_{z}\rangle$ (AMD along $\bm{c}$) is found for $\bm{N}\,\|\,[\bar{1}100]$, while it vanishes for $\bm{N}\,\|\,[11\bar{2}0]$ and symmetry-related axes. Their findings in this model require valence SOC, because the model does not include a core hole. However, they provide multiplet calculations where the Mn$^{2+}$ ($d^{5}$) ground state has $\langle T_{z}\rangle\approx 0$ and the XMCD arises from the final states, in line with the
dominant role of the core--valence interaction~\cite{hariki2024MnTe}.

To the authors' knowledge, no measurement of the XMCD has been made yet in VNb$_{3}$S$_{6}$. Thus,
our results here constitute a prediction for future experiments. For the experimentally established in-plane N\'eel vector along $[ 2\bar{1}\bar{1}0]$ (or its symmetry-related counterparts, with the symmetry of MPG $\mathtt{2'2'2}$), we expect a finite XMCD only for $\bm{\hat{k}}\,\|\,\bm{c}$, varying as $\cos(3\phi)$, so that it is suppressed as $\bm{N}$ rotates toward $[01\bar{1}0]$ and changes sign between domains related by a $60^{\circ}$ rotation. As in MnTe, $\bm{h}$ vanishes in the FVS limit and appears only at cubic order, so the signal should be governed by the core--valence multipole interaction and the valence SOC. Nevertheless, the symmetry constraints we find for $\bm{h}$ and $\bm{T}$ apply equally well to the anomalous Hall conductivity, the $\omega\to 0$ limit of $\bm{h}$, which has been measured. In bulk crystals with ABAB stacking, Ref.~\cite{Ray2025_VNb3S6_neutrons} finds by neutron diffraction a collinear order with the staggered moment along $[2\bar{1}\bar{1}0]$ and symmetry-related axes, and a spontaneous $\sigma_{xy}$ far larger than expected from the small net moment. Of the two symmetry-distinct moment configurations (corresponding to those in Fig.~\ref{fig:VNb3S6}), they identify the one that permits $\sigma_{xy}$ as the ground state, consistent with $h_{z}\neq 0$ for $\mathtt{2'2'2}$ and $h_{z}=0$ for $\bm{N}\,\|\,[01\bar{1}0]$ ($\mathtt{222}$). The smaller zero-field $\sigma_{yz}$ they observe between $10$ and $30~\mathrm{K}$, corresponding to $h_{x}$, is forbidden for either in-plane orientation of $\bm{N}$; they attribute it to an
in-plane ferromagnetic component associated with stacking faults. Ref.~\cite{Lygouras202_VNb3S6_MagStruct} confirms the magnetic space group $\mathtt{C2'2'2_{1}}$ for the ABAB polytype, in which only $\sigma_{xy}$ is allowed, and notes that the Hall vector would vanish for moments along $\bm{a}^{*}$, which is equivalent to $[01\bar{1}0]$, in agreement with our results.

\FloatBarrier

%%%%%%%%%%%%%%%%%%%%%%%%%%%%%%%%%%%%%%%%%%%%%%%%%%  DISCUSSION  %%%%%%%%%%%%
%%%%%%%%%%%%%%%%%%%%%%%%%%%%%%%%%%%%%%

\section{Discussion}

As a symmetry-sensitive probe of non-relativistic ground state order, XMCD --- especially at the L edges --- has proven to be a useful tool for studying a subset of collinear altermagnets. To accelerate the study of altermagnets using XMCD, we have developed a comprehensive symmetry framework to identify the AM systems which can be analyzed by XMCD, the measurement geometries relevant in each case, and how these considerations naturally arise from Landau theory. In particular, we have derived all symmetry-allowed octupolar contributions $\mathfrak{T}_{\alpha\beta\gamma}$ to the AMD $\bm{T}$ and the corresponding allowed Hall vector $\bm{h}$ (Tab.~\ref{tab:tvec_short} and Tab.~\ref{tab:Hall_AMD_Ttensor_full}), explained the experimental geometries maximizing the XMCD signal in each case, and we have identified the twelve magnetic point groups consistent with \emph{both} altermagnetism at zero propagation vector and with XMCD (Tab.~\ref{tab:AM_MPG}). This direct MPG approach is practical for experimental purposes, but is somewhat divorced from the rich non-relativistic physics dominating altermagnetic systems. 

To restore the connection with non-relativistic systems, we have extended the Landau theory for all 54 classes of spin-orbit--free collinear altermagnets \cite{schiff2025collinear} to predict the lowest order, linear or cubic, contributions to the AMD and the optical Hall vector once SOC is restored. We have found that a linear (symmetric) contribution (Tab.~\ref{tab:LT_linear_symm}) is consistent with the results of the free valence spin approximation \cite{kunes2025xmcd}, while the remaining linear antisymmetric (Tab.~\ref{tab:LT_linear_anti}) and cubic (Tab.~\ref{tab:LT_cubic}) contributions rely on either valence SOC or core-valence multipolar exchange \footnote{For the $2p$ core levels, it can be shown that the FVS is purely symmetric.}. By predicting these contributions, the XMCD predictions in all twelve altermagnetic MPGs can be fully accounted for, including cases where the FVS finds XMCD is forbidden. 

Finally, we have demonstrated the consistency of our framework on four altermagnetic materials whose XMCD has been studied (MnF$_{2}$, NiF$_{2}$, Fe$_{2}$O$_{3}$ and MnTe), compared against a single non-altermagnetic material (Cr$_{2}$O$_{3}$) to illustrate the specificity of XMCD to altermagnets within the class of collinear antiferromagnets, and applied this framework to make new predictions on VNb$_{3}$S$_{6}$ (Tab.~\ref{tab:material_properties}). These six examples help illuminate three kinds of ``zero'' that can be obtained in predicting XMCD. The first kind, demonstrated by Cr$_{2}$O$_{3}$ arises from non-altermagnetic antiferromagnets, characterized by an inversion-odd N\'eel vector (and irrep $\Gamma_{\bm{N}}$); such materials will never exhibit an odd-order coupling to inversion-even quantities like $\bm{h}$ and $\bm{T}$. The second kind, demonstrated by Fe$_{2}$O$_{3}$, MnTe and VNb$_{3}$S$_{6}$, signifies vanishing XMCD in the limit of zero valence SOC, as a consequence of the particular sublattice structure $\bm{N}$ develops. Here, $\bm{N}$ does not transform as any of the $\mathcal{Q}_{\alpha\beta}$ components appearing in Eq.~\ref{eq:T_N_Q}, and so the AMD in this limit (corresponding to the FVS approximation) vanishes, and linear antisymmetric or cubic couplings dictate the dichroic response. The third kind of zero is explained by the N\'eel vector orientation, exhibited across all AM classes, and extensively studied in the literature for the examples discussed here. 

It is worth highlighting that the specific N\'eel vector orientations leading to a non-zero signal can now be directly predicted \emph{without} the knowledge of the MPG: instead, the spin-orbit free ground state, furnished by the crystal point group and the irrep $\Gamma_{\bm{N}}$ describing the N\'eel vector, will predict the lowest order dependence of $\bm{h}$ on $\bm{N}$, or equivalently, the lowest order coupling between $\bm{N}$ and $\bm{T}$ in the free energy. In this way, the non-relativistic symmetries of altermagnets directly determine the expected XMCD signal, and are distinguished from ordinary AFMs in which the XMCD vanishes.

It is worth clarifying that the non-relativistic symmetries governing altermagnets can be destroyed in the XMCD experiment in two ways: by including valence spin-orbit coupling, or by including multipolar core-valence interactions. The former directly breaks the non-relativistic spin-rotation invariance of the valence electron sector Hamiltonian; the latter, on the other hand, couples the core spins (which are spin-orbit coupled) to the valence spins, thereby indirectly breaking the valence sector spin-rotation symmetry. This is made clear by the considerations of the FVS approximation \cite{kunes2025xmcd}, but is not a necessary distinction in the Landau theory technique, as it is based solely on symmetries of the ground state, where no core hole exists. 

This work establishes concrete experimental predictions in the altermagnet VNb$_{3}$S$_{6}$, and the analysis of Sec.~\ref{sec:LT} and Sec.~\ref{sec:material_examples} demonstrate how to generate analogous predictions in any altermagnet using the tables provided here, either using the direct MPG approach or the Landau theory framework. The former requires knowledge of the ground state MPG, whereas the latter requires the crystal point group and irrep $\Gamma_{\bm{N}}$ describing the transformation of $\bm{N}$. In this way, we have comprehensively predicted the symmetry-allowed XMCD across all altermagnets with zero propagation vector.

The tabulation of symmetry-allowed octupolar terms $\mathfrak{T}_{\alpha\beta\gamma}$ are also well-suited to support experiments following the proposal of Ref.~\cite{Mizumaki2026_XAS_XES_MCD}, which explores the possibility of measuring ferroically ordered magnetic octupoles directly with X-ray emission magnetic circular dichroism. Its sum rule contains the irreducible (spherical) rank-3 part of the $\mathfrak{T}_{\alpha\beta\gamma}$ tabulated for all MPGs (Tab.~\ref{tab:Hall_AMD_Ttensor_full}). In MPGs where no rank-1 (AMD and spin dipole) component is allowed, XES-MCD isolates this octupole. Therefore, these predictions offer an opportunity for future directions of work.

\begin{acknowledgments}
Work at the University of California, Irvine and the University of California, San Diego was supported in part by a University of California LFRP grant (L25CR8980).

This research used resources of the Advanced Light Source, a DOE Office of Science User Facility under Contract No. DE-AC02-05CH11231.

The authors acknowledge the use of Claude (Anthropic) for proofreading, checking the internal consistency of the manuscript and its symmetry tables, and formatting the bibliography. All scientific content, analysis, and conclusions are the authors' own, and the authors take full responsibility for the manuscript.

Certain software is identified in this paper in order to specify the procedure adequately. Such identification is not intended to imply recommendation or endorsement by NIST, nor is it intended to imply that the software identified is necessarily the best available for the purpose.
\end{acknowledgments}
%\clearpage
\bibliography{references}
%\clearpage
%%%%%%%%%%%%%%%%%%%%%%%%%%%%%%%%%%%%%%%%%%%%%%%%%%%%  APPENDIX  %%%%%%%%%%%%%
%%%%%%%%%%%%%%%%%%%%%%%%%%%%%%%%%%%%%%%
\appendix

%\section{The anisotropic magnetic dipole with spin rotation symmetry}\label{app:T_transf}

\section{XMCD within the Bethe-Salpeter equation approach}\label{app:BSeq_approach}

Two changes in {\sc ocean} were necessary to enable the calculation of XMCD. First, circular polarization was added to the electron-photon transition operator. 
The transition operator is carried out as a real-space integral between a core orbital, constructed from a radial function and a spherical harmonic, and a set of local basis functions. 
Within the dipole approximation, the angular part of this integral depends on the direction of the photon polarization, while the radial part is fixed for a given core and basis. For circularly polarized light, the polarization can be written as $1/\sqrt{2}(1,\pm i,0)$ for right ($-$) or left ($+$) polarized light traveling along the $+z$ direction. Within {\sc ocean} left/right polarization and the photon momentum direction is specified, determining the polarization used. This is set with respect to the same coordinate frame as the input lattice vectors.  

The second change is that heretofore, {\sc ocean} assumed that the conduction and valence spin were along $z$. 
The actual spin axis affects the form of the core-level spin orbit. In {\sc ocean}, the core levels are treated as an initially degenerate set of orbitals in the $m_l\sigma$ basis \cite{OCEAN2022}. Within this basis the core-level spin-orbit interaction is non-diagonal, and this calculation is carried out with both the orbital spherical harmonics and the spins aligned along $z$. For the former, this is required consistency with the previously discussed real-space transition matrix elements. For the latter, the core-level spin axis must be aligned through a rotation to match the spin axis of the valence and conduction bands. This is because the x-ray absorption event is spin conserving. We achieve this by rotating the spin blocks within the core-level spin-orbit Hamiltonian.

\subsection{Local moments}

Starting from Eq.~\ref{eq:T_definition},
\begin{equation}
    T_{\alpha} = \sum_a T^{(a)}_{\alpha} = \sum_\beta \sum_a  Q^{(a)}_{\alpha \beta}   S_\beta^{(a)}  \; ,
    %= \sum_a \int_{r\le r_c} d^3 r q_{\alpha \beta}(\hat{r}
\end{equation}
 we are taking the expectation of the quadrupole tensor and spin operator at each site $a$. This can be written as the real-space integral within the sphere defined by a cut-off $r_c$ around the site summed over all occupied states
\begin{equation}
    T^{(a)}_{\alpha} = \sum_\beta N_\mathbf{k}^{-1} \sum_{n \mathbf{k} \sigma}^{\mathrm{occ}} \langle \psi_{n \mathbf{k} \sigma} \vert \mathbf{Q}_{\alpha \beta}(\mathbf{r})\mathbf{S}_\beta(\mathbf{r}) \vert \psi_{n \mathbf{k} \sigma} \rangle_{r \le r_c}
    \label{eq:Appendix_T}
\end{equation}
where we normalize by the number of $\mathbf{k}$-points $N_\mathbf{k}$. Coming from collinear DFT, we can define a spin-density difference
\begin{align}
m(\mathbf{r}) &= n_{\uparrow}(\mathbf{r}) - n_{\downarrow}(\mathbf{r}) \\
\mathbf{S}(\mathbf{r}) &= \frac{1}{2}m(\mathbf{r}) \hat{N}
\end{align}
where $\hat{N}$ is the Néel vector. 
Substituting these two into Eq.~\ref{eq:Appendix_T}, 
\begin{align}
 T^{(a)}_{\alpha} = \frac{1}{2} \sum_\beta \frac{\hat{N}_\beta}{N_\mathbf{k}}  \sum_{n \mathbf{k} \sigma }^{\mathrm{occ}} \Big[ & \langle \psi_{n \mathbf{k} \sigma} \vert \mathbf{Q}_{\alpha \beta}(\mathbf{r})n_\uparrow(\mathbf{r}) \vert \psi_{n \mathbf{k} \sigma} \rangle_{r \le r_c}   \nonumber \\
-&    \langle \psi_{n \mathbf{k} \sigma} \vert \mathbf{Q}_{\alpha \beta}(\mathbf{r})n_\downarrow(\mathbf{r}) \vert \psi_{n \mathbf{k} \sigma} \rangle_{r \le r_c}  \Big] \nonumber \\
=\frac{1}{2} \sum_\beta \hat{N}_\beta & \left[ Q^{a(\uparrow)}_{\alpha \beta}-Q^{a(\downarrow)}_{\alpha \beta} \right] = \sum_\beta \hat{N}_\beta \bar{Q}^{(a)}_{\alpha \beta}
\end{align}
Which defines the local spin-resolved charge anisotropy tensor $Q^{(a)\sigma}$ and local difference $\bar{Q}^{(a)} = \frac{1}{2}[Q^{(a)\uparrow}- Q^{(a)\downarrow} ]$.
Within the radius $r_c$ around the metal site $a$ we can write the electron orbitals in terms of a complete set of local basis states $\phi_{\nu lm} (\mathbf{r})=R_{\nu l }(r)Y_{lm}(\hat{r})$  and coefficients $C$
\begin{equation}
    \psi_{nk \sigma}(\mathbf{r})_{|r| \le r_c} = \sum_{{\nu l m}} R_{\nu l}(r) Y_{lm}(\hat{r}) C^{nk \sigma}_{\nu l m} \; .
    \end{equation}
Allowing us to write the spin-resolved charge anisotropy as
\begin{align}
 %       Q^{a(\sigma)}_{ij} &= N_k^{-1}\sum_{nk} f_{nk\sigma} \langle \psi_{nk\sigma}\vert \delta_{ij}-3\hat{r}_i \hat{r}_j \vert \psi_{nk\sigma} \rangle _{|r| \le r_c} \nonumber \\
Q^{a(\sigma)}_{\alpha \beta} = N_\mathbf{k}^{-1} \sum^{\mathrm{occ}}_{nk}  & \sum_{\substack{\nu l m \\ \nu' l' m'}}  C^{nk \sigma}_{\nu l m} [C^{nk \sigma}_{\nu' l' m'}]^*
 \langle \phi_{\nu' l'} \vert \phi_{\nu l}  \rangle  \times \nonumber \\
 & \times \langle Y_{l'm'} \vert \delta_{\alpha \beta}-3\hat{r_\alpha} \hat{r_\beta} \vert Y_{lm} \rangle \;.
    \end{align}
We can sum over all sites separate from the Néel vector
\begin{equation}
    T_\alpha= \sum_a T^{(a)}_\alpha = \sum_\beta \hat{N}_\beta \sum_a \bar{Q}^{(a)}_{\alpha \beta} = \sum_\beta \hat{N}_\beta \bar{Q}_{\alpha \beta},
\end{equation}
defining the total charge anisotropy difference tensor $\bar{Q}$. 
In the case of MnF$_2$, we calculate 
\begin{align}
\label{eq:Tij_MnF2_site}
\bar{Q}^{(\pm)}_{\alpha \beta} \approx 
\begin{pmatrix}
\pm t  & t' & 0 \\
t' & \pm t & 0 \\
0 & 0 & \mp 2t
\end{pmatrix} 
\end{align}
where the $(+)$ denotes the spin up Mn site and $(-)$ the spin down. Finally, summing over sites gives 
\begin{align}
\bar{Q}_{\alpha \beta} \approx 
\begin{pmatrix}
0 & 2t' & 0 \\
2t' & 0 & 0 \\
0 & 0 & 0
\end{pmatrix} \,
\label{eq:Tij_MnF2}
\end{align}
where the values of $t$ and $t'$ are influenced in part by our choice of local basis. The form of Eq.~\ref{eq:Tij_MnF2} exemplifies the spin-orbit free multipole that serves as a pseudoprimary order parameter in the Landau theory for MnF$_{2}$ \cite{schiff2025collinear}. 
The same form of Eqs.~\ref{eq:Tij_MnF2_site} and \ref{eq:Tij_MnF2} is obtained for NiF$_2$. 

For the other materials considered here, we calculate $\bar{Q} \approx 0$. This is expected as they are $g$-wave altermagnets, or in the case of Cr$_2$O$_3$, antiferromagnetic, and have vanishing $\mathbf{Q}$ in the absence of valence spin-orbit coupling.

\if{0}
In the case of Fe$_2$O$_3$ and Cr$_2$O$_3$, we obtain
\begin{align}
\label{eq:Tij_Fe2O3_site}
\bar{T}^{(\pm)}_{\alpha \beta} \approx t
\begin{pmatrix}
\pm 1  & 0 & 0 \\
0 & \pm 1 & 0 \\
0 & 0 & \mp 2
\end{pmatrix} 
\end{align}
which clearly give zero for the full $\bar{T}$, reflecting the nature of both as $g$-wave AM. 
\fi

\subsection{Multipole expansion of the BSE terms}

\begin{figure}%[h!]
\includegraphics[width=\columnwidth]{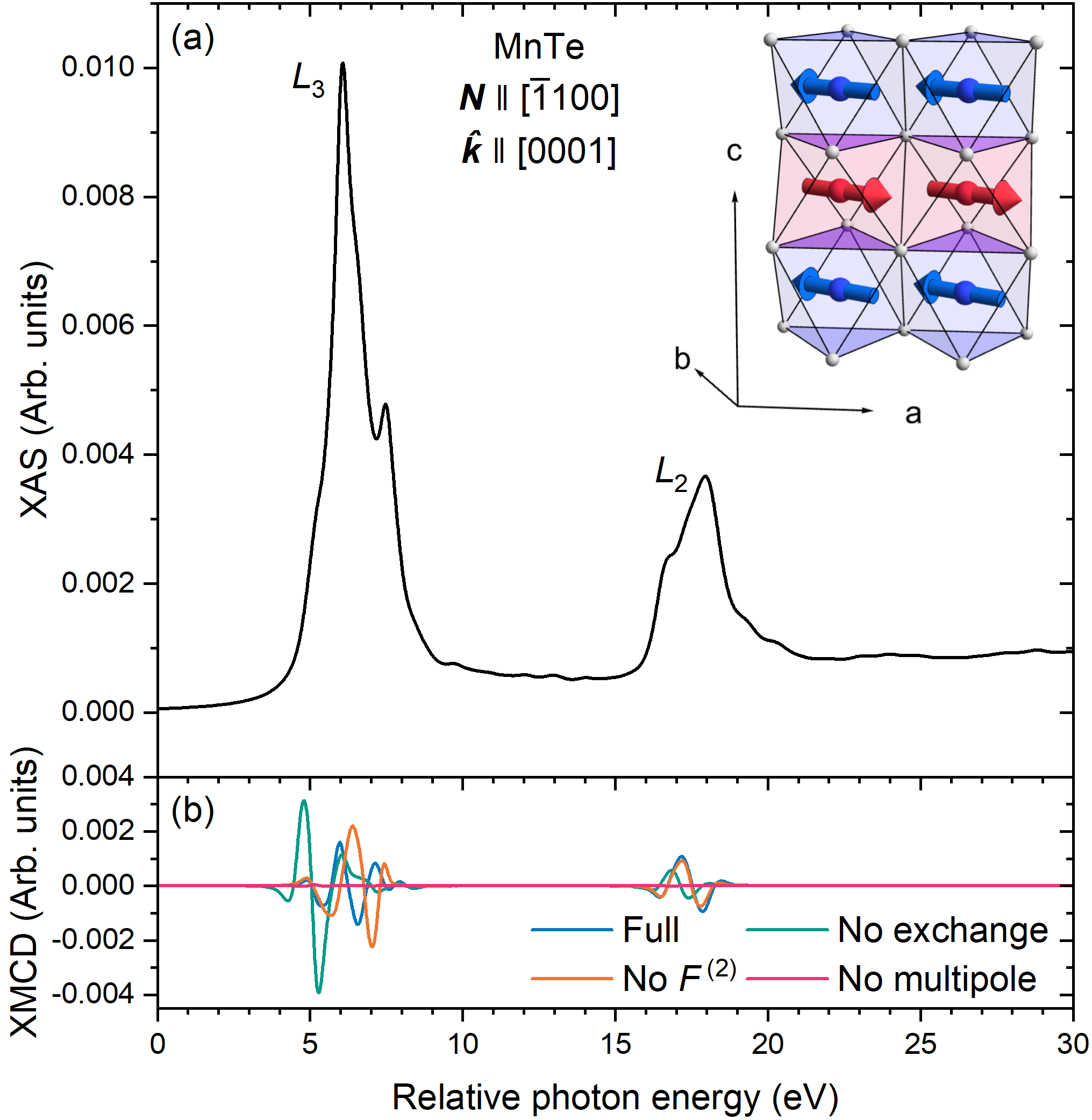}
\caption{ MnTe for $\mathbf{N} \parallel [\bar{1}100]$ and $\hat{\mathbf{k}} \parallel [0001]$ with various multipole parts of the BSE Hamiltonian disabled. When no multipole parts of the interaction are included, the XMCD signal disappears. However, only $F^{(0)}$ and $F^{(2)}$ (No exchange) or only $F^{(0)}$, $G^{(1)}$, and $G^{(3)}$ (No F$^{(2)}$) are both sufficient to yield an XMCD signal of reasonable magnitude. }  
\label{fig_MnT2_mult}
\end{figure}

Within {\sc ocean}, the lowest-order, static BSE is implemented, and core-level spin orbit is treated as an energy-correction term while valence-level spin-orbit is currently neglected entirely \cite{OCEAN2022,PhysRevB.86.195135}.
For the BSE, we have an excited electron $c=n\mathbf{k}\sigma$, denoted by band, $\mathbf{k}$-point, and spin, and a core hole $\alpha=m_ls$, denoted by magnetic quantum number and spin. 
The core hole is confined to a single atomic site, principal quantum number and angular quantum number, justified by the spatial locality of the core levels and the energy spacing of the x-ray edges, but azimuthal quantum numbers do mix, e.g., the L$_{2,3}$ edge is treated all together. 
\begin{align}
    H^{\mathrm{BSE}} &= H^0 + V_X - W \\
    H^0 &= \varepsilon_c \delta_{cc'} - (\varepsilon_{\alpha} \delta_{\alpha\alpha'} + \xi \mathbf{L}\cdot \mathbf{S} + X) 
\end{align}
where the spin-orbit interaction is given by $\mathbf{L \cdot S}$ and $X$ is the exchange energy of a given core level $m_l s$ due to the occupied valence states. Notably the spin-orbit interaction is not diagonal in our $\alpha$ basis, and the exchange is $\alpha$ dependent. 
The two interaction terms can be expanded around the localized core hole, giving a multipole expansion.
This is done in {\sc ocean} in a general way, using a complete local basis, but it is instructive to consider the specific case of an electron in states with $l=2$ given by $\psi_c(\mathbf{r}) = R_c(r)Y_{2m}(\hat{r})$ and the core hole in $l=1$ $\psi_\alpha(\mathbf{r})  = R_{\alpha}(r)Y_{1m}(\hat{r})$,
\begin{widetext}
\begin{align}
    -W+V_X = -\langle R_c \vert W_0 \vert R_{c'}\rangle  \nonumber &-\left[\sum_{m_k=-2}^{2}  \int drdr'\psi^\dagger_c(r)\psi_{c'}(r) \frac{r_{<}^2}{r_>^{3}} Y_{2 m_k}(r) Y^*_{2 m_k}(r') \psi_\alpha(r') \psi^\dagger_{\alpha '}(r')   \right]\delta_{\sigma \sigma'}\delta_{s s'} \\
    &+\left[\sum_{m_k=-1}^{1} \int drdr' \psi^\dagger_c(r)\psi_{c'}(r') \frac{r_{<}}{r_>^{2}} Y_{1 m_k}(r) Y^*_{1 m_k}(r') \psi_\alpha(r) \psi^\dagger_{\alpha '}(r') \nonumber  \right]\delta_{\sigma s}\delta_{\sigma' s'} \\
    &+\left[\sum_{m_k=-3}^{3} \int drdr' \psi^\dagger_c(r)\psi_{c'}(r') \frac{r_{<}^3}{r_>^{4}} Y_{3 m_k}(r) Y^*_{3 m_k}(r') \psi_\alpha(r) \psi^\dagger_{\alpha '}(r')  \right]\delta_{\sigma s}\delta_{\sigma' s'}
\end{align}
\end{widetext}
where $W_0$ is the spherically symmetric part of the direct ($F^{(0)}$). The $l=2$ direct ($F^{(2)}$) (second term) is taken to be unscreened due to its short-ranged nature. The third and fourth terms are the $l=1$ and $l=3$ parts of the exchange, ($G^{(1)}$ and $G^{(3)}$), respectively.
Without the multipole terms of the interaction (and without the valence spin orbit which we neglect) there is nothing to couple the spin and orbital directions (Fig.~\ref{fig_MnT2_mult}). 

\section{Group projector technique}\label{app:symm_methods}

Here we summarize the established group projector techniques~\cite{DamnjanovicSymm, MDamnjanovicModGProj,bradley2009mathematical,GTPackBook,KosterGroupProj, QGMGpThr} used to find the symmetry-adapted basis (SAB) of any representation. The obtained set of vectors $|\Gamma,t,m\rangle$ for each irrep $\Gamma$ in the decomposition of $D$, for each repetition $t$ (ranging from $1$ to the number of times $\Gamma$ is contained in $D$) have the property that 
$$D(g)|\Gamma,t,m\rangle = \sum_{m'=1}^{|\Gamma|} \Gamma_{m'm}(g)|\Gamma,t,m'\rangle,$$ where $|\Gamma|$
is the dimension of the irrep $\Gamma.$  A group projector $P_{11}^{\Gamma}(D)$ for a representation $D$ onto an irrep $\Gamma$ of the group $\mathbf{G}$ is given by 
\begin{equation}\label{eq:gpProj}
P_{11}^{\Gamma}(D) = \frac{|\Gamma|}{|\mathbf{G}|} \sum_{g\in\mathbf{G}} \Gamma_{11}^{*}(g) D(g),
\end{equation}
and it is non-zero whenever $\Gamma$ is contained in the decomposition of $D$. In Eq.~(\ref{eq:gpProj}), $\Gamma_{11}$ is the matrix element of $\Gamma$ in the first row and first column. If the dimension $|\Gamma|$ of $\Gamma$ is one, then the SAB of $D$ for the irrep $\Gamma$ is a basis in the image of the projector $P_{11}^{\Gamma}(D).$ However, if $\Gamma$ has dimension $|\Gamma| > 1$, then the SAB of $D$ for $\Gamma$ will be given by a basis the image of $P_{11}^{\Gamma}(D),$ and in addition, the vectors obtained by acting on them with the group operators with $m \in \{2,.., |\Gamma|\},$ $$P_{m1}^{\Gamma}(D) = \frac{|\Gamma|}{|\mathbf{G}|} \sum_{g\in\mathbf{G}} \Gamma_{m1}^{*}(g) D(g).$$ 
We use this technique in Sec.~\ref{sec:allowed_Th} to verify the scalars, that is, the SAB for the trivial (co-)irrep in each magnetic point group, for the time-odd axial vector and for the $aeV[V^{2}]$ representations. 

\clearpage
\onecolumngrid
\section{Tabulation of symmetry-allowed components of the Hall vector \texorpdfstring{$\bm{h}$}{h}, AMD \texorpdfstring{$\bm{T}$}{T}, and octupole  \texorpdfstring{$\bm{\mathfrak{T}}$}{T}}

For convenience, Tab.~\ref{tab:Hall_AMD_Ttensor_full} also lists six alternative settings, $\mathtt{\bar{4}m2}$, $\mathtt{\bar{4}m21'}$, $\mathtt{\bar{6}2m}$, $\mathtt{\bar{6}2m1'}$, $\mathtt{4'/m'mm'}$ and $\mathtt{6'/mm'm}$, which are related to $\mathtt{\bar{4}2m}$, $\mathtt{\bar{4}2m1'}$, $\mathtt{\bar{6}m2}$, $\mathtt{\bar{6}m21'}$, $\mathtt{4'/m'm'm}$ and $\mathtt{6'/mmm'}$, respectively, by a rotation about the principal axis, bringing the total number of entries to 128.

\FloatBarrier
\begin{center}
\refstepcounter{table}\label{tab:Hall_AMD_Ttensor_full}\footnotesize\textbf{TABLE~\thetable{}.} For each magnetic point group indicated in the first column, the second column indicates the allowed components $h_{i}$ of the Hall vector. Because XMCD $\propto \bm{h}(\omega)\cdot\bm{\hat{k}}$, this is also the beam direction maximizing the XMCD signal. The third column indicates the symmetry-adapted form of the AMD $\bm{T}$. The final column indicates the full symmetry-adapted octupole tensor $\mathfrak{T}_{ijk}$. Note that these terms are subject to the constraint $\sum_{\alpha} \mathfrak{T}_{\alpha\alpha\gamma} = 0$ due to the traceless feature of the quadrupole.

\footnotesize
\renewcommand{\arraystretch}{1.1}
\begin{tabular}{l l l l}
\toprule
MPG & Hall $\bm{h}$ & AMD $\bm{T}$ & Octupole \\
\midrule
$\mathtt{1}$ & $\begin{pmatrix} h_{x} \\ h_{y} \\ h_{z} \end{pmatrix}$ & $\begin{pmatrix} \mathfrak{T}_{xxx} + \mathfrak{T}_{xyy} + \mathfrak{T}_{xzz} \\ \mathfrak{T}_{yxx} + \mathfrak{T}_{yyy} + \mathfrak{T}_{yzz} \\ \mathfrak{T}_{zxx} + \mathfrak{T}_{zyy} + \mathfrak{T}_{zzz} \end{pmatrix}$ & $\begin{pmatrix} \mathfrak{T}_{xxx} & \mathfrak{T}_{yyx} & \mathfrak{T}_{zzx} & \mathfrak{T}_{yxx} & \mathfrak{T}_{zxx} & \mathfrak{T}_{yzx} \\ \mathfrak{T}_{xxy} & \mathfrak{T}_{yyy} & \mathfrak{T}_{zzy} & \mathfrak{T}_{xyy} & \mathfrak{T}_{xzy} & \mathfrak{T}_{zyy} \\ \mathfrak{T}_{xxz} & \mathfrak{T}_{yyz} & \mathfrak{T}_{zzz} & \mathfrak{T}_{xyz} & \mathfrak{T}_{xzz} & \mathfrak{T}_{yzz} \end{pmatrix}$ \\
$\mathtt{11'}$ & --- & --- & --- \\
$\mathtt{\bar{1}}$ & $\begin{pmatrix} h_{x} \\ h_{y} \\ h_{z} \end{pmatrix}$ & $\begin{pmatrix} \mathfrak{T}_{xxx} + \mathfrak{T}_{xyy} + \mathfrak{T}_{xzz} \\ \mathfrak{T}_{yxx} + \mathfrak{T}_{yyy} + \mathfrak{T}_{yzz} \\ \mathfrak{T}_{zxx} + \mathfrak{T}_{zyy} + \mathfrak{T}_{zzz} \end{pmatrix}$ & $\begin{pmatrix} \mathfrak{T}_{xxx} & \mathfrak{T}_{yyx} & \mathfrak{T}_{zzx} & \mathfrak{T}_{yxx} & \mathfrak{T}_{zxx} & \mathfrak{T}_{yzx} \\ \mathfrak{T}_{xxy} & \mathfrak{T}_{yyy} & \mathfrak{T}_{zzy} & \mathfrak{T}_{xyy} & \mathfrak{T}_{xzy} & \mathfrak{T}_{zyy} \\ \mathfrak{T}_{xxz} & \mathfrak{T}_{yyz} & \mathfrak{T}_{zzz} & \mathfrak{T}_{xyz} & \mathfrak{T}_{xzz} & \mathfrak{T}_{yzz} \end{pmatrix}$ \\
$\mathtt{\bar{1}1'}$ & --- & --- & --- \\
$\mathtt{\bar{1}'}$ & --- & --- & --- \\
$\mathtt{2}$ & $\begin{pmatrix} 0 \\ h_{y} \\ 0 \end{pmatrix}$ & $\begin{pmatrix} 0 \\ \mathfrak{T}_{yxx} + \mathfrak{T}_{yyy} + \mathfrak{T}_{yzz} \\ 0 \end{pmatrix}$ & $\begin{pmatrix} 0 & 0 & 0 & \mathfrak{T}_{yxx} & 0 & \mathfrak{T}_{yzx} \\ \mathfrak{T}_{xxy} & \mathfrak{T}_{yyy} & \mathfrak{T}_{zzy} & 0 & \mathfrak{T}_{xzy} & 0 \\ 0 & 0 & 0 & \mathfrak{T}_{xyz} & 0 & \mathfrak{T}_{yzz} \end{pmatrix}$ \\
$\mathtt{21'}$ & --- & --- & --- \\
$\mathtt{2'}$ & $\begin{pmatrix} h_{x} \\ 0 \\ h_{z} \end{pmatrix}$ & $\begin{pmatrix} \mathfrak{T}_{xxx} + \mathfrak{T}_{xyy} + \mathfrak{T}_{xzz} \\ 0 \\ \mathfrak{T}_{zxx} + \mathfrak{T}_{zyy} + \mathfrak{T}_{zzz} \end{pmatrix}$ & $\begin{pmatrix} \mathfrak{T}_{xxx} & \mathfrak{T}_{yyx} & \mathfrak{T}_{zzx} & 0 & \mathfrak{T}_{zxx} & 0 \\ 0 & 0 & 0 & \mathfrak{T}_{xyy} & 0 & \mathfrak{T}_{zyy} \\ \mathfrak{T}_{xxz} & \mathfrak{T}_{yyz} & \mathfrak{T}_{zzz} & 0 & \mathfrak{T}_{xzz} & 0 \end{pmatrix}$ \\
$\mathtt{m}$ & $\begin{pmatrix} 0 \\ h_{y} \\ 0 \end{pmatrix}$ & $\begin{pmatrix} 0 \\ \mathfrak{T}_{yxx} + \mathfrak{T}_{yyy} + \mathfrak{T}_{yzz} \\ 0 \end{pmatrix}$ & $\begin{pmatrix} 0 & 0 & 0 & \mathfrak{T}_{yxx} & 0 & \mathfrak{T}_{yzx} \\ \mathfrak{T}_{xxy} & \mathfrak{T}_{yyy} & \mathfrak{T}_{zzy} & 0 & \mathfrak{T}_{xzy} & 0 \\ 0 & 0 & 0 & \mathfrak{T}_{xyz} & 0 & \mathfrak{T}_{yzz} \end{pmatrix}$ \\
$\mathtt{m1'}$ & --- & --- & --- \\
$\mathtt{m'}$ & $\begin{pmatrix} h_{x} \\ 0 \\ h_{z} \end{pmatrix}$ & $\begin{pmatrix} \mathfrak{T}_{xxx} + \mathfrak{T}_{xyy} + \mathfrak{T}_{xzz} \\ 0 \\ \mathfrak{T}_{zxx} + \mathfrak{T}_{zyy} + \mathfrak{T}_{zzz} \end{pmatrix}$ & $\begin{pmatrix} \mathfrak{T}_{xxx} & \mathfrak{T}_{yyx} & \mathfrak{T}_{zzx} & 0 & \mathfrak{T}_{zxx} & 0 \\ 0 & 0 & 0 & \mathfrak{T}_{xyy} & 0 & \mathfrak{T}_{zyy} \\ \mathfrak{T}_{xxz} & \mathfrak{T}_{yyz} & \mathfrak{T}_{zzz} & 0 & \mathfrak{T}_{xzz} & 0 \end{pmatrix}$ \\
$\mathtt{2/m}$ & $\begin{pmatrix} 0 \\ h_{y} \\ 0 \end{pmatrix}$ & $\begin{pmatrix} 0 \\ \mathfrak{T}_{yxx} + \mathfrak{T}_{yyy} + \mathfrak{T}_{yzz} \\ 0 \end{pmatrix}$ & $\begin{pmatrix} 0 & 0 & 0 & \mathfrak{T}_{yxx} & 0 & \mathfrak{T}_{yzx} \\ \mathfrak{T}_{xxy} & \mathfrak{T}_{yyy} & \mathfrak{T}_{zzy} & 0 & \mathfrak{T}_{xzy} & 0 \\ 0 & 0 & 0 & \mathfrak{T}_{xyz} & 0 & \mathfrak{T}_{yzz} \end{pmatrix}$ \\
$\mathtt{2/m1'}$ & --- & --- & --- \\
$\mathtt{2'/m}$ & --- & --- & --- \\
$\mathtt{2/m'}$ & --- & --- & --- \\
$\mathtt{2'/m'}$ & $\begin{pmatrix} h_{x} \\ 0 \\ h_{z} \end{pmatrix}$ & $\begin{pmatrix} \mathfrak{T}_{xxx} + \mathfrak{T}_{xyy} + \mathfrak{T}_{xzz} \\ 0 \\ \mathfrak{T}_{zxx} + \mathfrak{T}_{zyy} + \mathfrak{T}_{zzz} \end{pmatrix}$ & $\begin{pmatrix} \mathfrak{T}_{xxx} & \mathfrak{T}_{yyx} & \mathfrak{T}_{zzx} & 0 & \mathfrak{T}_{zxx} & 0 \\ 0 & 0 & 0 & \mathfrak{T}_{xyy} & 0 & \mathfrak{T}_{zyy} \\ \mathfrak{T}_{xxz} & \mathfrak{T}_{yyz} & \mathfrak{T}_{zzz} & 0 & \mathfrak{T}_{xzz} & 0 \end{pmatrix}$ \\
$\mathtt{222}$ & --- & --- & $\begin{pmatrix} 0 & 0 & 0 & 0 & 0 & \mathfrak{T}_{yzx} \\ 0 & 0 & 0 & 0 & \mathfrak{T}_{xzy} & 0 \\ 0 & 0 & 0 & \mathfrak{T}_{xyz} & 0 & 0 \end{pmatrix}$ \\
$\mathtt{2221'}$ & --- & --- & --- \\
$\mathtt{2'2'2}$ & $\begin{pmatrix} 0 \\ 0 \\ h_{z} \end{pmatrix}$ & $\begin{pmatrix} 0 \\ 0 \\ \mathfrak{T}_{zxx} + \mathfrak{T}_{zyy} + \mathfrak{T}_{zzz} \end{pmatrix}$ & $\begin{pmatrix} 0 & 0 & 0 & 0 & \mathfrak{T}_{zxx} & 0 \\ 0 & 0 & 0 & 0 & 0 & \mathfrak{T}_{zyy} \\ \mathfrak{T}_{xxz} & \mathfrak{T}_{yyz} & \mathfrak{T}_{zzz} & 0 & 0 & 0 \end{pmatrix}$ \\
$\mathtt{mm2}$ & --- & --- & $\begin{pmatrix} 0 & 0 & 0 & 0 & 0 & \mathfrak{T}_{yzx} \\ 0 & 0 & 0 & 0 & \mathfrak{T}_{xzy} & 0 \\ 0 & 0 & 0 & \mathfrak{T}_{xyz} & 0 & 0 \end{pmatrix}$ \\
$\mathtt{mm21'}$ & --- & --- & --- \\
$\mathtt{m'm2'}$ & $\begin{pmatrix} 0 \\ h_{y} \\ 0 \end{pmatrix}$ & $\begin{pmatrix} 0 \\ \mathfrak{T}_{yxx} + \mathfrak{T}_{yyy} + \mathfrak{T}_{yzz} \\ 0 \end{pmatrix}$ & $\begin{pmatrix} 0 & 0 & 0 & \mathfrak{T}_{yxx} & 0 & 0 \\ \mathfrak{T}_{xxy} & \mathfrak{T}_{yyy} & \mathfrak{T}_{zzy} & 0 & 0 & 0 \\ 0 & 0 & 0 & 0 & 0 & \mathfrak{T}_{yzz} \end{pmatrix}$ \\
$\mathtt{m'm'2}$ & $\begin{pmatrix} 0 \\ 0 \\ h_{z} \end{pmatrix}$ & $\begin{pmatrix} 0 \\ 0 \\ \mathfrak{T}_{zxx} + \mathfrak{T}_{zyy} + \mathfrak{T}_{zzz} \end{pmatrix}$ & $\begin{pmatrix} 0 & 0 & 0 & 0 & \mathfrak{T}_{zxx} & 0 \\ 0 & 0 & 0 & 0 & 0 & \mathfrak{T}_{zyy} \\ \mathfrak{T}_{xxz} & \mathfrak{T}_{yyz} & \mathfrak{T}_{zzz} & 0 & 0 & 0 \end{pmatrix}$ \\
\bottomrule
\end{tabular}
\end{center}

\clearpage
\begin{center}
\footnotesize\textbf{TABLE~\ref{tab:Hall_AMD_Ttensor_full}} (continued)

\footnotesize
\renewcommand{\arraystretch}{1.1}
\begin{tabular}{l l l l}
\toprule
MPG & Hall $\bm{h}$ & AMD $\bm{T}$ & Octupole \\
\midrule
$\mathtt{mmm}$ & --- & --- & $\begin{pmatrix} 0 & 0 & 0 & 0 & 0 & \mathfrak{T}_{yzx} \\ 0 & 0 & 0 & 0 & \mathfrak{T}_{xzy} & 0 \\ 0 & 0 & 0 & \mathfrak{T}_{xyz} & 0 & 0 \end{pmatrix}$ \\
$\mathtt{mmm1'}$ & --- & --- & --- \\
$\mathtt{m'mm}$ & --- & --- & --- \\
$\mathtt{m'm'm}$ & $\begin{pmatrix} 0 \\ 0 \\ h_{z} \end{pmatrix}$ & $\begin{pmatrix} 0 \\ 0 \\ \mathfrak{T}_{zxx} + \mathfrak{T}_{zyy} + \mathfrak{T}_{zzz} \end{pmatrix}$ & $\begin{pmatrix} 0 & 0 & 0 & 0 & \mathfrak{T}_{zxx} & 0 \\ 0 & 0 & 0 & 0 & 0 & \mathfrak{T}_{zyy} \\ \mathfrak{T}_{xxz} & \mathfrak{T}_{yyz} & \mathfrak{T}_{zzz} & 0 & 0 & 0 \end{pmatrix}$ \\
$\mathtt{m'm'm'}$ & --- & --- & --- \\
$\mathtt{4}$ & $\begin{pmatrix} 0 \\ 0 \\ h_{z} \end{pmatrix}$ & $\begin{pmatrix} 0 \\ 0 \\ 2\mathfrak{T}_{zxx} + \mathfrak{T}_{zzz} \end{pmatrix}$ & $\begin{pmatrix} 0 & 0 & 0 & 0 & \mathfrak{T}_{zxx} & \mathfrak{T}_{yzx} \\ 0 & 0 & 0 & 0 & -\mathfrak{T}_{yzx} & \mathfrak{T}_{zxx} \\ \mathfrak{T}_{yyz} & \mathfrak{T}_{yyz} & \mathfrak{T}_{zzz} & 0 & 0 & 0 \end{pmatrix}$ \\
$\mathtt{41'}$ & --- & --- & --- \\
$\mathtt{4'}$ & --- & --- & $\begin{pmatrix} 0 & 0 & 0 & 0 & \mathfrak{T}_{zxx} & \mathfrak{T}_{yzx} \\ 0 & 0 & 0 & 0 & \mathfrak{T}_{yzx} & -\mathfrak{T}_{zxx} \\ -\mathfrak{T}_{yyz} & \mathfrak{T}_{yyz} & 0 & \mathfrak{T}_{xyz} & 0 & 0 \end{pmatrix}$ \\
$\mathtt{\bar{4}}$ & $\begin{pmatrix} 0 \\ 0 \\ h_{z} \end{pmatrix}$ & $\begin{pmatrix} 0 \\ 0 \\ 2\mathfrak{T}_{zxx} + \mathfrak{T}_{zzz} \end{pmatrix}$ & $\begin{pmatrix} 0 & 0 & 0 & 0 & \mathfrak{T}_{zxx} & \mathfrak{T}_{yzx} \\ 0 & 0 & 0 & 0 & -\mathfrak{T}_{yzx} & \mathfrak{T}_{zxx} \\ \mathfrak{T}_{yyz} & \mathfrak{T}_{yyz} & \mathfrak{T}_{zzz} & 0 & 0 & 0 \end{pmatrix}$ \\
$\mathtt{\bar{4}1'}$ & --- & --- & --- \\
$\mathtt{\bar{4}'}$ & --- & --- & $\begin{pmatrix} 0 & 0 & 0 & 0 & \mathfrak{T}_{zxx} & \mathfrak{T}_{yzx} \\ 0 & 0 & 0 & 0 & \mathfrak{T}_{yzx} & -\mathfrak{T}_{zxx} \\ -\mathfrak{T}_{yyz} & \mathfrak{T}_{yyz} & 0 & \mathfrak{T}_{xyz} & 0 & 0 \end{pmatrix}$ \\
$\mathtt{4/m}$ & $\begin{pmatrix} 0 \\ 0 \\ h_{z} \end{pmatrix}$ & $\begin{pmatrix} 0 \\ 0 \\ 2\mathfrak{T}_{zxx} + \mathfrak{T}_{zzz} \end{pmatrix}$ & $\begin{pmatrix} 0 & 0 & 0 & 0 & \mathfrak{T}_{zxx} & \mathfrak{T}_{yzx} \\ 0 & 0 & 0 & 0 & -\mathfrak{T}_{yzx} & \mathfrak{T}_{zxx} \\ \mathfrak{T}_{yyz} & \mathfrak{T}_{yyz} & \mathfrak{T}_{zzz} & 0 & 0 & 0 \end{pmatrix}$ \\
$\mathtt{4/m1'}$ & --- & --- & --- \\
$\mathtt{4'/m}$ & --- & --- & $\begin{pmatrix} 0 & 0 & 0 & 0 & \mathfrak{T}_{zxx} & \mathfrak{T}_{yzx} \\ 0 & 0 & 0 & 0 & \mathfrak{T}_{yzx} & -\mathfrak{T}_{zxx} \\ -\mathfrak{T}_{yyz} & \mathfrak{T}_{yyz} & 0 & \mathfrak{T}_{xyz} & 0 & 0 \end{pmatrix}$ \\
$\mathtt{4/m'}$ & --- & --- & --- \\
$\mathtt{4'/m'}$ & --- & --- & --- \\
$\mathtt{422}$ & --- & --- & $\begin{pmatrix} 0 & 0 & 0 & 0 & 0 & \mathfrak{T}_{yzx} \\ 0 & 0 & 0 & 0 & -\mathfrak{T}_{yzx} & 0 \\ 0 & 0 & 0 & 0 & 0 & 0 \end{pmatrix}$ \\
$\mathtt{4221'}$ & --- & --- & --- \\
$\mathtt{4'22'}$ & --- & --- & $\begin{pmatrix} 0 & 0 & 0 & 0 & 0 & \mathfrak{T}_{yzx} \\ 0 & 0 & 0 & 0 & \mathfrak{T}_{yzx} & 0 \\ 0 & 0 & 0 & \mathfrak{T}_{xyz} & 0 & 0 \end{pmatrix}$ \\
$\mathtt{42'2'}$ & $\begin{pmatrix} 0 \\ 0 \\ h_{z} \end{pmatrix}$ & $\begin{pmatrix} 0 \\ 0 \\ 2\mathfrak{T}_{zxx} + \mathfrak{T}_{zzz} \end{pmatrix}$ & $\begin{pmatrix} 0 & 0 & 0 & 0 & \mathfrak{T}_{zxx} & 0 \\ 0 & 0 & 0 & 0 & 0 & \mathfrak{T}_{zxx} \\ \mathfrak{T}_{yyz} & \mathfrak{T}_{yyz} & \mathfrak{T}_{zzz} & 0 & 0 & 0 \end{pmatrix}$ \\
$\mathtt{4mm}$ & --- & --- & $\begin{pmatrix} 0 & 0 & 0 & 0 & 0 & \mathfrak{T}_{yzx} \\ 0 & 0 & 0 & 0 & -\mathfrak{T}_{yzx} & 0 \\ 0 & 0 & 0 & 0 & 0 & 0 \end{pmatrix}$ \\
$\mathtt{4mm1'}$ & --- & --- & --- \\
$\mathtt{4'm'm}$ & --- & --- & $\begin{pmatrix} 0 & 0 & 0 & 0 & \mathfrak{T}_{zxx} & 0 \\ 0 & 0 & 0 & 0 & 0 & -\mathfrak{T}_{zxx} \\ -\mathfrak{T}_{yyz} & \mathfrak{T}_{yyz} & 0 & 0 & 0 & 0 \end{pmatrix}$ \\
$\mathtt{4m'm'}$ & $\begin{pmatrix} 0 \\ 0 \\ h_{z} \end{pmatrix}$ & $\begin{pmatrix} 0 \\ 0 \\ 2\mathfrak{T}_{zxx} + \mathfrak{T}_{zzz} \end{pmatrix}$ & $\begin{pmatrix} 0 & 0 & 0 & 0 & \mathfrak{T}_{zxx} & 0 \\ 0 & 0 & 0 & 0 & 0 & \mathfrak{T}_{zxx} \\ \mathfrak{T}_{yyz} & \mathfrak{T}_{yyz} & \mathfrak{T}_{zzz} & 0 & 0 & 0 \end{pmatrix}$ \\
$\mathtt{\bar{4}2m}$ & --- & --- & $\begin{pmatrix} 0 & 0 & 0 & 0 & 0 & \mathfrak{T}_{yzx} \\ 0 & 0 & 0 & 0 & -\mathfrak{T}_{yzx} & 0 \\ 0 & 0 & 0 & 0 & 0 & 0 \end{pmatrix}$ \\
$\mathtt{\bar{4}2m1'}$ & --- & --- & --- \\
\bottomrule
\end{tabular}
\end{center}

\clearpage
\begin{center}
\footnotesize\textbf{TABLE~\ref{tab:Hall_AMD_Ttensor_full}} (continued)

\footnotesize
\renewcommand{\arraystretch}{1.1}
\begin{tabular}{l l l l}
\toprule
MPG & Hall $\bm{h}$ & AMD $\bm{T}$ & Octupole \\
\midrule
$\mathtt{\bar{4}'2'm}$ & --- & --- & $\begin{pmatrix} 0 & 0 & 0 & 0 & \mathfrak{T}_{zxx} & 0 \\ 0 & 0 & 0 & 0 & 0 & -\mathfrak{T}_{zxx} \\ -\mathfrak{T}_{yyz} & \mathfrak{T}_{yyz} & 0 & 0 & 0 & 0 \end{pmatrix}$ \\
$\mathtt{\bar{4}'2m'}$ & --- & --- & $\begin{pmatrix} 0 & 0 & 0 & 0 & 0 & \mathfrak{T}_{yzx} \\ 0 & 0 & 0 & 0 & \mathfrak{T}_{yzx} & 0 \\ 0 & 0 & 0 & \mathfrak{T}_{xyz} & 0 & 0 \end{pmatrix}$ \\
$\mathtt{\bar{4}2'm'}$ & $\begin{pmatrix} 0 \\ 0 \\ h_{z} \end{pmatrix}$ & $\begin{pmatrix} 0 \\ 0 \\ 2\mathfrak{T}_{zxx} + \mathfrak{T}_{zzz} \end{pmatrix}$ & $\begin{pmatrix} 0 & 0 & 0 & 0 & \mathfrak{T}_{zxx} & 0 \\ 0 & 0 & 0 & 0 & 0 & \mathfrak{T}_{zxx} \\ \mathfrak{T}_{yyz} & \mathfrak{T}_{yyz} & \mathfrak{T}_{zzz} & 0 & 0 & 0 \end{pmatrix}$ \\
$\mathtt{\bar{4}m2}$ & --- & --- & $\begin{pmatrix} 0 & 0 & 0 & 0 & 0 & \mathfrak{T}_{yzx} \\ 0 & 0 & 0 & 0 & -\mathfrak{T}_{yzx} & 0 \\ 0 & 0 & 0 & 0 & 0 & 0 \end{pmatrix}$ \\
$\mathtt{\bar{4}m21'}$ & --- & --- & --- \\
$\mathtt{4/mmm}$ & --- & --- & $\begin{pmatrix} 0 & 0 & 0 & 0 & 0 & \mathfrak{T}_{yzx} \\ 0 & 0 & 0 & 0 & -\mathfrak{T}_{yzx} & 0 \\ 0 & 0 & 0 & 0 & 0 & 0 \end{pmatrix}$ \\
$\mathtt{4/mmm1'}$ & --- & --- & --- \\
$\mathtt{4/m'mm}$ & --- & --- & --- \\
$\mathtt{4'/mm'm}$ & --- & --- & $\begin{pmatrix} 0 & 0 & 0 & 0 & \mathfrak{T}_{zxx} & 0 \\ 0 & 0 & 0 & 0 & 0 & -\mathfrak{T}_{zxx} \\ -\mathfrak{T}_{yyz} & \mathfrak{T}_{yyz} & 0 & 0 & 0 & 0 \end{pmatrix}$ \\
$\mathtt{4'/m'm'm}$ & --- & --- & --- \\
$\mathtt{4'/m'mm'}$ & --- & --- & --- \\
$\mathtt{4/mm'm'}$ & $\begin{pmatrix} 0 \\ 0 \\ h_{z} \end{pmatrix}$ & $\begin{pmatrix} 0 \\ 0 \\ 2\mathfrak{T}_{zxx} + \mathfrak{T}_{zzz} \end{pmatrix}$ & $\begin{pmatrix} 0 & 0 & 0 & 0 & \mathfrak{T}_{zxx} & 0 \\ 0 & 0 & 0 & 0 & 0 & \mathfrak{T}_{zxx} \\ \mathfrak{T}_{yyz} & \mathfrak{T}_{yyz} & \mathfrak{T}_{zzz} & 0 & 0 & 0 \end{pmatrix}$ \\
$\mathtt{4/m'm'm'}$ & --- & --- & --- \\
$\mathtt{3}$ & $\begin{pmatrix} 0 \\ 0 \\ h_{z} \end{pmatrix}$ & $\begin{pmatrix} 0 \\ 0 \\ 2\mathfrak{T}_{zxx} + \mathfrak{T}_{zzz} \end{pmatrix}$ & $\begin{pmatrix} -\mathfrak{T}_{yyx} & \mathfrak{T}_{yyx} & 0 & \mathfrak{T}_{yxx} & \mathfrak{T}_{zxx} & \mathfrak{T}_{yzx} \\ \mathfrak{T}_{yxx} & -\mathfrak{T}_{yxx} & 0 & \mathfrak{T}_{yyx} & -\mathfrak{T}_{yzx} & \mathfrak{T}_{zxx} \\ \mathfrak{T}_{yyz} & \mathfrak{T}_{yyz} & \mathfrak{T}_{zzz} & 0 & 0 & 0 \end{pmatrix}$ \\
$\mathtt{31'}$ & --- & --- & --- \\
$\mathtt{\bar{3}}$ & $\begin{pmatrix} 0 \\ 0 \\ h_{z} \end{pmatrix}$ & $\begin{pmatrix} 0 \\ 0 \\ 2\mathfrak{T}_{zxx} + \mathfrak{T}_{zzz} \end{pmatrix}$ & $\begin{pmatrix} -\mathfrak{T}_{yyx} & \mathfrak{T}_{yyx} & 0 & \mathfrak{T}_{yxx} & \mathfrak{T}_{zxx} & \mathfrak{T}_{yzx} \\ \mathfrak{T}_{yxx} & -\mathfrak{T}_{yxx} & 0 & \mathfrak{T}_{yyx} & -\mathfrak{T}_{yzx} & \mathfrak{T}_{zxx} \\ \mathfrak{T}_{yyz} & \mathfrak{T}_{yyz} & \mathfrak{T}_{zzz} & 0 & 0 & 0 \end{pmatrix}$ \\
$\mathtt{\bar{3}1'}$ & --- & --- & --- \\
$\mathtt{\bar{3}'}$ & --- & --- & --- \\
$\mathtt{32}$ & --- & --- & $\begin{pmatrix} 0 & 0 & 0 & \mathfrak{T}_{yxx} & 0 & \mathfrak{T}_{yzx} \\ \mathfrak{T}_{yxx} & -\mathfrak{T}_{yxx} & 0 & 0 & -\mathfrak{T}_{yzx} & 0 \\ 0 & 0 & 0 & 0 & 0 & 0 \end{pmatrix}$ \\
$\mathtt{321'}$ & --- & --- & --- \\
$\mathtt{32'}$ & $\begin{pmatrix} 0 \\ 0 \\ h_{z} \end{pmatrix}$ & $\begin{pmatrix} 0 \\ 0 \\ 2\mathfrak{T}_{zxx} + \mathfrak{T}_{zzz} \end{pmatrix}$ & $\begin{pmatrix} -\mathfrak{T}_{yyx} & \mathfrak{T}_{yyx} & 0 & 0 & \mathfrak{T}_{zxx} & 0 \\ 0 & 0 & 0 & \mathfrak{T}_{yyx} & 0 & \mathfrak{T}_{zxx} \\ \mathfrak{T}_{yyz} & \mathfrak{T}_{yyz} & \mathfrak{T}_{zzz} & 0 & 0 & 0 \end{pmatrix}$ \\
$\mathtt{3m}$ & --- & --- & $\begin{pmatrix} 0 & 0 & 0 & \mathfrak{T}_{yxx} & 0 & \mathfrak{T}_{yzx} \\ \mathfrak{T}_{yxx} & -\mathfrak{T}_{yxx} & 0 & 0 & -\mathfrak{T}_{yzx} & 0 \\ 0 & 0 & 0 & 0 & 0 & 0 \end{pmatrix}$ \\
$\mathtt{3m1'}$ & --- & --- & --- \\
$\mathtt{3m'}$ & $\begin{pmatrix} 0 \\ 0 \\ h_{z} \end{pmatrix}$ & $\begin{pmatrix} 0 \\ 0 \\ 2\mathfrak{T}_{zxx} + \mathfrak{T}_{zzz} \end{pmatrix}$ & $\begin{pmatrix} -\mathfrak{T}_{yyx} & \mathfrak{T}_{yyx} & 0 & 0 & \mathfrak{T}_{zxx} & 0 \\ 0 & 0 & 0 & \mathfrak{T}_{yyx} & 0 & \mathfrak{T}_{zxx} \\ \mathfrak{T}_{yyz} & \mathfrak{T}_{yyz} & \mathfrak{T}_{zzz} & 0 & 0 & 0 \end{pmatrix}$ \\
$\mathtt{\bar{3}m}$ & --- & --- & $\begin{pmatrix} 0 & 0 & 0 & \mathfrak{T}_{yxx} & 0 & \mathfrak{T}_{yzx} \\ \mathfrak{T}_{yxx} & -\mathfrak{T}_{yxx} & 0 & 0 & -\mathfrak{T}_{yzx} & 0 \\ 0 & 0 & 0 & 0 & 0 & 0 \end{pmatrix}$ \\
$\mathtt{\bar{3}m1'}$ & --- & --- & --- \\
$\mathtt{\bar{3}'m}$ & --- & --- & --- \\
$\mathtt{\bar{3}'m'}$ & --- & --- & --- \\
$\mathtt{\bar{3}m'}$ & $\begin{pmatrix} 0 \\ 0 \\ h_{z} \end{pmatrix}$ & $\begin{pmatrix} 0 \\ 0 \\ 2\mathfrak{T}_{zxx} + \mathfrak{T}_{zzz} \end{pmatrix}$ & $\begin{pmatrix} -\mathfrak{T}_{yyx} & \mathfrak{T}_{yyx} & 0 & 0 & \mathfrak{T}_{zxx} & 0 \\ 0 & 0 & 0 & \mathfrak{T}_{yyx} & 0 & \mathfrak{T}_{zxx} \\ \mathfrak{T}_{yyz} & \mathfrak{T}_{yyz} & \mathfrak{T}_{zzz} & 0 & 0 & 0 \end{pmatrix}$ \\
\bottomrule
\end{tabular}
\end{center}

\clearpage
\begin{center}
\footnotesize\textbf{TABLE~\ref{tab:Hall_AMD_Ttensor_full}} (continued)

\footnotesize
\renewcommand{\arraystretch}{1.1}
\begin{tabular}{l l l l}
\toprule
MPG & Hall $\bm{h}$ & AMD $\bm{T}$ & Octupole \\
\midrule
$\mathtt{6}$ & $\begin{pmatrix} 0 \\ 0 \\ h_{z} \end{pmatrix}$ & $\begin{pmatrix} 0 \\ 0 \\ 2\mathfrak{T}_{zxx} + \mathfrak{T}_{zzz} \end{pmatrix}$ & $\begin{pmatrix} 0 & 0 & 0 & 0 & \mathfrak{T}_{zxx} & \mathfrak{T}_{yzx} \\ 0 & 0 & 0 & 0 & -\mathfrak{T}_{yzx} & \mathfrak{T}_{zxx} \\ \mathfrak{T}_{yyz} & \mathfrak{T}_{yyz} & \mathfrak{T}_{zzz} & 0 & 0 & 0 \end{pmatrix}$ \\
$\mathtt{61'}$ & --- & --- & --- \\
$\mathtt{6'}$ & --- & --- & $\begin{pmatrix} -\mathfrak{T}_{yyx} & \mathfrak{T}_{yyx} & 0 & \mathfrak{T}_{yxx} & 0 & 0 \\ \mathfrak{T}_{yxx} & -\mathfrak{T}_{yxx} & 0 & \mathfrak{T}_{yyx} & 0 & 0 \\ 0 & 0 & 0 & 0 & 0 & 0 \end{pmatrix}$ \\
$\mathtt{\bar{6}}$ & $\begin{pmatrix} 0 \\ 0 \\ h_{z} \end{pmatrix}$ & $\begin{pmatrix} 0 \\ 0 \\ 2\mathfrak{T}_{zxx} + \mathfrak{T}_{zzz} \end{pmatrix}$ & $\begin{pmatrix} 0 & 0 & 0 & 0 & \mathfrak{T}_{zxx} & \mathfrak{T}_{yzx} \\ 0 & 0 & 0 & 0 & -\mathfrak{T}_{yzx} & \mathfrak{T}_{zxx} \\ \mathfrak{T}_{yyz} & \mathfrak{T}_{yyz} & \mathfrak{T}_{zzz} & 0 & 0 & 0 \end{pmatrix}$ \\
$\mathtt{\bar{6}1'}$ & --- & --- & --- \\
$\mathtt{\bar{6}'}$ & --- & --- & $\begin{pmatrix} -\mathfrak{T}_{yyx} & \mathfrak{T}_{yyx} & 0 & \mathfrak{T}_{yxx} & 0 & 0 \\ \mathfrak{T}_{yxx} & -\mathfrak{T}_{yxx} & 0 & \mathfrak{T}_{yyx} & 0 & 0 \\ 0 & 0 & 0 & 0 & 0 & 0 \end{pmatrix}$ \\
$\mathtt{6/m}$ & $\begin{pmatrix} 0 \\ 0 \\ h_{z} \end{pmatrix}$ & $\begin{pmatrix} 0 \\ 0 \\ 2\mathfrak{T}_{zxx} + \mathfrak{T}_{zzz} \end{pmatrix}$ & $\begin{pmatrix} 0 & 0 & 0 & 0 & \mathfrak{T}_{zxx} & \mathfrak{T}_{yzx} \\ 0 & 0 & 0 & 0 & -\mathfrak{T}_{yzx} & \mathfrak{T}_{zxx} \\ \mathfrak{T}_{yyz} & \mathfrak{T}_{yyz} & \mathfrak{T}_{zzz} & 0 & 0 & 0 \end{pmatrix}$ \\
$\mathtt{6/m1'}$ & --- & --- & --- \\
$\mathtt{6'/m}$ & --- & --- & --- \\
$\mathtt{6/m'}$ & --- & --- & --- \\
$\mathtt{6'/m'}$ & --- & --- & $\begin{pmatrix} -\mathfrak{T}_{yyx} & \mathfrak{T}_{yyx} & 0 & \mathfrak{T}_{yxx} & 0 & 0 \\ \mathfrak{T}_{yxx} & -\mathfrak{T}_{yxx} & 0 & \mathfrak{T}_{yyx} & 0 & 0 \\ 0 & 0 & 0 & 0 & 0 & 0 \end{pmatrix}$ \\
$\mathtt{622}$ & --- & --- & $\begin{pmatrix} 0 & 0 & 0 & 0 & 0 & \mathfrak{T}_{yzx} \\ 0 & 0 & 0 & 0 & -\mathfrak{T}_{yzx} & 0 \\ 0 & 0 & 0 & 0 & 0 & 0 \end{pmatrix}$ \\
$\mathtt{6221'}$ & --- & --- & --- \\
$\mathtt{6'22'}$ & --- & --- & $\begin{pmatrix} -\mathfrak{T}_{yyx} & \mathfrak{T}_{yyx} & 0 & 0 & 0 & 0 \\ 0 & 0 & 0 & \mathfrak{T}_{yyx} & 0 & 0 \\ 0 & 0 & 0 & 0 & 0 & 0 \end{pmatrix}$ \\
$\mathtt{62'2'}$ & $\begin{pmatrix} 0 \\ 0 \\ h_{z} \end{pmatrix}$ & $\begin{pmatrix} 0 \\ 0 \\ 2\mathfrak{T}_{zxx} + \mathfrak{T}_{zzz} \end{pmatrix}$ & $\begin{pmatrix} 0 & 0 & 0 & 0 & \mathfrak{T}_{zxx} & 0 \\ 0 & 0 & 0 & 0 & 0 & \mathfrak{T}_{zxx} \\ \mathfrak{T}_{yyz} & \mathfrak{T}_{yyz} & \mathfrak{T}_{zzz} & 0 & 0 & 0 \end{pmatrix}$ \\
$\mathtt{6mm}$ & --- & --- & $\begin{pmatrix} 0 & 0 & 0 & 0 & 0 & \mathfrak{T}_{yzx} \\ 0 & 0 & 0 & 0 & -\mathfrak{T}_{yzx} & 0 \\ 0 & 0 & 0 & 0 & 0 & 0 \end{pmatrix}$ \\
$\mathtt{6mm1'}$ & --- & --- & --- \\
$\mathtt{6'mm'}$ & --- & --- & $\begin{pmatrix} -\mathfrak{T}_{yyx} & \mathfrak{T}_{yyx} & 0 & 0 & 0 & 0 \\ 0 & 0 & 0 & \mathfrak{T}_{yyx} & 0 & 0 \\ 0 & 0 & 0 & 0 & 0 & 0 \end{pmatrix}$ \\
$\mathtt{6m'm'}$ & $\begin{pmatrix} 0 \\ 0 \\ h_{z} \end{pmatrix}$ & $\begin{pmatrix} 0 \\ 0 \\ 2\mathfrak{T}_{zxx} + \mathfrak{T}_{zzz} \end{pmatrix}$ & $\begin{pmatrix} 0 & 0 & 0 & 0 & \mathfrak{T}_{zxx} & 0 \\ 0 & 0 & 0 & 0 & 0 & \mathfrak{T}_{zxx} \\ \mathfrak{T}_{yyz} & \mathfrak{T}_{yyz} & \mathfrak{T}_{zzz} & 0 & 0 & 0 \end{pmatrix}$ \\
$\mathtt{\bar{6}m2}$ & --- & --- & $\begin{pmatrix} 0 & 0 & 0 & 0 & 0 & \mathfrak{T}_{yzx} \\ 0 & 0 & 0 & 0 & -\mathfrak{T}_{yzx} & 0 \\ 0 & 0 & 0 & 0 & 0 & 0 \end{pmatrix}$ \\
$\mathtt{\bar{6}m21'}$ & --- & --- & --- \\
$\mathtt{\bar{6}'m'2}$ & --- & --- & $\begin{pmatrix} 0 & 0 & 0 & \mathfrak{T}_{yxx} & 0 & 0 \\ \mathfrak{T}_{yxx} & -\mathfrak{T}_{yxx} & 0 & 0 & 0 & 0 \\ 0 & 0 & 0 & 0 & 0 & 0 \end{pmatrix}$ \\
$\mathtt{\bar{6}'m2'}$ & --- & --- & $\begin{pmatrix} -\mathfrak{T}_{yyx} & \mathfrak{T}_{yyx} & 0 & 0 & 0 & 0 \\ 0 & 0 & 0 & \mathfrak{T}_{yyx} & 0 & 0 \\ 0 & 0 & 0 & 0 & 0 & 0 \end{pmatrix}$ \\
$\mathtt{\bar{6}m'2'}$ & $\begin{pmatrix} 0 \\ 0 \\ h_{z} \end{pmatrix}$ & $\begin{pmatrix} 0 \\ 0 \\ 2\mathfrak{T}_{zxx} + \mathfrak{T}_{zzz} \end{pmatrix}$ & $\begin{pmatrix} 0 & 0 & 0 & 0 & \mathfrak{T}_{zxx} & 0 \\ 0 & 0 & 0 & 0 & 0 & \mathfrak{T}_{zxx} \\ \mathfrak{T}_{yyz} & \mathfrak{T}_{yyz} & \mathfrak{T}_{zzz} & 0 & 0 & 0 \end{pmatrix}$ \\
$\mathtt{\bar{6}2m}$ & --- & --- & $\begin{pmatrix} 0 & 0 & 0 & 0 & 0 & \mathfrak{T}_{yzx} \\ 0 & 0 & 0 & 0 & -\mathfrak{T}_{yzx} & 0 \\ 0 & 0 & 0 & 0 & 0 & 0 \end{pmatrix}$ \\
\bottomrule
\end{tabular}
\end{center}

\clearpage
\begin{center}
\footnotesize\textbf{TABLE~\ref{tab:Hall_AMD_Ttensor_full}} (continued)

\footnotesize
\renewcommand{\arraystretch}{1.1}
\begin{tabular}{l l l l}
\toprule
MPG & Hall $\bm{h}$ & AMD $\bm{T}$ & Octupole \\
\midrule
$\mathtt{\bar{6}2m1'}$ & --- & --- & --- \\
$\mathtt{6/mmm}$ & --- & --- & $\begin{pmatrix} 0 & 0 & 0 & 0 & 0 & \mathfrak{T}_{yzx} \\ 0 & 0 & 0 & 0 & -\mathfrak{T}_{yzx} & 0 \\ 0 & 0 & 0 & 0 & 0 & 0 \end{pmatrix}$ \\
$\mathtt{6/mmm1'}$ & --- & --- & --- \\
$\mathtt{6/m'mm}$ & --- & --- & --- \\
$\mathtt{6'/mm'm}$ & --- & --- & --- \\
$\mathtt{6'/mmm'}$ & --- & --- & --- \\
$\mathtt{6'/m'mm'}$ & --- & --- & $\begin{pmatrix} -\mathfrak{T}_{yyx} & \mathfrak{T}_{yyx} & 0 & 0 & 0 & 0 \\ 0 & 0 & 0 & \mathfrak{T}_{yyx} & 0 & 0 \\ 0 & 0 & 0 & 0 & 0 & 0 \end{pmatrix}$ \\
$\mathtt{6/mm'm'}$ & $\begin{pmatrix} 0 \\ 0 \\ h_{z} \end{pmatrix}$ & $\begin{pmatrix} 0 \\ 0 \\ 2\mathfrak{T}_{zxx} + \mathfrak{T}_{zzz} \end{pmatrix}$ & $\begin{pmatrix} 0 & 0 & 0 & 0 & \mathfrak{T}_{zxx} & 0 \\ 0 & 0 & 0 & 0 & 0 & \mathfrak{T}_{zxx} \\ \mathfrak{T}_{yyz} & \mathfrak{T}_{yyz} & \mathfrak{T}_{zzz} & 0 & 0 & 0 \end{pmatrix}$ \\
$\mathtt{6/m'm'm'}$ & --- & --- & --- \\
$\mathtt{23}$ & --- & --- & $\begin{pmatrix} 0 & 0 & 0 & 0 & 0 & \mathfrak{T}_{yzx} \\ 0 & 0 & 0 & 0 & \mathfrak{T}_{yzx} & 0 \\ 0 & 0 & 0 & \mathfrak{T}_{yzx} & 0 & 0 \end{pmatrix}$ \\
$\mathtt{231'}$ & --- & --- & --- \\
$\mathtt{m\bar{3}}$ & --- & --- & $\begin{pmatrix} 0 & 0 & 0 & 0 & 0 & \mathfrak{T}_{yzx} \\ 0 & 0 & 0 & 0 & \mathfrak{T}_{yzx} & 0 \\ 0 & 0 & 0 & \mathfrak{T}_{yzx} & 0 & 0 \end{pmatrix}$ \\
$\mathtt{m\bar{3}1'}$ & --- & --- & --- \\
$\mathtt{m'\bar{3}'}$ & --- & --- & --- \\
$\mathtt{432}$ & --- & --- & --- \\
$\mathtt{4321'}$ & --- & --- & --- \\
$\mathtt{4'32'}$ & --- & --- & $\begin{pmatrix} 0 & 0 & 0 & 0 & 0 & \mathfrak{T}_{yzx} \\ 0 & 0 & 0 & 0 & \mathfrak{T}_{yzx} & 0 \\ 0 & 0 & 0 & \mathfrak{T}_{yzx} & 0 & 0 \end{pmatrix}$ \\
$\mathtt{\bar{4}3m}$ & --- & --- & --- \\
$\mathtt{\bar{4}3m1'}$ & --- & --- & --- \\
$\mathtt{\bar{4}'3m'}$ & --- & --- & $\begin{pmatrix} 0 & 0 & 0 & 0 & 0 & \mathfrak{T}_{yzx} \\ 0 & 0 & 0 & 0 & \mathfrak{T}_{yzx} & 0 \\ 0 & 0 & 0 & \mathfrak{T}_{yzx} & 0 & 0 \end{pmatrix}$ \\
$\mathtt{m\bar{3}m}$ & --- & --- & --- \\
$\mathtt{m\bar{3}m1'}$ & --- & --- & --- \\
$\mathtt{m'\bar{3}'m}$ & --- & --- & --- \\
$\mathtt{m\bar{3}m'}$ & --- & --- & $\begin{pmatrix} 0 & 0 & 0 & 0 & 0 & \mathfrak{T}_{yzx} \\ 0 & 0 & 0 & 0 & \mathfrak{T}_{yzx} & 0 \\ 0 & 0 & 0 & \mathfrak{T}_{yzx} & 0 & 0 \end{pmatrix}$ \\
$\mathtt{m'\bar{3}'m'}$ & --- & --- & --- \\
\bottomrule
\end{tabular}
\end{center}

%\twocolumngrid
\section{Couplings between the N\'eel vector, AMD and optical Hall vector}\label{app:LT_couplings}

The following tables list the lowest order couplings between $\bm{N}$ and $\bm{T}$, as well as the leading order polynomial of $\bm{N}$ contributing to $\bm{h}$. The three cases are described in Sec.~\ref{sec:LT}. We use the following conventions on the axes: the monoclinic unique axis is $\hat y$; we take orthorhombic two-folds along $\bm{\hat x},\bm{\hat y},\bm{\hat z}$; the tetragonal $C_2'\parallel\bm{\hat x}$ ($\mathtt{\bar{4}2m}$: $C_2\parallel\bm{\hat x}$; $\mathtt{\bar{4}m2}$: $m\perp\bm{\hat x}$); in the hexagonal and trigonal classes $\bm{\hat x}\parallel\bm a_1$, $\bm{\hat z}\parallel\bm c$, the frame in which MnTe (Mn on $2a$ of $P6_3/mmc$, sublattice-preserving two-folds along $\bm a_1$) has $\Gamma_{\mathbf N}=B_{2g}$, i.e.\ $B_1$/$B_2$ subscripts are even/odd under the two-folds along $\bm{\hat y}$; the trigonal groups $\mathtt{32}$, $\mathtt{3m}$, $\mathtt{\bar{3}m}$ are taken with their two-folds along $\hat y$ (mirrors $\perp\bm{\hat y}$), so that $\mathtt{6/mmm}$ $B_{2g}$ restricts to $\mathtt{\bar{3}m}$ $A_{2g}$; $\mathtt{\bar{6}m2}$: $C_2\parallel\bm{\hat y}$; $\mathtt{\bar{6}2m}$: $C_2\parallel\bm{\hat x}$. These axis conventions align with those of Tab.~\ref{tab:tvec_short}, Tab.~\ref{tab:AM_MPG}, as well as Ref.~\cite{schiff2025collinear}. We emphasize that the irrep labels are taken from the \textsc{Bilbao crystallographic Server}, and occasionally do not match the standard Mulliken notation (e.g. $\mathtt{\bar{6}m2}$ has $A_{1}^{\prime\prime}$ even under the $m_{001}$ mirror, while in $\mathtt{\bar{6}2m}$ $A_{1}^{\prime\prime}$ is odd under this mirror). Specifically, our MPG irrep data are sourced from the \emph{Corepresentations of Magnetic Point Groups} package on the server \cite{BilbaoIrreps,Xu2020_MPGCoIrrCite}.%With these axes the SO-free multipoles of Table~XIV transform as labelled, e.g.\ $\mathtt{4/mmm}$ $B_{2g}\{xy\}$, $\mathtt{\bar{3}m}$ $A_{2g}\{y(y^2-3x^2)z\}$, $\mathtt{6/mmm}$ $B_{2g}\{yz(y^2-3x^2)\}$. $^{\ddagger}$Standard (character-table) label. Matched by basis polynomial, Table~XIV of Schiff \emph{et al.} uses: $\mathtt{\bar{4}m2}$: $A_2\{xyz\}\to B_2$, $B_1\{xy\}\to A_2$, $B_2\{z\}\to B_1$; $\mathtt{\bar{6}m2}$: $A_1''\{xz(x^2-3y^2)\}\to A_2'$, $A_2'\{x(3y^2-x^2)\}\to A_1''$.

% table_linear_symmetric -- requires only amsmath
\begin{table*}[tp]
\caption{Linear, symmetric XMCD response for each crystalline point group $\mathtt{F}$ and $\Gamma_{\mathbf{N}}$ describing the spin-orbit free AM order, along with the minimal $n$ such that $\Gamma_{\mathbf N}\subset[V^n]$. The $\bm{T}$--$\mathbf N$ columns list independent invariant couplings between the AMD, $\bm{T}$, and the N\'eel vector, $\mathbf N$. The $\bm{h}(\mathbf{N})$ columns list  independent linear terms in the expansion of $\bm{h}$ in $\mathbf{N}.$ $^{\ast}$Also carries an antisymmetric channel (Table~\ref{tab:LT_linear_anti}).}\label{tab:LT_linear_symm}
\centering\footnotesize\renewcommand{\arraystretch}{1.2}\setlength{\tabcolsep}{3.5pt}
\begin{tabular*}{\textwidth}{@{\extracolsep{\fill}}l l c l l l l c l l@{}}
\hline\hline
$\mathtt{F}$ & $\Gamma_{\mathbf{N}}$ & $n$ & $\bm{T}$--$\mathbf{N}$ coupling & $\bm{h}(\mathbf{N})$ & $\mathtt{F}$ & $\Gamma_{\mathbf{N}}$ & $n$ & $\bm{T}$--$\mathbf{N}$ coupling & $\bm{h}(\mathbf{N})$ \\
\hline
$\mathtt{2}$ & $B$$^{\ast}$ & $1$ & \begin{tabular}[c]{@{}l@{}}$T_{x}N_{y}+T_{y}N_{x}$\\$T_{y}N_{z}+T_{z}N_{y}$\end{tabular} & $a_{1}\!\begin{pmatrix}N_{y}\\N_{x}\\0\end{pmatrix}+a_{2}\!\begin{pmatrix}0\\N_{z}\\N_{y}\end{pmatrix}$ & $\mathtt{4}$ & $B$ & $2$ & \begin{tabular}[c]{@{}l@{}}$T_{x}N_{x}-T_{y}N_{y}$\\$T_{x}N_{y}+T_{y}N_{x}$\end{tabular} & $a_{1}\!\begin{pmatrix}N_{x}\\-N_{y}\\0\end{pmatrix}+a_{2}\!\begin{pmatrix}N_{y}\\N_{x}\\0\end{pmatrix}$ \\
\noalign{\vskip 3pt}
$\mathtt{m}$ & $A^{\prime\prime}$$^{\ast}$ & $1$ & \begin{tabular}[c]{@{}l@{}}$T_{x}N_{y}+T_{y}N_{x}$\\$T_{y}N_{z}+T_{z}N_{y}$\end{tabular} & $a_{1}\!\begin{pmatrix}N_{y}\\N_{x}\\0\end{pmatrix}+a_{2}\!\begin{pmatrix}0\\N_{z}\\N_{y}\end{pmatrix}$ & $\mathtt{\bar{4}}$ & $B$ & $1$ & \begin{tabular}[c]{@{}l@{}}$T_{x}N_{x}-T_{y}N_{y}$\\$T_{x}N_{y}+T_{y}N_{x}$\end{tabular} & $a_{1}\!\begin{pmatrix}N_{x}\\-N_{y}\\0\end{pmatrix}+a_{2}\!\begin{pmatrix}N_{y}\\N_{x}\\0\end{pmatrix}$ \\
\noalign{\vskip 3pt}
$\mathtt{2/m}$ & $B_g$$^{\ast}$ & $2$ & \begin{tabular}[c]{@{}l@{}}$T_{x}N_{y}+T_{y}N_{x}$\\$T_{y}N_{z}+T_{z}N_{y}$\end{tabular} & $a_{1}\!\begin{pmatrix}N_{y}\\N_{x}\\0\end{pmatrix}+a_{2}\!\begin{pmatrix}0\\N_{z}\\N_{y}\end{pmatrix}$ & $\mathtt{4/m}$ & $B_g$ & $2$ & \begin{tabular}[c]{@{}l@{}}$T_{x}N_{x}-T_{y}N_{y}$\\$T_{x}N_{y}+T_{y}N_{x}$\end{tabular} & $a_{1}\!\begin{pmatrix}N_{x}\\-N_{y}\\0\end{pmatrix}+a_{2}\!\begin{pmatrix}N_{y}\\N_{x}\\0\end{pmatrix}$ \\
\noalign{\vskip 3pt}
$\mathtt{222}$ & $B_{1}$$^{\ast}$ & $1$ & \begin{tabular}[c]{@{}l@{}}$T_{x}N_{y}+T_{y}N_{x}$\end{tabular} & $a\!\begin{pmatrix}N_{y}\\N_{x}\\0\end{pmatrix}$ & $\mathtt{422}$ & $B_{1}$ & $2$ & \begin{tabular}[c]{@{}l@{}}$T_{x}N_{x}-T_{y}N_{y}$\end{tabular} & $a\!\begin{pmatrix}N_{x}\\-N_{y}\\0\end{pmatrix}$ \\
\noalign{\vskip 3pt}
 & $B_{2}$$^{\ast}$ & $1$ & \begin{tabular}[c]{@{}l@{}}$T_{x}N_{z}+T_{z}N_{x}$\end{tabular} & $a\!\begin{pmatrix}N_{z}\\0\\N_{x}\end{pmatrix}$ &  & $B_{2}$ & $2$ & \begin{tabular}[c]{@{}l@{}}$T_{x}N_{y}+T_{y}N_{x}$\end{tabular} & $a\!\begin{pmatrix}N_{y}\\N_{x}\\0\end{pmatrix}$ \\
\noalign{\vskip 3pt}
 & $B_{3}$$^{\ast}$ & $1$ & \begin{tabular}[c]{@{}l@{}}$T_{y}N_{z}+T_{z}N_{y}$\end{tabular} & $a\!\begin{pmatrix}0\\N_{z}\\N_{y}\end{pmatrix}$ & $\mathtt{4mm}$ & $B_1$ & $2$ & \begin{tabular}[c]{@{}l@{}}$T_{x}N_{x}-T_{y}N_{y}$\end{tabular} & $a\!\begin{pmatrix}N_{x}\\-N_{y}\\0\end{pmatrix}$ \\
\noalign{\vskip 3pt}
$\mathtt{mm2}$ & $A_2$$^{\ast}$ & $2$ & \begin{tabular}[c]{@{}l@{}}$T_{x}N_{y}+T_{y}N_{x}$\end{tabular} & $a\!\begin{pmatrix}N_{y}\\N_{x}\\0\end{pmatrix}$ &  & $B_2$ & $2$ & \begin{tabular}[c]{@{}l@{}}$T_{x}N_{y}+T_{y}N_{x}$\end{tabular} & $a\!\begin{pmatrix}N_{y}\\N_{x}\\0\end{pmatrix}$ \\
\noalign{\vskip 3pt}
 & $B_1$$^{\ast}$ & $1$ & \begin{tabular}[c]{@{}l@{}}$T_{x}N_{z}+T_{z}N_{x}$\end{tabular} & $a\!\begin{pmatrix}N_{z}\\0\\N_{x}\end{pmatrix}$ & $\mathtt{\bar{4}2m}$ & $B_1$ & $2$ & \begin{tabular}[c]{@{}l@{}}$T_{x}N_{x}-T_{y}N_{y}$\end{tabular} & $a\!\begin{pmatrix}N_{x}\\-N_{y}\\0\end{pmatrix}$ \\
\noalign{\vskip 3pt}
 & $B_2$$^{\ast}$ & $1$ & \begin{tabular}[c]{@{}l@{}}$T_{y}N_{z}+T_{z}N_{y}$\end{tabular} & $a\!\begin{pmatrix}0\\N_{z}\\N_{y}\end{pmatrix}$ &  & $B_2$ & $1$ & \begin{tabular}[c]{@{}l@{}}$T_{x}N_{y}+T_{y}N_{x}$\end{tabular} & $a\!\begin{pmatrix}N_{y}\\N_{x}\\0\end{pmatrix}$ \\
\noalign{\vskip 3pt}
$\mathtt{mmm}$ & $B_{1g}$$^{\ast}$ & $2$ & \begin{tabular}[c]{@{}l@{}}$T_{x}N_{y}+T_{y}N_{x}$\end{tabular} & $a\!\begin{pmatrix}N_{y}\\N_{x}\\0\end{pmatrix}$ & $\mathtt{\bar{4}m2}$ & $B_1$ & $1$ & \begin{tabular}[c]{@{}l@{}}$T_{x}N_{x}-T_{y}N_{y}$\end{tabular} & $a\!\begin{pmatrix}N_{x}\\-N_{y}\\0\end{pmatrix}$ \\
\noalign{\vskip 3pt}
 & $B_{2g}$$^{\ast}$ & $2$ & \begin{tabular}[c]{@{}l@{}}$T_{x}N_{z}+T_{z}N_{x}$\end{tabular} & $a\!\begin{pmatrix}N_{z}\\0\\N_{x}\end{pmatrix}$ &  & $A_2$ & $2$ & \begin{tabular}[c]{@{}l@{}}$T_{x}N_{y}+T_{y}N_{x}$\end{tabular} & $a\!\begin{pmatrix}N_{y}\\N_{x}\\0\end{pmatrix}$ \\
\noalign{\vskip 3pt}
 & $B_{3g}$$^{\ast}$ & $2$ & \begin{tabular}[c]{@{}l@{}}$T_{y}N_{z}+T_{z}N_{y}$\end{tabular} & $a\!\begin{pmatrix}0\\N_{z}\\N_{y}\end{pmatrix}$ & $\mathtt{4/mmm}$ & $B_{1g}$ & $2$ & \begin{tabular}[c]{@{}l@{}}$T_{x}N_{x}-T_{y}N_{y}$\end{tabular} & $a\!\begin{pmatrix}N_{x}\\-N_{y}\\0\end{pmatrix}$ \\
\noalign{\vskip 3pt}
 & & & &  &  & $B_{2g}$ & $2$ & \begin{tabular}[c]{@{}l@{}}$T_{x}N_{y}+T_{y}N_{x}$\end{tabular} & $a\!\begin{pmatrix}N_{y}\\N_{x}\\0\end{pmatrix}$ \\
\hline\hline
\end{tabular*}
\end{table*}

% table_linear_antisymmetric -- requires only amsmath
\begin{table*}[tp]
\caption{Linear, antisymmetric XMCD response. Columns as in Table~\ref{tab:LT_linear_symm}. $^{\ast}$Also carries a symmetric channel (Table~\ref{tab:LT_linear_symm}); all other rows have $\bm{h}\propto\mathbf N\times\hat{\mathbf z}$ as the leading term.}\label{tab:LT_linear_anti}
\centering\footnotesize\renewcommand{\arraystretch}{1.2}\setlength{\tabcolsep}{3.5pt}
\begin{tabular*}{\textwidth}{@{\extracolsep{\fill}}l l c l l l l c l l@{}}
\hline\hline
$\mathtt{F}$ & $\Gamma_{\mathbf{N}}$ & $n$ & $\bm{T}$--$\mathbf{N}$ coupling & $\bm{h}(\mathbf{N})$ & $\mathtt{F}$ & $\Gamma_{\mathbf{N}}$ & $n$ & $\bm{T}$--$\mathbf{N}$ coupling & $\bm{h}(\mathbf{N})$ \\
\hline
$\mathtt{2}$ & $B$$^{\ast}$ & $1$ & \begin{tabular}[c]{@{}l@{}}$T_{x}N_{y}-T_{y}N_{x}$\\$T_{y}N_{z}-T_{z}N_{y}$\end{tabular} & $c_{1}\!\begin{pmatrix}N_{y}\\-N_{x}\\0\end{pmatrix}+c_{2}\!\begin{pmatrix}0\\N_{z}\\-N_{y}\end{pmatrix}$ & $\mathtt{422}$ & $A_{2}$ & $1$ & \begin{tabular}[c]{@{}l@{}}$T_{x}N_{y}-T_{y}N_{x}$\end{tabular} & $c\!\begin{pmatrix}N_{y}\\-N_{x}\\0\end{pmatrix}$ \\
\noalign{\vskip 3pt}
$\mathtt{m}$ & $A^{\prime\prime}$$^{\ast}$ & $1$ & \begin{tabular}[c]{@{}l@{}}$T_{x}N_{y}-T_{y}N_{x}$\\$T_{y}N_{z}-T_{z}N_{y}$\end{tabular} & $c_{1}\!\begin{pmatrix}N_{y}\\-N_{x}\\0\end{pmatrix}+c_{2}\!\begin{pmatrix}0\\N_{z}\\-N_{y}\end{pmatrix}$ & $\mathtt{4mm}$ & $A_2$ & $4$ & \begin{tabular}[c]{@{}l@{}}$T_{x}N_{y}-T_{y}N_{x}$\end{tabular} & $c\!\begin{pmatrix}N_{y}\\-N_{x}\\0\end{pmatrix}$ \\
\noalign{\vskip 3pt}
$\mathtt{2/m}$ & $B_g$$^{\ast}$ & $2$ & \begin{tabular}[c]{@{}l@{}}$T_{x}N_{y}-T_{y}N_{x}$\\$T_{y}N_{z}-T_{z}N_{y}$\end{tabular} & $c_{1}\!\begin{pmatrix}N_{y}\\-N_{x}\\0\end{pmatrix}+c_{2}\!\begin{pmatrix}0\\N_{z}\\-N_{y}\end{pmatrix}$ & $\mathtt{\bar{4}2m}$ & $A_2$ & $3$ & \begin{tabular}[c]{@{}l@{}}$T_{x}N_{y}-T_{y}N_{x}$\end{tabular} & $c\!\begin{pmatrix}N_{y}\\-N_{x}\\0\end{pmatrix}$ \\
\noalign{\vskip 3pt}
$\mathtt{222}$ & $B_{1}$$^{\ast}$ & $1$ & \begin{tabular}[c]{@{}l@{}}$T_{x}N_{y}-T_{y}N_{x}$\end{tabular} & $c\!\begin{pmatrix}N_{y}\\-N_{x}\\0\end{pmatrix}$ & $\mathtt{\bar{4}m2}$ & $B_2$ & $3$ & \begin{tabular}[c]{@{}l@{}}$T_{x}N_{y}-T_{y}N_{x}$\end{tabular} & $c\!\begin{pmatrix}N_{y}\\-N_{x}\\0\end{pmatrix}$ \\
\noalign{\vskip 3pt}
 & $B_{2}$$^{\ast}$ & $1$ & \begin{tabular}[c]{@{}l@{}}$T_{x}N_{z}-T_{z}N_{x}$\end{tabular} & $c\!\begin{pmatrix}N_{z}\\0\\-N_{x}\end{pmatrix}$ & $\mathtt{4/mmm}$ & $A_{2g}$ & $4$ & \begin{tabular}[c]{@{}l@{}}$T_{x}N_{y}-T_{y}N_{x}$\end{tabular} & $c\!\begin{pmatrix}N_{y}\\-N_{x}\\0\end{pmatrix}$ \\
\noalign{\vskip 3pt}
 & $B_{3}$$^{\ast}$ & $1$ & \begin{tabular}[c]{@{}l@{}}$T_{y}N_{z}-T_{z}N_{y}$\end{tabular} & $c\!\begin{pmatrix}0\\N_{z}\\-N_{y}\end{pmatrix}$ & $\mathtt{32}$ & $A_2$ & $1$ & \begin{tabular}[c]{@{}l@{}}$T_{x}N_{y}-T_{y}N_{x}$\end{tabular} & $c\!\begin{pmatrix}N_{y}\\-N_{x}\\0\end{pmatrix}$ \\
\noalign{\vskip 3pt}
$\mathtt{mm2}$ & $A_2$$^{\ast}$ & $2$ & \begin{tabular}[c]{@{}l@{}}$T_{x}N_{y}-T_{y}N_{x}$\end{tabular} & $c\!\begin{pmatrix}N_{y}\\-N_{x}\\0\end{pmatrix}$ & $\mathtt{3m}$ & $A_2$ & $3$ & \begin{tabular}[c]{@{}l@{}}$T_{x}N_{y}-T_{y}N_{x}$\end{tabular} & $c\!\begin{pmatrix}N_{y}\\-N_{x}\\0\end{pmatrix}$ \\
\noalign{\vskip 3pt}
 & $B_1$$^{\ast}$ & $1$ & \begin{tabular}[c]{@{}l@{}}$T_{x}N_{z}-T_{z}N_{x}$\end{tabular} & $c\!\begin{pmatrix}N_{z}\\0\\-N_{x}\end{pmatrix}$ & $\mathtt{\bar{3}m}$ & $A_{2g}$ & $4$ & \begin{tabular}[c]{@{}l@{}}$T_{x}N_{y}-T_{y}N_{x}$\end{tabular} & $c\!\begin{pmatrix}N_{y}\\-N_{x}\\0\end{pmatrix}$ \\
\noalign{\vskip 3pt}
 & $B_2$$^{\ast}$ & $1$ & \begin{tabular}[c]{@{}l@{}}$T_{y}N_{z}-T_{z}N_{y}$\end{tabular} & $c\!\begin{pmatrix}0\\N_{z}\\-N_{y}\end{pmatrix}$ & $\mathtt{622}$ & $A_{2}$ & $1$ & \begin{tabular}[c]{@{}l@{}}$T_{x}N_{y}-T_{y}N_{x}$\end{tabular} & $c\!\begin{pmatrix}N_{y}\\-N_{x}\\0\end{pmatrix}$ \\
\noalign{\vskip 3pt}
$\mathtt{mmm}$ & $B_{1g}$$^{\ast}$ & $2$ & \begin{tabular}[c]{@{}l@{}}$T_{x}N_{y}-T_{y}N_{x}$\end{tabular} & $c\!\begin{pmatrix}N_{y}\\-N_{x}\\0\end{pmatrix}$ & $\mathtt{6mm}$ & $A_2$ & $6$ & \begin{tabular}[c]{@{}l@{}}$T_{x}N_{y}-T_{y}N_{x}$\end{tabular} & $c\!\begin{pmatrix}N_{y}\\-N_{x}\\0\end{pmatrix}$ \\
\noalign{\vskip 3pt}
 & $B_{2g}$$^{\ast}$ & $2$ & \begin{tabular}[c]{@{}l@{}}$T_{x}N_{z}-T_{z}N_{x}$\end{tabular} & $c\!\begin{pmatrix}N_{z}\\0\\-N_{x}\end{pmatrix}$ & $\mathtt{\bar{6}m2}$ & $A_1^{\prime\prime}$ & $3$ & \begin{tabular}[c]{@{}l@{}}$T_{x}N_{y}-T_{y}N_{x}$\end{tabular} & $c\!\begin{pmatrix}N_{y}\\-N_{x}\\0\end{pmatrix}$ \\
\noalign{\vskip 3pt}
 & $B_{3g}$$^{\ast}$ & $2$ & \begin{tabular}[c]{@{}l@{}}$T_{y}N_{z}-T_{z}N_{y}$\end{tabular} & $c\!\begin{pmatrix}0\\N_{z}\\-N_{y}\end{pmatrix}$ & $\mathtt{\bar{6}2m}$ & $A_2^{\prime}$ & $3$ & \begin{tabular}[c]{@{}l@{}}$T_{x}N_{y}-T_{y}N_{x}$\end{tabular} & $c\!\begin{pmatrix}N_{y}\\-N_{x}\\0\end{pmatrix}$ \\
\noalign{\vskip 3pt}
 & & & &  & $\mathtt{6/mmm}$ & $A_{2g}$ & $6$ & \begin{tabular}[c]{@{}l@{}}$T_{x}N_{y}-T_{y}N_{x}$\end{tabular} & $c\!\begin{pmatrix}N_{y}\\-N_{x}\\0\end{pmatrix}$ \\
\hline\hline
\end{tabular*}
\end{table*}

% table_cubic -- requires only amsmath
\begin{table*}[tp]
\caption{Cases where the leading order XMCD response is cubic. Columns as in Table~\ref{tab:LT_linear_symm}. Coefficients $\tilde{b}_{i}$ are not inherited from the guaranteed tensor of Ref.~\cite{schiff2025collinear}.}\label{tab:LT_cubic}
\centering\footnotesize\renewcommand{\arraystretch}{1.25}
\begin{tabular*}{\textwidth}{@{\extracolsep{\fill}}l l c l l@{}}
\hline\hline
$\mathtt{F}$ & $\Gamma_{\mathbf{N}}$ & $n$ & $\bm{T}$--$\mathbf{N}$ coupling & $\bm{h}(\mathbf{N})$ \\
\hline
$\mathtt{6}$ & $B$ & $3$ & \begin{tabular}[c]{@{}l@{}}$N_{z}\,[T_{x}(N_{x}^{2} - N_{y}^{2})-2T_{y}N_{x} N_{y}]$\\$N_{z}\,[2T_{x}N_{x} N_{y}+T_{y}(N_{x}^{2} - N_{y}^{2})]$\\$N_{x}T_{z}(N_{x}^{2} - 3 N_{y}^{2})$\\$N_{y}T_{z}(3 N_{x}^{2} - N_{y}^{2})$\end{tabular} & \begin{tabular}[c]{@{}l@{}}$b_{1}\!\begin{pmatrix}N_{z}(N_{x}^{2} - N_{y}^{2})\\-2N_{x} N_{y} N_{z}\\0\end{pmatrix}+b_{2}\!\begin{pmatrix}2N_{x} N_{y} N_{z}\\N_{z}(N_{x}^{2} - N_{y}^{2})\\0\end{pmatrix}$\\$+\tilde{b}_{3}\!\begin{pmatrix}0\\0\\N_{x}(N_{x}^{2} - 3 N_{y}^{2})\end{pmatrix}+\tilde{b}_{4}\!\begin{pmatrix}0\\0\\N_{y}(3 N_{x}^{2} - N_{y}^{2})\end{pmatrix}$\end{tabular} \\
\noalign{\vskip 4pt}
$\mathtt{\bar{6}}$ & $A^{\prime\prime}$ & $1$ & \begin{tabular}[c]{@{}l@{}}$N_{z}\,[T_{x}(N_{x}^{2} - N_{y}^{2})-2T_{y}N_{x} N_{y}]$\\$N_{z}\,[2T_{x}N_{x} N_{y}+T_{y}(N_{x}^{2} - N_{y}^{2})]$\\$N_{x}T_{z}(N_{x}^{2} - 3 N_{y}^{2})$\\$N_{y}T_{z}(3 N_{x}^{2} - N_{y}^{2})$\end{tabular} & \begin{tabular}[c]{@{}l@{}}$b_{1}\!\begin{pmatrix}N_{z}(N_{x}^{2} - N_{y}^{2})\\-2N_{x} N_{y} N_{z}\\0\end{pmatrix}+b_{2}\!\begin{pmatrix}2N_{x} N_{y} N_{z}\\N_{z}(N_{x}^{2} - N_{y}^{2})\\0\end{pmatrix}$\\$+b_{3}\!\begin{pmatrix}0\\0\\N_{x}(N_{x}^{2} - 3 N_{y}^{2})\end{pmatrix}+b_{4}\!\begin{pmatrix}0\\0\\N_{y}(3 N_{x}^{2} - N_{y}^{2})\end{pmatrix}$\end{tabular} \\
\noalign{\vskip 4pt}
$\mathtt{6/m}$ & $B_g$ & $4$ & \begin{tabular}[c]{@{}l@{}}$N_{z}\,[T_{x}(N_{x}^{2} - N_{y}^{2})-2T_{y}N_{x} N_{y}]$\\$N_{z}\,[2T_{x}N_{x} N_{y}+T_{y}(N_{x}^{2} - N_{y}^{2})]$\\$N_{x}T_{z}(N_{x}^{2} - 3 N_{y}^{2})$\\$N_{y}T_{z}(3 N_{x}^{2} - N_{y}^{2})$\end{tabular} & \begin{tabular}[c]{@{}l@{}}$b_{1}\!\begin{pmatrix}N_{z}(N_{x}^{2} - N_{y}^{2})\\-2N_{x} N_{y} N_{z}\\0\end{pmatrix}+b_{2}\!\begin{pmatrix}2N_{x} N_{y} N_{z}\\N_{z}(N_{x}^{2} - N_{y}^{2})\\0\end{pmatrix}$\\$+b_{3}\!\begin{pmatrix}0\\0\\N_{x}(N_{x}^{2} - 3 N_{y}^{2})\end{pmatrix}+b_{4}\!\begin{pmatrix}0\\0\\N_{y}(3 N_{x}^{2} - N_{y}^{2})\end{pmatrix}$\end{tabular} \\
\noalign{\vskip 4pt}
$\mathtt{622}$ & $B_{1}$ & $3$ & \begin{tabular}[c]{@{}l@{}}$N_{z}\,[T_{x}(N_{x}^{2} - N_{y}^{2})-2T_{y}N_{x} N_{y}]$\\$N_{x}T_{z}(N_{x}^{2} - 3 N_{y}^{2})$\end{tabular} & $b_{1}\!\begin{pmatrix}N_{z}(N_{x}^{2} - N_{y}^{2})\\-2N_{x} N_{y} N_{z}\\0\end{pmatrix}+\tilde{b}_{2}\!\begin{pmatrix}0\\0\\N_{x}(N_{x}^{2} - 3 N_{y}^{2})\end{pmatrix}$ \\
\noalign{\vskip 4pt}
 & $B_{2}$ & $3$ & \begin{tabular}[c]{@{}l@{}}$N_{z}\,[2T_{x}N_{x} N_{y}+T_{y}(N_{x}^{2} - N_{y}^{2})]$\\$N_{y}T_{z}(3 N_{x}^{2} - N_{y}^{2})$\end{tabular} & $b_{1}\!\begin{pmatrix}2N_{x} N_{y} N_{z}\\N_{z}(N_{x}^{2} - N_{y}^{2})\\0\end{pmatrix}+\tilde{b}_{2}\!\begin{pmatrix}0\\0\\N_{y}(3 N_{x}^{2} - N_{y}^{2})\end{pmatrix}$ \\
\noalign{\vskip 4pt}
$\mathtt{6mm}$ & $B_1$ & $3$ & \begin{tabular}[c]{@{}l@{}}$N_{z}\,[2T_{x}N_{x} N_{y}+T_{y}(N_{x}^{2} - N_{y}^{2})]$\\$N_{y}T_{z}(3 N_{x}^{2} - N_{y}^{2})$\end{tabular} & $b_{1}\!\begin{pmatrix}2N_{x} N_{y} N_{z}\\N_{z}(N_{x}^{2} - N_{y}^{2})\\0\end{pmatrix}+\tilde{b}_{2}\!\begin{pmatrix}0\\0\\N_{y}(3 N_{x}^{2} - N_{y}^{2})\end{pmatrix}$ \\
\noalign{\vskip 4pt}
 & $B_2$ & $3$ & \begin{tabular}[c]{@{}l@{}}$N_{z}\,[T_{x}(N_{x}^{2} - N_{y}^{2})-2T_{y}N_{x} N_{y}]$\\$N_{x}T_{z}(N_{x}^{2} - 3 N_{y}^{2})$\end{tabular} & $b_{1}\!\begin{pmatrix}N_{z}(N_{x}^{2} - N_{y}^{2})\\-2N_{x} N_{y} N_{z}\\0\end{pmatrix}+\tilde{b}_{2}\!\begin{pmatrix}0\\0\\N_{x}(N_{x}^{2} - 3 N_{y}^{2})\end{pmatrix}$ \\
\noalign{\vskip 4pt}
$\mathtt{\bar{6}m2}$ & $A_2^{\prime}$ & $4$ & \begin{tabular}[c]{@{}l@{}}$N_{z}\,[T_{x}(N_{x}^{2} - N_{y}^{2})-2T_{y}N_{x} N_{y}]$\\$N_{x}T_{z}(N_{x}^{2} - 3 N_{y}^{2})$\end{tabular} & $b_{1}\!\begin{pmatrix}N_{z}(N_{x}^{2} - N_{y}^{2})\\-2N_{x} N_{y} N_{z}\\0\end{pmatrix}+b_{2}\!\begin{pmatrix}0\\0\\N_{x}(N_{x}^{2} - 3 N_{y}^{2})\end{pmatrix}$ \\
\noalign{\vskip 4pt}
 & $A_2^{\prime\prime}$ & $1$ & \begin{tabular}[c]{@{}l@{}}$N_{z}\,[2T_{x}N_{x} N_{y}+T_{y}(N_{x}^{2} - N_{y}^{2})]$\\$N_{y}T_{z}(3 N_{x}^{2} - N_{y}^{2})$\end{tabular} & $b_{1}\!\begin{pmatrix}2N_{x} N_{y} N_{z}\\N_{z}(N_{x}^{2} - N_{y}^{2})\\0\end{pmatrix}+\tilde{b}_{2}\!\begin{pmatrix}0\\0\\N_{y}(3 N_{x}^{2} - N_{y}^{2})\end{pmatrix}$ \\
\noalign{\vskip 4pt}
$\mathtt{\bar{6}2m}$ & $A_1^{\prime\prime}$ & $4$ & \begin{tabular}[c]{@{}l@{}}$N_{z}\,[2T_{x}N_{x} N_{y}+T_{y}(N_{x}^{2} - N_{y}^{2})]$\\$N_{y}T_{z}(3 N_{x}^{2} - N_{y}^{2})$\end{tabular} & $b_{1}\!\begin{pmatrix}2N_{x} N_{y} N_{z}\\N_{z}(N_{x}^{2} - N_{y}^{2})\\0\end{pmatrix}+b_{2}\!\begin{pmatrix}0\\0\\N_{y}(3 N_{x}^{2} - N_{y}^{2})\end{pmatrix}$ \\
\noalign{\vskip 4pt}
 & $A_2^{\prime\prime}$ & $1$ & \begin{tabular}[c]{@{}l@{}}$N_{z}\,[T_{x}(N_{x}^{2} - N_{y}^{2})-2T_{y}N_{x} N_{y}]$\\$N_{x}T_{z}(N_{x}^{2} - 3 N_{y}^{2})$\end{tabular} & $b_{1}\!\begin{pmatrix}N_{z}(N_{x}^{2} - N_{y}^{2})\\-2N_{x} N_{y} N_{z}\\0\end{pmatrix}+\tilde{b}_{2}\!\begin{pmatrix}0\\0\\N_{x}(N_{x}^{2} - 3 N_{y}^{2})\end{pmatrix}$ \\
 $\mathtt{6/mmm}$ & $B_{1g}$ & $4$ & \begin{tabular}[c]{@{}l@{}}$N_{z}\,[T_{x}(N_{x}^{2} - N_{y}^{2})-2T_{y}N_{x} N_{y}]$\\$N_{x}T_{z}(N_{x}^{2} - 3 N_{y}^{2})$\end{tabular} & $b_{1}\!\begin{pmatrix}N_{z}(N_{x}^{2} - N_{y}^{2})\\-2N_{x} N_{y} N_{z}\\0\end{pmatrix}+b_{2}\!\begin{pmatrix}0\\0\\N_{x}(N_{x}^{2} - 3 N_{y}^{2})\end{pmatrix}$ \\
\noalign{\vskip 4pt}
 & $B_{2g}$ & $4$ & \begin{tabular}[c]{@{}l@{}}$N_{z}\,[2T_{x}N_{x} N_{y}+T_{y}(N_{x}^{2} - N_{y}^{2})]$\\$N_{y}T_{z}(3 N_{x}^{2} - N_{y}^{2})$\end{tabular} & $b_{1}\!\begin{pmatrix}2N_{x} N_{y} N_{z}\\N_{z}(N_{x}^{2} - N_{y}^{2})\\0\end{pmatrix}+b_{2}\!\begin{pmatrix}0\\0\\N_{y}(3 N_{x}^{2} - N_{y}^{2})\end{pmatrix}$ \\
%\noalign{\vskip 4pt}
\hline\hline
\end{tabular*}
\end{table*}

\begin{table*}[tp]
\addtocounter{table}{-1}
\caption{(Continued.) Cases where the leading order XMCD response is cubic. Columns as in Table~\ref{tab:LT_linear_symm}.}
\centering\footnotesize\renewcommand{\arraystretch}{1.25}
\begin{tabular*}{\textwidth}{@{\extracolsep{\fill}}l l c l l@{}}
\hline\hline
$\mathtt{F}$ & $\Gamma_{\mathbf{N}}$ & $n$ & $\bm{T}$--$\mathbf{N}$ coupling & $\bm{h}(\mathbf{N})$ \\
\hline
$\mathtt{432}$ & $A_2$ & $3$ & \begin{tabular}[c]{@{}l@{}}$T_{x}N_{x}(N_{y}^{2} - N_{z}^{2})-T_{y}N_{y}(N_{x}^{2} - N_{z}^{2})+T_{z}N_{z}(N_{x}^{2} - N_{y}^{2})$\end{tabular} & $b\!\begin{pmatrix}N_{x}(N_{y}^{2} - N_{z}^{2})\\-N_{y}(N_{x}^{2} - N_{z}^{2})\\N_{z}(N_{x}^{2} - N_{y}^{2})\end{pmatrix}$ \\
\noalign{\vskip 4pt}
$\mathtt{\bar{4}3m}$ & $A_2$ & $6$ & \begin{tabular}[c]{@{}l@{}}$T_{x}N_{x}(N_{y}^{2} - N_{z}^{2})-T_{y}N_{y}(N_{x}^{2} - N_{z}^{2})+T_{z}N_{z}(N_{x}^{2} - N_{y}^{2})$\end{tabular} & $b\!\begin{pmatrix}N_{x}(N_{y}^{2} - N_{z}^{2})\\-N_{y}(N_{x}^{2} - N_{z}^{2})\\N_{z}(N_{x}^{2} - N_{y}^{2})\end{pmatrix}$ \\
\noalign{\vskip 4pt}
$\mathtt{m\bar{3}m}$ & $A_{2g}$ & $6$ & \begin{tabular}[c]{@{}l@{}}$T_{x}N_{x}(N_{y}^{2} - N_{z}^{2})-T_{y}N_{y}(N_{x}^{2} - N_{z}^{2})+T_{z}N_{z}(N_{x}^{2} - N_{y}^{2})$\end{tabular} & $b\!\begin{pmatrix}N_{x}(N_{y}^{2} - N_{z}^{2})\\-N_{y}(N_{x}^{2} - N_{z}^{2})\\N_{z}(N_{x}^{2} - N_{y}^{2})\end{pmatrix}$ \\
\hline\hline
\end{tabular*}
\end{table*}

\FloatBarrier
\twocolumngrid

\end{document}